\documentclass{article}

\usepackage[final]{ProbNum26}

\usepackage{microtype}
\usepackage{graphicx}
\usepackage{subfigure}
\usepackage{subcaption}
\usepackage{booktabs}
\usepackage{multirow}
\usepackage{hyperref}    
\usepackage{tabularx}
\usepackage{csquotes}
\usepackage{adjustbox}
\usepackage{url}
\usepackage{amsmath, amssymb, amsfonts, mathtools}
\usepackage{pdflscape}
\usepackage{bm}
\usepackage{subcaption}
\usepackage{xcolor}
\usepackage{float}
\usepackage{placeins}
\usepackage{natbib}

\usepackage[capitalise,nameinlink]{cleveref}

\usepackage{kantlipsum}

\probnumtitle{Modelling Gas-Phase Reaction Kinetics
with Guided Particle Diffusion Sampling}
\probnumauthors{%
\name{Andrew Millard}%
\affiliation{Division of Statistics and Machine Learning (STIMA), Linköping University, Sweden}
\and%
\name{Zheng Zhao}%
\affiliation{STIMA, Linköping University, Sweden}
\and%
\name{Henrik Pedersen}%
\affiliation{Division of Chemistry, Linköping University, Sweden}
}

\probnumabstract{
Physics-guided sampling with diffusion priors has recently shown strong performance in solving complex systems of partial differential equations (PDEs) from sparse observations. However, these methods are typically evaluated on benchmark problems that do not fully demonstrate their ability to generate temporally consistent solutions of time-dependent PDEs, often focusing instead on reconstructing a single snapshot. In this work, we apply these methods to gas-phase reaction kinetics problems governed by the advection-reaction-diffusion (ARD) equation, providing a setting that more closely reflects realistic laboratory experiments. We demonstrate that guided sampling can be used to reconstruct full spatiotemporal trajectories, rather than isolated states. Furthermore, we show that these methods generalise to previously unseen parameter regimes, highlighting their potential for real-world applications.

}

\begin{document}

\section{Introduction and Relevant Work} \label{sec:introduction}

Studies of chemical kinetics \citep{atkinson1984kinetics}, i.e., the rates at which chemical reactions proceed, are fundamental to understanding both chemical processes and their underlying mechanisms. By quantifying how fast reactions occur and how they respond to environmental conditions, kinetic models enable the prediction and control of complex chemical systems. A well-known example is the role of reaction kinetics in explaining the depletion of the ozone layer and the formation of ozone holes over the polar regions. Chemical kinetics is also central to catalysis and has played a key role in enabling large-scale processes that support global food production. More broadly, it underpins a wide range of applications, from modelling atmospheric chemistry \citep{herrmann2003kinetics} to simulating chemical vapour deposition (CVD) processes \citep{danielsson2020systematic}, a widely used technique for producing thin films in electronic devices and protective coatings.

Given a set of reaction mechanisms and their constituent steps, the rate of each participating species (molecule, atom, ion, or radical) can be expressed, via kinetic rate theory, as a function of species concentrations and reaction rate constants \citep{atkins2023atkins}. The resulting set of reaction rate expressions forms a system of differential equations, the solution of which describes the temporal evolution of species concentrations. Classical numerical methods for solving partial differential equations (PDEs), such as finite element methods~\citep[FEMs,][]{monk2003finite, kreiss1974finite}, finite difference methods~\citep[FDMs,][]{leveque2007finite} and spectral methods \citep{trefethen1996finite} provide reliable approximations to these systems, but can become computationally expensive as the dynamics grow more complex or the dimensionality increases. Therefore, application of these methods to environments requiring high-resolution simulations or studies requiring exploration of large parameter spaces can become prohibitively expensive.

This motivates the exploration of machine learning methods as potential surrogates for classical numerical solvers. Machine learning approaches have been used to model reaction rates \citep{doi:10.1021/acs.chemrev.1c00033}, photochemical reactions \cite{Staszak+2023+51+72}, and reaction networks \citep{stocker2020machine}. However, these approaches typically focus on modelling specific components of gas-phase chemistry, rather than providing a holistic simulation of the full chemical reaction process. End-to-end machine learning approaches have, however, seen significant success in other domains such as structural biology \citep{jumper2021highly, abramson2024accurate}, where models take an amino acid sequence as input and directly predict the corresponding three-dimensional protein structure. The success of such approaches, particularly the inclusion of generative components to model complex, high-dimensional structures, as incorporated in \citet{abramson2024accurate}, suggests that similar techniques may be promising for end-to-end modelling of PDE-governed systems.

Recent advances in deep generative models, including diffusion models \citep{DBLP:journals/corr/abs-2505-17351, yang2023denoising}, have demonstrated strong capability in approximating the spatiotemporal structure of complex PDE systems. Furthermore, guided sampling approaches \citep{huang2024diffusionpdegenerativepdesolvingpartial, millard2026particleguideddiffusionmodelspartial, yao2025guided, jacobsen2025cocogen} incorporate physical constraints such as the governing PDE, directly into the sampling process by incorporating it as part of a likelihood term, thereby steering generated solutions toward physically consistent states. However, most existing work in this area focuses on idealised benchmarks  \cite{millard2026particleguideddiffusionmodelspartial, huang2024diffusionpdegenerativepdesolvingpartial, yao2025guided}. 
It is yet unclear whether they can apply to practical and realistic laboratory scenarios, and this is our main focus here. 

Our contributions to this field are threefold: 
\begin{itemize}
    \item We present a series of experiments designed to evaluate whether guided sampling techniques generalize effectively to chemistry scenarios by evaluating our methods on simulations closer to real-world scenarios.
    \item We demonstrate that guided diffusion-based methods remain effective under these more complex and physically relevant conditions, and generalise well to previously unseen parameter regimes. 
    \item We show that guided sampling from diffusion priors can capture complex chemical reaction dynamics across a broad range of temporal scales and accurately reconstruct spatiotemporal chemical species concentration fields from sparse observations.
\end{itemize}

The remainder of this paper adopts a mathematical structure similar to that of \citet{millard2026particleguideddiffusionmodelspartial}, with closely aligned notation and presentation.

\section{Background} \label{sec:background}

We begin by a generic class of time-dependent PDEs and then in this section we explain how a generative diffusion model can be applied to numerically solve them. 

For a spatial domain $\Omega$ and time horizon $[0,\mathcal{T}]$, a general time-dependent PDE can be written as
\begin{equation}
    \begin{split}
        f(u, c, \tau,a) &= 0, \quad c \in \Omega,\quad \tau \in [0,\mathcal{T}], \\
        u(c,\tau) &= g(c,\tau), \quad c \in \partial \Omega,\quad \tau \in [0,\mathcal{T}], \\
        u(c,0) &= a(c),
    \end{split}
\end{equation}
where $c$ denotes the spatial coordinate, $\tau$ denotes physical (PDE) time (distinct from diffusion time $t$), and $u \in \mathcal{U}$ is the PDE solution field. The function $a \in \mathcal{A}$ specifies the initial condition (or, more generally, fixed PDE parameters), and $g$ denotes the boundary conditions.

Throughout this work, the abstract solution field $u(c,\tau)$ represents a spatiotemporal physical field. In later sections \ref{sec:methodology} we denote this field as a concentration field $\mathcal{C}(c,\tau)$ governed by a reaction-transport PDE. The notation $\mathcal{C}(\cdot,\tau)$ refers to the spatial concentration field over $\Omega$ at a fixed PDE time $\tau$.

\textbf{Sampling for Generative Diffusion Models} 

Generative diffusion models \citep{sohl2015deep, song2021scorebasedgenerativemodelingstochastic, ho2020denoising, cao2024survey, karras2022elucidatingdesignspacediffusionbased} provide a probabilistic framework for sampling from complex, high-dimensional data distributions $p_{\mathrm{data}}(x)$. 

The forward process progressively corrupts $p_{\mathrm{data}}(x)$ with Gaussian noise according to a predetermined noise schedule $\sigma_t \in [0,\sigma_{\max}]$ \citep{song2021scorebasedgenerativemodelingstochastic, karras2022elucidatingdesignspacediffusionbased, ho2020denoising}, $t \in [0,T]$, yielding a reference distribution $p(x_T) \approx \mathcal{N}(0,\sigma_{\max}^2 I)$. Sampling proceeds by drawing $x_T$ from this reference distribution and reversing the noising process to obtain $x_0 \sim p_{\mathrm{data}}(x)$. 

The probability flow ordinary differential equation (ODE) describes how a sample drawn from the reference distribution can be gradually denoised in reverse time \citep{karras2022elucidatingdesignspacediffusionbased, song2021scorebasedgenerativemodelingstochastic} in order to recover a sample from the data distribution:
\begin{equation}
    d x_t = -\dot{\sigma}_t \sigma_t \nabla_{x_t} \log p_t(x_t, \sigma_t)\, dt, \quad t\in[0, T], 
    \label{eq:prob_ODE_noise}
\end{equation}
where $\log p_t(x_t, \sigma_t)$ is the score function. \citet{karras2022elucidatingdesignspacediffusionbased} propose learning a denoising function $\delta_\theta(x_t,\sigma_t)$ so that the score function can be estimated as
\begin{equation}
    \nabla_{x_t} \log p_t(x_t, \sigma_t) \approx 
    \nabla_{x_t} \log p_t^\theta(x_t,\sigma_t)
    =
    \frac{x_t - \delta_\theta(x_t,\sigma_t)}{\sigma_t^2}, 
\end{equation}
where $\delta_\theta$ is a neural network trained to denoise the sample at a given noise level $\sigma_t$. This reverse-time denoising process can also be simulated using a corresponding stochastic differential equation (SDE) which recovers the same marginal distributions as Equation~\eqref{eq:prob_ODE_noise} \citep{song2021scorebasedgenerativemodelingstochastic}:
\begin{equation}
    d x_t
    =
    -2 \dot{\sigma}_t \sigma_t \nabla_x \log p_t(x_t,\sigma_t)\, dt
    + \sqrt{2 \dot{\sigma}_t \sigma_t}\, dW_t,
    \label{eq:reverse_sde_noise_form}
\end{equation}
where $W_t$ is a Brownian motion and $\dot{\sigma}_t$ denotes the time derivative of the noise schedule.

\textbf{Guided Sampling} 

We consider the setting in which physical parameters are fixed and known, and a neural network is trained over realisations of the PDE solution field. This neural network represents the diffusion model prior $p_\theta(x)$, and diffusion samples therefore take the form
\begin{equation}
    x \in \mathcal{X} = \mathcal{U},
\end{equation}
where $x$ represents a discretised realisation of the PDE solution field, which in this work corresponds to a concentration field at a single physical time. Sampling from the diffusion model by solving Equation~\eqref{eq:prob_ODE_noise} or \eqref{eq:reverse_sde_noise_form} corresponds to sampling from the prior distribution without incorporating additional information, such as the governing PDE equation or sparse observations. 

To incorporate such information, we condition the diffusion process on auxiliary data $y$. In this case, we aim to sample from the posterior distribution
\begin{equation}
    p_\theta(x \mid y) \propto p(y \mid x)\, p_\theta(x). 
    \label{eq:cond_posterior}
\end{equation}

When using the reverse-time SDE, sampling from this distribution can be achieved by modifying Equation~\eqref{eq:reverse_sde_noise_form} as
\begin{equation}
    \begin{split}
        dx_t 
        &= -2\dot{\sigma}_t \sigma_t \nabla_{x_t} \log p_t^\theta(x_t, \sigma_t)\,dt \\
        &-2\dot{\sigma}_t \sigma_t \nabla_{x_t} \log p_t^\theta(y \mid x_t, \sigma_t)\,dt 
        + \sqrt{2 \dot{\sigma}_t \sigma_t} \, dW_t,
        \label{eq:guidance_sde_noise}
    \end{split}
\end{equation}
where the additional gradient term is known as the \emph{guidance term}. This term encourages samples to be consistent with the conditional information, which in our setting includes sparse observations and the governing PDE dynamics.

\textbf{Sequential Monte Carlo}

Sequential Monte Carlo (SMC) methods~\citep[e.g.,][]{chopin2020introduction,NaessethLS:2019a} provide a principled method for sampling from a sequence of distributions and therefore, are well suited for the sequential denoising process associated with generative diffusion models. Previous work explores the use of SMC as a framework for conditional sampling of diffusion priors \citep{wu2023-TDS, cardoso2023monte, stevens2025sequential, dou2024diffusion, zhao2025conditional} for problems such as inpainting \citep{pmlr-v267-ekstrom-kelvinius25b}, motif-scaffolding for proteins \citep{Trippe2023diffusion} and many others. In previous subsections, we have used $t$ when discussing continuous time processes. In the following sections, we discuss simulating these processes in discrete time and therefore denote this discrete time as $k$. 

SMC simulates propagating an ensemble of $N$ particles $\{x_k^{(i)}\}_{i=1}^N$ across $k=0,1,\ldots, K$ iterations/time steps. In this setting, the particles are first sampled from the diffusion prior $M_K$ and weighted according to an initial potential $G_K$: 
\begin{equation}
    x^{(i)}_K \sim M_K = p(x_K) \approx \mathcal{N}(0, I), \quad 
    w^{(i)}_K \propto G_K(x^{(i)}_K). 
    \label{equ:init_weight}
\end{equation}

For $K$ iterations, samples are propagated via a Markov kernel/proposal distribution $M_{k-1}$
\begin{equation}
    x^{(i)}_{k-1} \sim M_{k-1}(x^{(i)}_{k-1} \mid x^{(i)}_{k}), 
\end{equation}
and reweighted
\begin{equation}
    w^{(i)}_{k-1} \propto w^{(i)}_{k} G_{k-1}(x^{(i)}_k, x^{(i)}_{k-1}). 
\end{equation}

We resample  \citep{douc2005comparison} the particles if our effective sample size (ESS) drops below a certain threshold $N_{\text{eff}}$. At each iteration $k$, we have a weighted ensemble $\{(x_k^{(i)}, w_k^{(i)}) \}_{i=1}^N$ that provides an approximation of the target distribution $\nu$
\begin{align}
    \nu_k(x_{k:K}) &\propto G_K(x_K) M_K(x_K) \nonumber \\
    &\times\prod_{j={k+1}}^K G_{j-1}(x_j, x_{j-1}) M_{j-1}(x_{j-1} \mid x_{j}).
    \label{eq:FK_formula}
\end{align}

Equation~\eqref{eq:FK_formula} is known as the Feynman-Kac (FK) formula and describes the time evolution of stochastic processes. We use this SMC framework in order to sample from the posterior distribution given by Equation~\eqref{eq:cond_posterior}. To do this, we choose proposals and potential functions $\{M_k, G_k \}_{k=0}^K$  such that the final time \emph{marginal} distribution is the conditional target distribution i.e. $\nu_0(x_0) = p_\theta(x_0 \mid y)$. Appendix~\ref{app:smc_psuedo} gives the pseudocode for the SMC framework. 

\textbf{Stochastic Sampling}

We approximate the reverse process described by Equation~\eqref{eq:reverse_sde_noise_form} using a guided Euler-Maruyama (GEM) update \citep{millard2026particleguideddiffusionmodelspartial} 
\begin{align}
    x_{k-1}
    =
    x_k
    +
    (\sigma_{k-1}^2 - \sigma_k^2)
    \frac{x_k - \delta_\theta(x_k,\sigma_k)}{\sigma_k^2} \nonumber \\
    +
    (\sigma_{k-1}^2 - \sigma_k^2)
    \nabla_{x_k} \log \tilde{p}_\theta(y \mid x_k)
    +
    \sqrt{\sigma_{k-1}^2 - \sigma_k^2}\, \zeta,
    \label{eq:EM_denoiser_guided_constant}
\end{align}
with $\zeta \sim \mathcal{N}(0,I)$ and $\tilde{p}_\theta$ denotes a likelihood function whose formulation is detailed in later sections. The time interval $\Delta t$ is implicitly accounted for in the noise scheduler. The pseudocode for this can be found in Appendix~\ref{app:sto_solvers}, Algorithm~\ref{alg:gem_proposal}. 

\textbf{Twisted SMC} 

The GEM update in Equation~\eqref{eq:EM_denoiser_guided_constant} gives a Markov kernel $\Tilde{p}_{\theta}(x_{k-1} \mid x_{k}, y) = \mathcal{N}(x_k \mid \mu_{\text{prop}}, \, (\sigma_{k-1}^2 - \sigma_{k}^2) \,  I)$ which can be used to propagate samples within SMC. If we define our proposal distribution $M_{k-1}(x_{k-1} \mid x_{k}) = \Tilde{p}_{\theta}(x_{k-1} \mid x_{k}, y)$ and the corresponding weighting function as
\begin{align}
    G_{k-1}(x_{k}, x_{k-1}) =  \,
    \frac{\Tilde{p}_{\theta}(y \mid x_{k-1})}{\Tilde{p}_{\theta}(y \mid x_{k})} \,
    \frac{p_\theta(x_{k-1} \mid x_{k})}{\Tilde{p}_{\theta}(x_{k-1} \mid x_{k}, y)}, 
    \label{eq:weight_update_full} 
\end{align}
then this corresponds to the SMC framework used in the Twisted Diffusion Sampler~\citep[TDS,][]{wu2023-TDS}. We can verify that these choices of potentials and proposals ensure that we target the correct final time marginal by substituting them into the FK formula given by Equation~\eqref{eq:FK_formula}: 
\begin{align}
    \nu_{k-1}(x_{k-1:K}) &\propto \Tilde{p}_{\theta}(y \mid x_{k-1}) p_\theta(x_{k-1:K}) \nonumber \\
    \nu_{0:K}(x_{0:K}) &= p_\theta(x_{0:K} \mid y).
    \label{eq:FK_target_normal}
\end{align}

\textbf{Second-Order Stochastic Sampling}

\citet{karras2022elucidatingdesignspacediffusionbased} propose using second-order stochastic sampling process. First, the noise scale is jittered and then the ODE dynamics via the denoiser are run, i.e.\ we run the following set of
equations:
\begin{align}
    &\hat{\sigma}_{k} \leftarrow \sigma_{k} + \gamma_{k}\,\sigma_{k}, \\
    &\hat{x}_{k} = x_{k} + \sqrt{\hat{\sigma}^{2}_{k} - \sigma^{2}_{k}}\,\psi,
       \quad \psi \sim \mathcal{N}(0, I), \\
    &x_{k-1} = \hat{x}_{k} + \bigl(\sigma^{2}_{k-1} - \hat{\sigma}^{2}_{k}\bigr)
       \frac{\hat{x}_{k} - \delta_{\theta}(\hat{x}_{k}, \hat{\sigma}_{k})}{\hat{\sigma}^{2}_{k}},
       \label{eq:edm_original_sub}
\end{align}
where $\gamma_{k}$ is a constant used to reach a higher noise level and a second order correction is used to improve sampling quality. The full details of this can be found in \citet{karras2022elucidatingdesignspacediffusionbased}. \citet{millard2026particleguideddiffusionmodelspartial} propose including guidance information after the second order correction such that:
\begin{align}
    x_{k-1} &= \hat{x}_{k} + \bigl(\sigma^{2}_{k-1} - \hat{\sigma}^{2}_{k}\bigr)
    \frac{\hat{x}_{k} - \delta_{\theta}(\hat{x}_{k}, \hat{\sigma}_{k})}{\hat{\sigma}^{2}_{k}} \nonumber \\
    & - \alpha \nabla_{\hat{x}_{k}} \log \Tilde{p}_\theta(y \mid \hat{x}_{k}),
    \label{eq:SOSaG_proposal}
\end{align}
where $\alpha$ is a guidance weighting parameter. The resulting algorithm is called the Second-Order Stochastic Guided (SOSaG) update. The pseudocode for this algorithm is given in Appendix~\ref{app:sto_solvers}, Algorithm~\ref{alg:sosag_proposal}. Further details of this can be found in \citet{millard2026particleguideddiffusionmodelspartial}. 

\textbf{pBS SMC Framework} 

\citet{millard2026particleguideddiffusionmodelspartial} use the SOSaG proposal within a different SMC framework, the so-called pseudo-bootstrap (pBS) method. 

It is not obvious how to incorporate SOSaG as a proposal distribution within the TDS framework as the particles generated by it are not point-evaluable proposal density. The particles are jittered before the denoising process and therefore are not Gaussian distributed like the samples generated by GEM. Instead, they define a new FK model $\overline{\nu}_{0:K}$ using a similar structure to the boostrap particle filter \citep{gordon1993novel} 
\begin{equation*}
    \begin{split}
        \overline{M}_{k-1}(x_{k-1} \mid x_{k}) &= \overline{p}_\theta(x_{k-1} \mid x_{k}, y), \\
        \overline{G}_{k-1}(x_k, x_{k-1}) &= \frac{\Tilde{p}_{\theta}(y \mid x_{k-1})}{\Tilde{p}_{\theta}(y \mid x_{k})},
    \end{split}
\end{equation*}
where $\overline{p}_\theta(x_{k-1} \mid x_{k}, y)$ is the proposal distribution of the SOSaG method given in Equation~\eqref{eq:SOSaG_proposal}. This new FK model recovers a target distribution $\overline{\nu}_0(x_0) = \overline{p}_\theta(x_0 \mid y) \, p_{\theta}(y\mid x_0)$ at $k=0$. Although this does not target the same distribution as Equation~\eqref{eq:FK_target_normal}, it has been found to give more physically consistent results~\citep{millard2026particleguideddiffusionmodelspartial}. 
\section{Methodology} \label{sec:methodology}

\begin{figure*}
    \centering
    \includegraphics[width=\textwidth]{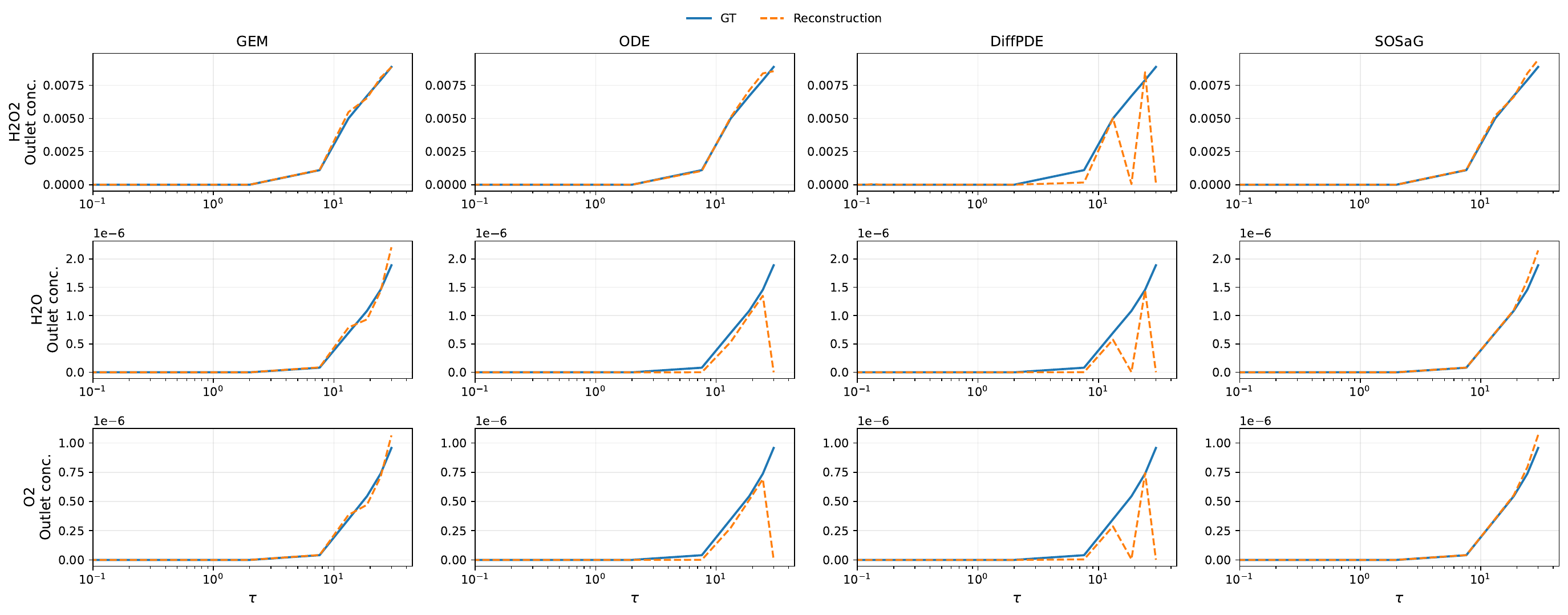}
    \caption{Species outlet comparison for H$_2$O$_2$ Decomposition.}
    \label{fig:outlet_comparison_h2o2}
\end{figure*}

We consider transport-reaction systems governed by advection-reaction-diffusion (ARD) equations \citep{RUBIO200890, ChrisCosner2014DiscreteandContinuousDynamicalSystems}. In this setting, a diffusion model sample (at diffusion time $k=0$)
\[
x \equiv \mathcal{C}(\cdot,\tau_n)
\]
represents a discretised concentration field at a given physical time $\tau_n \in [0, \mathcal{T}]$, corresponding to the solution of the underlying PDE $u(\cdot,\tau_n)$ introduced in Section~\ref{sec:background}. In this work, $\mathcal{C}$ denotes the vector of $S$ species concentrations, given by
\[
\mathcal{C} = (\mathcal{C}_1,\dots,\mathcal{C}_S).
\]

We focus on an axisymmetric cylindrical reactor of length $L$ and radius $A$ (see, Figure~\ref{fig:experiment_diagram}). The concentration of species $s$, denoted by $\mathcal{C}_s(r,z,\tau)$, evolves according to
\begin{equation}
\frac{\partial \mathcal{C}_s}{\partial \tau}
+ u(r)\,\frac{\partial \mathcal{C}_s}{\partial z}
=
D \nabla^2 \mathcal{C}_s
+ R_s(\mathcal{C}),
\label{eq:ard_general}
\end{equation}
where $r \in [0,A]$ and $z \in [0,L]$ denote the radial and axial coordinates, respectively. Here, $D$ is the molecular diffusivity and $R_s(\mathcal{C})$ represents the reaction source term \citep{doi:https://doi.org/10.1002/0471461296.ch3}.

The axial velocity field $u$ is assumed to follow a laminar parabolic profile,
\[
u(r) = 2U_{\mathrm{avg}}\!\left(1 - \left(\frac{r}{A}\right)^2\right),
\]
where $U_{\mathrm{avg}}$ denotes the cross-sectional average velocity, which is taken to be identical across all species.

In our experiments, all physical parameters of the ARD system are assumed to be known and fixed. Consequently, the inference task is restricted to recovering the spatiotemporal concentration fields $\lbrace \mathcal{C}_s \rbrace_{s=1}^S$.

%The specific chemical system and experimental configuration are described in Section~\ref{sec:experiments}.

\textbf{ARD Residual-Based Likelihood for Time-Series Reconstruction} 

A sequence of concentration fields is reconstructed by combining stochastic sampling proposals with the corresponding SMC framework \citep[see, e.g.,][]{wu2023-TDS, zhao2025conditional} with a diffusion prior applied independently at each physical time step $\tau_n$. For each $\tau_n$, the diffusion process is initialised from Gaussian noise and an SMC-guided denoising procedure is used to obtain a reconstruction of $\mathcal{C}(\cdot,\tau_n)$. During this process, candidate samples are evaluated according to a likelihood that enforces both agreement with sparse observations and consistency with the underlying ARD dynamics.

The likelihood is defined through a combination of observation and PDE residual terms. Specifically, the log-likelihood is given by
\begin{align}
\log p(y \mid x)
&=
-\frac{1}{\sigma_{\mathrm{obs}}}
\frac{1}{N_{\mathrm{obs}}}
\big\|\mathcal{C}_{\mathrm{obs}} - \mathcal{M} \odot x \big\|_2^2 \nonumber \\
&\;-\;
\frac{1}{\sigma_{\mathrm{pde}}}
\frac{1}{A L}
\big\|\mathcal{R}_n(\mathcal{C}; \mathcal{C}^{\tau_{n-1}})\big\|_2^2 + \text{const.},
\label{eq:likelihood_final}
\end{align}
where $\mathcal{M}$ is a binary mask indicating observed locations, $\mathcal{C}$ denotes the reconstruction of the physical concentration field corresponding to $x$, $\mathcal{C}_{\mathrm{obs}}$ represents sparse observations of the ground truth field and $\mathcal{R}_n$ is the PDE residual which is defined later in this section. The parameters $\sigma_{\mathrm{obs}}$ and $\sigma_{\mathrm{pde}}$ control the relative weighting of the observation and PDE terms, while $N_{\mathrm{obs}}$ denotes the number of observations.

Following \citet{wu2023-TDS}, the intermediate diffusion likelihood is approximated by evaluating the likelihood at the denoised estimate,
\[
p_t^\theta(y \mid x_t, \sigma_t)
\approx
p\!\left(y \mid x_0 = \delta_\theta(x_t,\sigma_t)\right)
=:
\tilde{p}_\theta(y \mid x_t).
\]

To quantify consistency with the governing dynamics, we introduce the spatial ARD operator
\begin{align}
\mathcal{F}(\mathcal{C})
&=
u(r)\,\frac{\partial \mathcal{C}}{\partial z}
-
D\left(
\frac{\partial^2 \mathcal{C}}{\partial z^2}
+
\frac{1}{r}\frac{\partial}{\partial r}
\left(
r\,\frac{\partial \mathcal{C}}{\partial r}
\right)
\right)
-
R(\mathcal{C}),
\end{align}
where $R(\mathcal{C})$ denotes the nonlinear reaction source term applied component-wise to the concentration vector $\mathcal{C}$. $R(\mathcal{C})$ is derived from the corresponding chemical reaction mechanism and kinetic rate laws. 

The PDE residual at discrete time points $\{\tau_n\}$, can be for example modelled by a first-order finite difference
\begin{align}
\mathcal{R}_n(\mathcal{C}^{\tau_n}; \mathcal{C}^{\tau_{n-1}})
&=
\frac{\mathcal{C}^{\tau_n} - \mathcal{C}^{\tau_{n-1}}}{\Delta \tau_n}
+
\mathcal{F}(\mathcal{C}^{\tau_n}), \nonumber \\
\Delta \tau_n &\coloneqq \tau_n - \tau_{n-1}, \quad n \geq 1.
\end{align}

Here, $\mathcal{C}^{\tau_{n-1}}$ denotes the reconstructed concentration field at the previous time step. At the initial time $\tau_0 = 0$, the concentration field is set to $\mathcal{C}^{\mathrm{init}} = 0$, corresponding to an initially empty reactor.
\section{Experimental Setup and Results} \label{sec:experiments}

\begin{figure}[H]
    \centering
    \includegraphics[width=\linewidth]{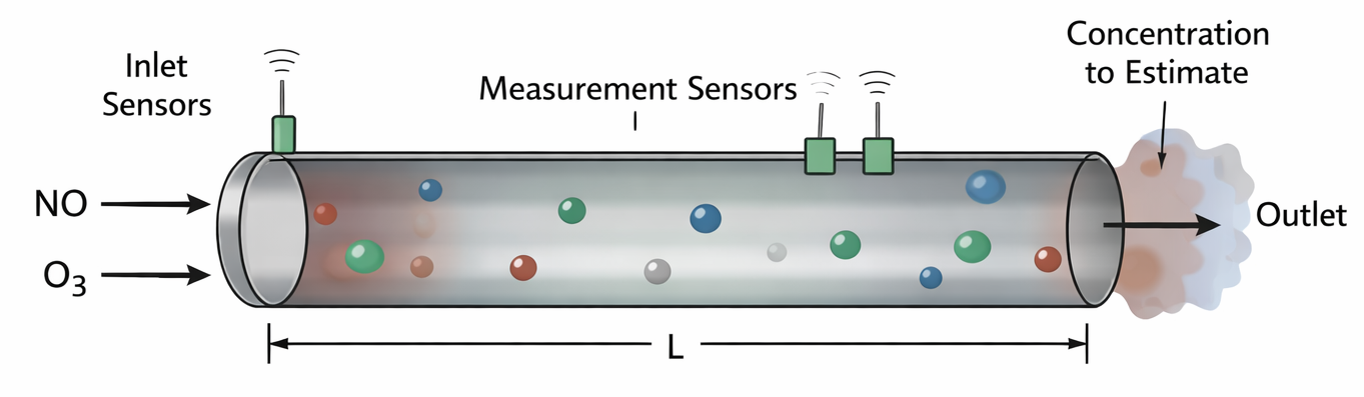}
    \caption{Diagram of the experimental set up being simulated.}
    \label{fig:experiment_diagram}
\end{figure}

The aim of these experiments is to simulate gas-phase reaction kinetics within a cylindrical tube reactor. Inlet gases enter from the left-hand side of the reactor, where they mix and react as they flow downstream toward the outlet on the right. Along the length of the reactor, sparse sensors provide local measurements of species concentrations. A schematic of this setup is shown in Figure~\ref{fig:experiment_diagram}.

The objective is twofold: first, to reconstruct the spatial distribution of chemical species concentrations throughout the reactor $\{\mathcal{C}_s(r,z,\tau)\}_{s=1}^{S}$, and second, to predict the temporal evolution of species concentrations at the reactor outlet $\{\mathcal{C}_s(r,L,\tau_n)\}_{s=1,\dots,S}^{n=0,\dots,\mathcal{T}}$. We compare against the first and second order ODE and stochastic methods metnioned previously. Both stochastic methods are used within an SMC framework. 

We consider six reduced reaction mechanisms spanning atmospheric chemistry, catalytic reaction systems, and simplified combustion kinetics.

\paragraph{NO-O$_3$ to NO$_2$.}
This mechanism represents a minimal atmospheric reaction system describing ozone consumption by nitric oxide:
\begin{equation}
\mathrm{NO + O_3 \rightarrow NO_2 + O_2}. \nonumber
\end{equation}
It provides a simple benchmark for nonlinear reactive transport with a single irreversible reaction.

\paragraph{Water-gas shift.}
This mechanism models the reversible water-gas shift reaction:
\begin{align}
\mathrm{CO + H_2O} &\mathrm{\rightarrow CO_2 + H_2}, \nonumber \\
\mathrm{CO_2 + H_2} &\mathrm{\rightarrow CO + H_2O}. \nonumber
\end{align}

\paragraph{Hydrogen peroxide decomposition.}
This mechanism describes the decomposition of hydrogen peroxide into water and oxygen:
\begin{equation}
\mathrm{2\,H_2O_2 \rightarrow 2\,H_2O + O_2}.
\end{equation}

\paragraph{Two-step methane oxidation.}
This mechanism approximates methane combustion using a reduced two-step description:
\begin{align}
\mathrm{2\,CH_4 + 3\,O_2} &\mathrm{\rightarrow 2\,CO + 4\,H_2O}, \nonumber \\
\mathrm{2\,CO + O_2} &\mathrm{\rightarrow 2\,CO_2}. \nonumber
\end{align}

\paragraph{Ammonia oxidation.}
This mechanism models a reduced form of ammonia oxidation:
\begin{align}
\mathrm{4\,NH_3 + 5\,O_2} &\mathrm{\rightarrow 4\,NO + 6\,H_2O}, \nonumber \\
\mathrm{2\,NO + O_2} &\mathrm{\rightarrow 2\,NO_2}. \nonumber
\end{align}

\paragraph{Hydrogen oxidation subset.}
This mechanism is a reduced radical subset of hydrogen combustion chemistry:
\begin{align}
\mathrm{H_2 + O_2} &\mathrm{\rightarrow 2\,OH}, \nonumber \\
\mathrm{H_2 + OH} &\mathrm{\rightarrow H_2O + H}, \nonumber \\
\mathrm{H + O_2} &\mathrm{\rightarrow HO_2}, \nonumber \\
\mathrm{HO_2 + H} &\mathrm{\rightarrow 2\,OH}. \nonumber
\end{align}

\textbf{Reaction Rates}

\begin{table*}[t]
\centering
\tiny
\caption{Gas-phase reaction kinetics full-field and outlet absolute errors across all experiments and methods (non-transformed / physical space). We give the mean $\pm$ one standard deviation, the best results are bolded.}
\label{tab:realchem_absolute_error_raw}
\begin{tabular}{cccccc}
\midrule
Method & RMSE & MAE & Outlet RMSE & Outlet MAE & Time (s) \\
\midrule
\multicolumn{6}{c}{\textbf{H$_2$O$_2$ Decomposition}} \\
\midrule
GEM & $(1.49 \pm 0.58) \times 10^{-4}$ & $(3.19 \pm 1.57) \times 10^{-5}$ & $(1.22 \pm 0.39) \times 10^{-4}$ & $(2.39 \pm 1.02) \times 10^{-5}$ & $(1.55 \pm 0.01) \times 10^{3}$\\
SOSaG & $\bm{(1.47 \pm 0.57) \times 10^{-4}}$ & $\bm{(3.10 \pm 1.49) \times 10^{-5}}$ & $\bm{(1.09 \pm 0.32) \times 10^{-4}}$ & $\bm{(2.17 \pm 0.81) \times 10^{-5}}$ & $(1.55 \pm 0.01) \times 10^{3}$  \\
ODE & $(3.29 \pm 1.66) \times 10^{-3}$ & $(5.30 \pm 3.38) \times 10^{-4}$ & $(2.28 \pm 1.20) \times 10^{-3}$ & $(3.98 \pm 2.48) \times 10^{-4}$ & $\bm{(1.46 \pm 0.02) \times 10^{3}}$ \\
DiffPDE & $(3.09 \pm 1.43) \times 10^{-3}$ & $(5.42 \pm 3.05) \times 10^{-4}$ & $(2.35 \pm 1.18) \times 10^{-3}$ & $(4.05 \pm 2.62) \times 10^{-4}$ & $(2.12 \pm 0.09) \times 10^{3}$ \\

\midrule

\multicolumn{6}{c}{\textbf{NO + O$_3$ $\rightarrow$ NO$_2$}} \\
\midrule
GEM & $(2.11 \pm 1.41) \times 10^{-4}$ & $(6.04 \pm 4.50) \times 10^{-5}$ & $\bm{(1.49 \pm 1.03) \times 10^{-4}}$ & $\bm{(4.19 \pm 3.45) \times 10^{-5}}$ & $(1.59 \pm 0.00) \times 10^{3}$ \\
SOSaG & $\bm{(2.03 \pm 1.39) \times 10^{-4}}$ & $\bm{(5.87 \pm 4.40) \times 10^{-5}}$ & $(1.57 \pm 1.26) \times 10^{-4}$ & $(4.77 \pm 4.61) \times 10^{-5}$ & $(2.36 \pm 0.01) \times 10^{3}$ \\
ODE & $(6.67 \pm 4.11) \times 10^{-3}$ & $(1.18 \pm 0.95) \times 10^{-3}$ & $(2.78 \pm 2.50) \times 10^{-3}$ & $(6.48 \pm 5.96) \times 10^{-4}$ & $\bm{(1.39 \pm 0.01) \times 10^{3}}$  \\
DiffPDE & $(3.32 \pm 2.62) \times 10^{-3}$ & $(9.07 \pm 7.77) \times 10^{-4}$ & $(1.73 \pm 1.36) \times 10^{-3}$ & $(5.10 \pm 4.21) \times 10^{-4}$ & $(2.07 \pm 0.01) \times 10^{3}$ \\

\midrule

\multicolumn{6}{c}{\textbf{Ammonia Oxidation}} \\
\midrule
GEM & $\bm{(3.48 \pm 2.17) \times 10^{-4}}$ & $\bm{(7.84 \pm 5.93) \times 10^{-5}}$ & $\bm{(4.14 \pm 2.97) \times 10^{-4}}$ & $\bm{(8.61 \pm 7.01) \times 10^{-5}}$ & $(1.58 \pm 0.00) \times 10^{3}$\\
SOSaG & $(3.88 \pm 2.44) \times 10^{-4}$ & $(9.39 \pm 7.39) \times 10^{-5}$ & $(4.35 \pm 3.14) \times 10^{-4}$ & $(9.67 \pm 8.57) \times 10^{-5}$ & $(2.40 \pm 0.01) \times 10^{3}$ \\
ODE & $(3.77 \pm 2.63) \times 10^{-3}$ & $(1.01 \pm 0.77) \times 10^{-3}$ & $(2.94 \pm 2.21) \times 10^{-3}$ & $(6.98 \pm 5.47) \times 10^{-4}$ & $\bm{(1.43 \pm 0.00) \times 10^{3}}$ \\
DiffPDE & $(3.77 \pm 2.65) \times 10^{-3}$ & $(9.91 \pm 7.57) \times 10^{-4}$ & $(2.82 \pm 2.09) \times 10^{-3}$ & $(6.60 \pm 5.10) \times 10^{-4}$ & $(2.10 \pm 0.01) \times 10^{3}$  \\

\midrule

\multicolumn{6}{c}{\textbf{Hydrogen Oxidation Subset}} \\
\midrule
GEM & $(2.51 \pm 3.51) \times 10^{-4}$ & $\bm{(3.59 \pm 4.35) \times 10^{-5}}$ & $\bm{(1.59 \pm 1.98) \times 10^{-4}}$ & $\bm{(3.24 \pm 4.79) \times 10^{-5}}$ & $(1.65 \pm 0.00) \times 10^{3}$ \\
SOSaG & $\bm{(2.21 \pm 2.85) \times 10^{-4}}$ & $(3.65 \pm 4.40) \times 10^{-5}$ & $(1.99 \pm 3.20) \times 10^{-4}$ & $(4.11 \pm 7.63) \times 10^{-5}$ & $(2.43 \pm 0.01) \times 10^{3}$\\
ODE & $(2.72 \pm 1.75) \times 10^{-3}$ & $(5.81 \pm 5.34) \times 10^{-4}$ & $(1.92 \pm 1.62) \times 10^{-3}$ & $(3.87 \pm 3.78) \times 10^{-4}$ & $\bm{(1.47 \pm 0.01) \times 10^{3}}$ \\
DiffPDE & $(2.65 \pm 1.59) \times 10^{-3}$ & $(5.49 \pm 4.48) \times 10^{-4}$ & $(1.78 \pm 1.22) \times 10^{-3}$ & $(3.83 \pm 3.38) \times 10^{-4}$ & $(2.14 \pm 0.01) \times 10^{3}$\\

\midrule

\multicolumn{6}{c}{\textbf{Two-step Methane Oxidation}} \\
\midrule
GEM & $\bm{(1.09 \pm 0.83) \times 10^{-4}}$ & $\bm{(2.40 \pm 1.75) \times 10^{-5}}$ & $\bm{(0.95 \pm 1.05) \times 10^{-4}}$ & $(1.93 \pm 1.98) \times 10^{-5}$ & $(1.59 \pm 0.00) \times 10^{3}$ \\
SOSaG & $(1.17 \pm 0.80) \times 10^{-4}$ & $(2.58 \pm 1.66) \times 10^{-5}$ & $(9.49 \pm 6.06) \times 10^{-5}$ & $\bm{(1.85 \pm 1.24) \times 10^{-5}}$ & $(2.41 \pm 0.00) \times 10^{3}$ \\
ODE & $(3.95 \pm 2.57) \times 10^{-3}$ & $(9.59 \pm 7.02) \times 10^{-4}$ & $(2.61 \pm 2.10) \times 10^{-3}$ & $(5.92 \pm 4.68) \times 10^{-4}$ & $\bm{(1.43 \pm 0.00) \times 10^{3}}$ \\
DiffPDE & $(3.41 \pm 2.78) \times 10^{-3}$ & $(9.30 \pm 7.31) \times 10^{-4}$ & $(2.37 \pm 1.94) \times 10^{-3}$ & $(5.42 \pm 4.17) \times 10^{-4}$ & $(2.11 \pm 0.00) \times 10^{3}$ \\

\midrule

\multicolumn{6}{c}{\textbf{Water Gas Shift}} \\
\midrule
GEM & $\bm{(7.47 \pm 5.25) \times 10^{-5}}$ & $(1.76 \pm 1.47) \times 10^{-5}$ & $\bm{(6.09 \pm 6.19) \times 10^{-5}}$ & $\bm{(1.48 \pm 1.63) \times 10^{-5}}$ & $(1.59 \pm 0.00) \times 10^{3}$ \\
SOSaG & $(7.61 \pm 4.96) \times 10^{-5}$ & $\bm{(1.76 \pm 1.39) \times 10^{-5}}$ & $(6.50 \pm 5.43) \times 10^{-5}$ & $(1.50 \pm 1.35) \times 10^{-5}$ & $(2.37 \pm 0.00) \times 10^{3}$ \\
ODE & $(5.96 \pm 4.08) \times 10^{-4}$ & $(7.93 \pm 6.29) \times 10^{-5}$ & $(3.26 \pm 3.50) \times 10^{-4}$ & $(5.92 \pm 5.62) \times 10^{-5}$ & $\bm{(1.40 \pm 0.00) \times 10^{3}}$ \\
DiffPDE & $(5.66 \pm 3.51) \times 10^{-4}$ & $(7.25 \pm 4.90) \times 10^{-5}$ & $(4.02 \pm 4.52) \times 10^{-4}$ & $(6.68 \pm 6.98) \times 10^{-5}$ & $(2.10 \pm 0.00) \times 10^{3}$ \\
\bottomrule
\end{tabular}
\end{table*}

A crucial coefficient in our ARD equation is the reaction rate constants. 
They are computed using Arrhenius kinetics,
\begin{equation}
\kappa_l(\mathcal{K}) = A_l \exp\!\left(-\frac{E_{a,l}}{R_{\mathrm{gas}}\mathcal{K}}\right),
\end{equation}
where $\mathcal{K}$ is the sampled temperature, $\kappa_l(\mathcal{K})$ is the temperature dependent rate constant for reaction $l$, $A_l$ is the pre-exponential factor, $E_{a,l}$ is the activation energy, and $R_{\mathrm{gas}}$ is the universal gas constant. 

Given a vector of species concentrations $\mathcal{C} = (\mathcal{C}_1,\dots,\mathcal{C}_S)$, the
mass-action kinetics defines the reaction rate of reaction $j$ as
\[
\chi_l(\mathcal{C}) = \kappa_l(\mathcal{K})\prod_{i=1}^{S} \mathcal{C}_i^{\alpha_{il}},
\]
where $\alpha_{il}$ denotes the stoichiometric coefficient of species $i$ as a reactant in reaction $l$. The nonlinear reaction source term $R(\mathcal{C})$ in the ARD system \eqref{eq:ard_general} is then obtained by combining the reaction rates with the corresponding product and reactant stoichiometry.

For example, in the NO-O$_3$ mechanism with species
\[
\mathcal{C} = (\mathcal{C}_{\mathrm{NO}}, \mathcal{C}_{\mathrm{O}_3}, \mathcal{C}_{\mathrm{NO}_2}, \mathcal{C}_{\mathrm{O}_2}),
\]
we consider a single reaction, and therefore write $r_l = r$. The reaction
\[
\mathrm{NO + O_3 \rightarrow NO_2 + O_2}
\]
has reaction rate
\[
\chi(\mathcal{C}) = \kappa(\mathcal{K})\, \mathcal{C}_{\mathrm{NO}} \mathcal{C}_{\mathrm{O}_3},
\]
which yields the source term
\[
R(\mathcal{C}) =
(-\chi(\mathcal{C}),-\chi(\mathcal{C}),\chi(\mathcal{C}),\chi(\mathcal{C})).
\]
Since only a single reaction is considered, NO and O$_3$ are consumed while NO$_2$ and O$_2$ are produced, so that the source term reflects the corresponding gain and loss of each species. 

\textbf{Data Generation}

Training and test data are generated by numerically solving the resulting ARD system in an axisymmetric cylindrical reactor on a structured $(r, z)$ grid of size $N_r \times N_z = 64\times 64$. For the advection and diffusion we discretised them via finite difference and the reaction terms were solved separately using operator splitting and explicit time integration. 

We leverage the axisymmetric structure of the reactor geometry and model only the radial and axial directions. Species concentration are fixed at the inlet $z=0$ with a zero-gradient condition imposed at the outlet $z=L$ to allow the species to exit freely. Symmetry is imposed at the centre line $r=0$  and a no flux condition is applied at the reactor wall $r=A$. 

For each simulated experiment, we sample the temperature $\mathcal{K}$, average axial velocity $U_{\mathrm{avg}}$, a base diffusivity $D_{\mathrm{base}}$, and inlet concentrations for the feed species. Species diffusivities are derived as $D_s = D_{\mathrm{base}} f_s$, where $f_s$ is a mechanism-specific diffusion factor. Product-species inlet concentrations are fixed to zero unless produced by reaction. 

For the training data we generate 2,000 simulations, each of which have 32 time snapshots meaning we have a total of 64,000 images to train our model on. Our test dataset consists of another 50 simulations. Further details of the dataset generation and model training are provided in Appendix~\ref{app:dataset_generation}.

As some of the species concentrations are negligible compared to the more dominate species, the data has to be transformed before training the diffusion model in order to be able to capture the dynamics of the spatiotemporal structure for all of the species. To do this, a geometric mean of all strictly positive values is used to calculate a per species constant $c_{0, s}$
\begin{equation}
    c_{0, s} = \exp\left(\frac{1}{N^+} \sum_{i: \mathcal{C}_i > 0} \log \mathcal{C}_i \right),
\end{equation}
where $N^+$ is the number of positive values. Then the data is transformed
\begin{equation}
    \mathcal{C}_s^X = \log \left(1 + \frac{\mathcal{C}_s}{c_{0, s}}\right). 
\end{equation}

Once the data is transformed, the data is normalised using a mean $\mu_X$ and variance $\sigma_X$ that is computed over all values of all species. The diffusion model is then trained on this transformed and normalised data. The test data is transformed using the same species coefficients and normalisation statistics calculated in training and sampling is performed on this transformed and normalised test data. We can then convert the sampled data back to physical space to get the estimated species concentration using the following inverse transform
\begin{equation}
    \mathcal{C}_s = c_{0, s}(\exp(\mathcal{C}_s^X \sigma_X + \mu_X) - 1). 
\end{equation}

We compare four sampling algorithms: the GEM and SOSaG methods used as proposals within a sequential Monte Carlo framework, the DiffusionPDE method \citep{huang2024diffusionpdegenerativepdesolvingpartial}, and a first-order guided ODE solver. Further details are provided in Appendix~\ref{app:ode_solvers}. For the two stochastic methods, $N=4$ particles were used as opposed to the $N=1$ for the ODE samplers and all methods used $K=400$ iterations for the discrete time process.

\section{Discussion}

\begin{table*}[t]
\centering
\tiny
\caption{Gas-phase reaction kinetics full-field and outlet relative percentage errors across all experiments and methods in transformed (normalised) and physical spaces. Sensitivity threshold $\phi = 10^{-8}$. }
\label{tab:realchem_relative_error_combined_2}
\begin{tabular}{ccccc}
\toprule
& \multicolumn{2}{c}{\textbf{Transformed Space}} & \multicolumn{2}{c}{\textbf{Physical Space}} \\
\cmidrule(lr){2-3} \cmidrule(lr){4-5}
\textbf{Method} & \textbf{Rel. (\%)} & \textbf{Outlet (\%)} & \textbf{Rel. (\%)} & \textbf{Outlet (\%)} \\
\midrule

\multicolumn{5}{c}{\textbf{H$_2$O$_2$ Decomposition}} \\
\midrule
GEM     & $\bm{3.99 \pm 1.85}$ & $0.743 \pm 0.725$ & $2.09 \pm 2.06$ & $\bm{2.38 \pm 0.27}$ \\
SOSaG   & $5.04 \pm 4.85$ & $\bm{0.619 \pm 0.429}$ & $\bm{1.52 \pm 0.26}$ & $2.84 \pm 0.31$ \\
ODE     & $63.4 \pm 58.8$ & $18.5 \pm 12.9$ & $(1.96 \pm 2.81)\times10^{4}$ & $(0.72 \pm 1.44)\times10^{5}$ \\
DiffPDE & $43.3 \pm 60.0$ & $18.4 \pm 7.2$ & $(2.01 \pm 1.73)\times10^{4}$ & $(2.21 \pm 3.00)\times10^{4}$ \\

\midrule

\multicolumn{5}{c}{\textbf{NO + O$_3$ $\rightarrow$ NO$_2$}} \\
\midrule
GEM     & $\bm{7.3 \pm 10.1}$ & $0.909 \pm 0.668$ & $5.18 \pm 8.39$ & $1.78 \pm 0.85$ \\
SOSaG   & $9.2 \pm 17.6$ & $\bm{0.673 \pm 0.307}$ & $\bm{4.20 \pm 5.51}$ & $\bm{1.60 \pm 0.86}$ \\
ODE     & $66.9 \pm 18.5$ & $45.2 \pm 25.0$ & $(5.01 \pm 4.94)\times10^{4}$ & $(2.99 \pm 3.45)\times10^{5}$ \\
DiffPDE & $58.2 \pm 24.2$ & $33.2 \pm 21.8$ & $(2.99 \pm 2.86)\times10^{4}$ & $(3.45 \pm 8.64)\times10^{5}$ \\

\midrule

\multicolumn{5}{c}{\textbf{Ammonia Oxidation}} \\
\midrule
GEM     & $\bm{6.63 \pm 3.01}$ & $\bm{4.82 \pm 5.49}$ & $19.5 \pm 41.9$ & $4.40 \pm 2.87$ \\
SOSaG   & $7.67 \pm 3.69$ & $5.30 \pm 4.07$ & $\bm{6.04 \pm 8.14}$ & $\bm{3.91 \pm 1.85}$ \\
ODE     & $53.9 \pm 20.7$ & $52.7 \pm 25.8$ & $(3.24 \pm 6.10)\times10^{3}$ & $(0.62 \pm 1.40)\times10^{3}$ \\
DiffPDE & $55.1 \pm 27.0$ & $46.4 \pm 18.3$ & $(4.40 \pm 6.49)\times10^{3}$ & $(1.03 \pm 1.78)\times10^{3}$ \\

\midrule

\multicolumn{5}{c}{\textbf{Hydrogen Oxidation Subset}} \\
\midrule
GEM     & $3.54 \pm 1.08$ & $1.74 \pm 1.40$ & $\bm{(2.05 \pm 3.07)\times10^{2}}$ & $(0.90 \pm 2.85)\times10^{4}$ \\
SOSaG   & $\bm{3.17 \pm 1.33}$ & $\bm{1.65 \pm 2.97}$ & $(2.95 \pm 9.04)\times10^{2}$ & $\bm{5.93 \pm 3.46}$ \\
ODE     & $56.2 \pm 31.5$ & $68.6 \pm 52.8$ & $(3.38 \pm 2.10)\times10^{4}$ & $(1.87 \pm 1.08)\times10^{5}$ \\
DiffPDE & $61.3 \pm 53.4$ & $77.7 \pm 83.2$ & $(3.80 \pm 3.25)\times10^{4}$ & $(2.34 \pm 1.90)\times10^{5}$ \\

\midrule

\multicolumn{5}{c}{\textbf{2 Step Methane Oxidation}} \\
\midrule
GEM     & $\bm{6.35 \pm 4.28}$ & $\bm{1.46 \pm 1.02}$ & $\bm{20.6 \pm 50.4}$ & $\bm{2.55 \pm 0.95}$ \\
SOSaG   & $9.6 \pm 14.4$ & $5.2 \pm 12.0$ & $(2.20 \pm 6.82)\times10^{2}$ & $(0.70 \pm 2.22)\times10^{4}$ \\
ODE     & $(1.38 \pm 1.75)\times10^{2}$ & $71.2 \pm 46.8$ & $(3.00 \pm 5.74)\times10^{4}$ & $(2.54 \pm 3.58)\times10^{5}$ \\
DiffPDE & $(1.22 \pm 1.49)\times10^{2}$ & $52.3 \pm 23.6$ & $(1.26 \pm 0.91)\times10^{4}$ & $(1.79 \pm 1.92)\times10^{5}$ \\

\midrule

\multicolumn{5}{c}{\textbf{Water Gas Shift}} \\
\midrule
GEM     & $3.45 \pm 1.40$ & $\bm{1.54 \pm 1.78}$ & $2.59 \pm 1.07$ & $\bm{2.40 \pm 0.66}$ \\
SOSaG   & $\bm{3.02 \pm 1.11}$ & $1.84 \pm 2.35$ & $\bm{2.24 \pm 0.69}$ & $2.67 \pm 0.70$ \\
ODE     & $27.3 \pm 7.9$ & $20.0 \pm 22.1$ & $(9.90 \pm 4.59)\times10^{4}$ & $(5.03 \pm 4.15)\times10^{4}$ \\
DiffPDE & $27.9 \pm 7.1$ & $19.8 \pm 16.2$ & $(9.33 \pm 7.29)\times10^{4}$ & $(4.51 \pm 5.46)\times10^{4}$ \\

\bottomrule
\end{tabular}
\end{table*}

Figure~\ref{fig:outlet_comparison_h2o2} gives a comparison of the predicted outlet estimates for the H$_2$O$_2$ Decomposition experiment for an example trajectory run. We can see that the stochastic methods are able simulate and estimate the species concentration better than the alternative methods. Corresponding figures for the other experiments for an example trajectory are given in Appendix~\ref{app:outlet_concentration_graphs}. 

Absolute errors for species concentrations and outlet concentrations across all timesteps, together with the average wall-clock time required to sample a full trajectory, are reported in Table~\ref{tab:realchem_absolute_error_raw}. The stochastic methods consistently outperform the deterministic alternatives, often by approximately an order of magnitude. Despite using multiple samples, the vectorised implementation allows the stochastic methods to achieve wall-clock times comparable to those of the first and second-order deterministic solvers.

Relative percentage errors for both the transformed and physical concentration fields are reported in Table~\ref{tab:realchem_relative_error_combined_2}. Again, the stochastic methods significantly outperform the alternative approaches in both spaces. However, error metrics in physical space can be strongly influenced by species whose concentrations are very small relative to the dominant species in the system.

For example, in the H$_2$O$_2$ decomposition experiment, the concentrations of O$_2$ and H$_2$O remain several orders of magnitude smaller than that of H$_2$O$_2$ throughout the trajectory. As a result, small deviations in the transformed space can lead to very large relative errors after applying the inverse transformation to recover physical concentrations. Consequently, relatively large percentage errors in physical space do not necessarily correspond to physically meaningful discrepancies. For instance, estimating a concentration of $1\times10^{-8}$ when the true value is $1\times10^{-10}$ produces a large relative error, even though such a difference would be far below the detection limits of typical laboratory instrumentation. For this reason, we set a minimum threshold $\phi$ when calculating the relative percentage which would correspond to a conservative estimate of the sensitivity of a real world sensor measurement. We have discussed this further in Appendix~\ref{app:relative_error_extras}. 

Therefore, accurate estimation of the dominant species can still yield large relative percentage errors when low-concentration species are included in the metric. This effect is illustrated in Appendix~\ref{app:field_recon_graphs} where we provide an example trajectory reconstruction for each method over select time indices. The visual reconstructions demonstrate that the stochastic methods are typically able to recover the concentration fields from sparse observations and produce solutions that closely match the ARD dynamics in transformed space. After applying the inverse transformation, the resulting concentration fields in physical space remain visually consistent with the underlying dynamics, with minor deviations occurring primarily in low-concentration species. Nevertheless, even small discrepancies in transformed space can amplify into large relative errors when expressed in physical concentration units.

\section{Conclusion}
In this work we demonstrate, across multiple chemical kinetics datasets that closely reflect real-world laboratory conditions, that guided stochastic sampling methods are capable of reconstructing full spatiotemporal trajectories of time-varying PDE systems. Both qualitative and quantitative results indicate that the generated solutions are physically admissible.

Although relatively large errors are reported in Table~\ref{tab:realchem_relative_error_combined_2}, the visualisations in Appendix~\ref{app:field_recon_graphs} show that species concentrations are accurately reconstructed in the transformed space, which in turn leads to meaningful reconstructions in physical space. Furthermore, the outlet concentration profiles presented in Figure~\ref{fig:outlet_comparison_h2o2} and Appendix~\ref{app:outlet_concentration_graphs} demonstrate that the proposed methods effectively capture the temporal evolution of species concentrations at the reactor outlet, as defined by the experimental setup in Figure~\ref{fig:experiment_diagram}.

\textbf{Probabilistic Numerics Link and Uncertainty Quantification}

Traditionally, the field of probabilistic numerics quantifies uncertainty by framing numerical problem-solving as a statistical inference task. This typically yields approximate solutions that carry structural uncertainty when applied to complex systems. Our work departs from this traditional approach: instead, we leverage the probabilistic setting of multimodal diffusion priors and stochastic processes to aim to produce plausible solutions to complex systems in more tractable time than classical solving methods allow.

Despite this difference, our work remains firmly situated within probabilistic numerics, since one of its recognized application areas is precisely the setting we address: ``Typical numerical tasks to which PN may be applied include optimization, integration, the solution of ordinary and \emph{partial differential equations}''\footnote{\url{https://www.probabilistic-numerics.org/}}. Accordingly, we believe this work contributes to the field by highlighting a lesser-known application of probabilistic numerics for tractable equation solving.

Nevertheless, understanding the errors of our methods is still an important aspect that needs to be discussed. The main sources of error come from the prior, the measurement noise and the PDE operator discretisation. To give an example of understanding the uncertainty for the latter case, we have provided standard deviation heatmaps in pixel space for the DiffusionPDE and SOSaG methods (Appendix~\ref{app:error_map}) for different resolutions. We can see from these that the SOSaG method gives a much lower standard deviation than the DiffusionPDE method, showing that the estimates are closer to the ground truth with a smaller deviation across runs. We can also see from the accompanying line plots that SOSaG is better at decreasing the standard deviation as the resolution increases.

\textbf{Future Work}

Although the ARD equation is commonly used to model reaction kinetics, it is not limited to this area. Previous work has applied them to areas such as evolutionary biology \citep{lam2022introduction}, environmental ecosystems \cite{cantin2021mathematical} and many others. In this work, we present a general likelihood for the ARD equation and therefore, with minor modifications, this framework can be applied to a wide range of areas outside of chemical kinetics. 

Another promising direction for future work is the development of a single, generalised diffusion prior capable of learning the spatiotemporal structure of chemical kinetics systems. In this work, a separate diffusion prior is trained for each experiment. In contrast, future approaches could aim to train a single model that captures the underlying ARD dynamics across a range of parameter values. At test time, guided sampling could then be applied to real-world chemical datasets. Success in this direction would demonstrate that diffusion priors, combined with guided sampling, can generalise across entire families of PDE systems and parameter regimes, rather than requiring bespoke models for each individual chemical mixture or experimental setup.

\section*{Acknowledgements}

This work was partially supported by the Swedish Research Council and the Wallenberg AI, Autonomous Systems and Software Program (WASP) funded by the Knut and Alice Wallenberg (KAW) Foundation. Computations were enabled by the supercomputing resource Berzelius provided by National Supercomputer Centre at Link\"{o}ping University and the KAW foundation.

\bibliography{references}

%\newpage

\appendix

\newpage
\section{SMC Pseudocode} \label{app:smc_psuedo}

\begin{algorithm}[H]
\caption{Sequential Monte Carlo (SMC)}
\label{alg:smc}
\begin{algorithmic}[1]
\Require Number of particles $N$, time steps $K$; proposals $M_K(x_K), M_k(x_{k-1} \mid x_{k})$; incremental weight functions $G_K(x_K), G_k(x_{k-1},x_k)$; resampling threshold $N_{eff} \in (0,1]$.
\State \textbf{Initialization:}
\For{$n = 1 \rightarrow N$}
    \State $x_K^{(i)} \sim M_K(x^{(i)}_K)$
    \State $w_K^{(i)} = G_K(x^{(i)}_K)$
\EndFor
\For{$k = K \rightarrow 1$}
    \For{$i = 1 \rightarrow N$}
        \State \textbf{Propagate:} $x_{k-1}^{(i)} \sim M_k(x^{(i)}_{k-1} \mid x^{(i)}_{k})$ 
        \State \textbf{Weight update:} $\tilde{w}_{k-1}^{(i)} = w^{(i)}_{k} G_k(x_{k-1}^{(i)}, x_k^{(i)})$
    \EndFor
    \For{$i = 1 \rightarrow N$}
        \State \textbf{Normalize:} $w_{k-1}^{(i)} = \tilde{w}_{k-1}^{(i)} \Big/ \sum_{j=1}^N \tilde{w}_{k-1}^{(j)}$
    \EndFor
    \State \textbf{Compute ESS:} $\mathrm{ESS}_k = \frac{1}{\sum_{i=1}^N (w_{k-1}^{(i)})^2}$
    \If{$\mathrm{ESS}_k \leq \tau$}
        \State \textbf{Resampling (Multinomial):}
        \For{$i = 1 \rightarrow N$}
            \State $a^{(i)} \sim \mathrm{Cat}(w_{k-1}^{(1)}, \dots, w_{k-1}^{(N)})$
            \State $x_{k-1}^{(i)} = x_{k-1}^{(a^{(i)})}$
            \State $w_{k-1}^{(i)} = \frac{1}{N}$
        \EndFor
    \EndIf
\EndFor
\State \textbf{Return:} $\{(x_0^{(i)}, w_0^{(i)})\}_{i=1}^N$
\end{algorithmic}
\end{algorithm}

\section{Stochastic Solvers Pseudocode} \label{app:sto_solvers}

\subsection{GEM}

\begin{algorithm}[h!]
\caption{GEM Algorithm}
\label{alg:gem_proposal}
\begin{algorithmic}[1]
\Require{$\delta_{\theta}(x; \sigma),\, \sigma_k, \sigma_{k-1}, x_{k}$}
    \State Sample $\epsilon_{k} \sim \mathcal{N}(0, I)$
    \State $d_{k} \leftarrow \dfrac{x_{k} - \delta_{\theta}(x_{k}, \sigma_{k})}{\sigma_{k}}$
    \State $x_{k-1} \leftarrow x_{k} + (\sigma_{k-1}^2 - \sigma_{k}^2) d_{k} + \sqrt{\sigma_{k-1}^2 - \sigma_{k}^2} \epsilon_{k}$
    \State $x_{k-1} \leftarrow x_{k-1} - (\sigma_{k-1}^2 - \sigma_{k}^2) \nabla_{x_{k}}\log \Tilde{p}_\theta(y \mid x_k)$
    \State \textbf{return} $x_{k-1}$
\end{algorithmic}
\end{algorithm}

\subsection{SOSaG}

\begin{algorithm}[H]
\caption{SOSaG Algorithm}
\label{alg:sosag_proposal}
\begin{algorithmic}[1]
\Require{$\delta_{\theta}(x,\sigma),\, \sigma_{k}, \sigma_{k-1}, x_{k}, \gamma_{k}, \alpha$}
    \State Sample $\epsilon_{k} \sim \mathcal{N}(0, I)$
    \State $\hat{\sigma}_{k} \leftarrow \sigma_{k} + \gamma_{k}\,\sigma_{k}$
    \State $\hat{x}_{k} \leftarrow x_{k} +
           \sqrt{\hat{\sigma}^{2}_{k} - \sigma^{2}_{k}}\,\epsilon_{k}$
    \State $d_{k} \leftarrow
           \dfrac{\hat{x}_{k} - \delta_{\theta}(\hat{x}_{k},\hat{\sigma}_{k})}
           {\hat{\sigma}_{k}}$
    \State $x_{k-1} \leftarrow \hat{x}_{k} +
           (\sigma_{k-1} - \hat{\sigma}_{k})\,d_{k}$
    \If{$\sigma_{k} \neq 0$}
        \State $d_{k-1} \leftarrow
        \dfrac{x_{k-1} - \delta_{\theta}(x_{k-1},\sigma_{k-1})}{\sigma_{k-1}}$
        \State $x_{k-1} \leftarrow \hat{x}_{k} +
        (\sigma_{k-1} - \hat{\sigma}_{k})\left(\tfrac{1}{2}d_{k} + \tfrac{1}{2}d_{k-1}\right)$
    \EndIf
    \State $x_{k-1} \leftarrow x_{k-1} - \alpha \nabla_{\hat{x}_{k}} \log \Tilde{p}_\theta (y \mid \hat{x}_{k})$
    \State \textbf{return} $x_{k-1}$
\end{algorithmic}
\end{algorithm}

\section{ODE Solvers} \label{app:ode_solvers}

\subsection{Guided ODE}
The Euler discretisation to numerically approximate the solution to the ODE in Equation~\eqref{eq:prob_ODE_noise} is given by
\begin{align}
    x_{k-1}
    =
    x_k
    +
    \frac{1}{2}(\sigma_{k-1}^2 - \sigma_k^2)
    \frac{x_k - \delta_\theta(x_k,\sigma_k)}{\sigma_k^2} 
    \label{eq:euler_discrete}
\end{align}
and the corresponding guided-Euler (GE) discretisation method is given by 
\begin{align}
    x_{k-1}
    =
    x_k
    +
    \frac{1}{2}(\sigma_{k-1}^2 - \sigma_k^2)
    \frac{x_k - \delta_\theta(x_k,\sigma_k)}{\sigma_k^2} \nonumber \\
    +
    \frac{1}{2}(\sigma_{k-1}^2 - \sigma_k^2)
    \nabla_{x_k} \log \tilde{p}_\theta(y \mid x_k). 
    \label{eq:euler_discrete_guided}
\end{align}
The pseudocode for Equation~\eqref{eq:euler_discrete_guided} is given in Algorithm~\ref{alg:ge_proposal}

\begin{algorithm}[h!]
\caption{GE Algorithm}
\label{alg:ge_proposal}
\begin{algorithmic}[1]
\Require{$\delta_{\theta}(x; \sigma),\, \sigma_k, \sigma_{k-1}, x_{k}$}
    \State Sample $\epsilon_{k} \sim \mathcal{N}(0, I)$
    \State $d_{k} \leftarrow \dfrac{x_{k} - \delta_{\theta}(x_{k}, \sigma_{k})}{\sigma_{k}}$
    \State $x_{k-1} \leftarrow x_{k} + (\sigma_{k-1}^2 - \sigma_{k}^2) d_{k} + \sqrt{\sigma_{k-1}^2 - \sigma_{k}^2} \epsilon_{k}$
    \State $x_{k-1} \leftarrow x_{k-1} - (\sigma_{k-1}^2 - \sigma_{k}^2) \nabla_{x_{k}}\log \Tilde{p}_\theta(y \mid x_k)$
    \State \textbf{return} $x_{k-1}$
\end{algorithmic}
\end{algorithm}

\subsection{DiffPDE}

Influenced by \citet{karras2022elucidatingdesignspacediffusionbased}, \citet{huang2024diffusionpdegenerativepdesolvingpartial} extend the GE proposal in Equation~\eqref{eq:euler_discrete_guided} by incorporating a second-order correction term. This discretisation scheme is given in Algorithm~\ref{alg:diffpde_proposal}. 

\begin{algorithm}[H]
\caption{DiffPDE Algorithm}
\label{alg:diffpde_proposal}
\begin{algorithmic}[1]
\Require{$\delta_{\theta}(x,\sigma),\, \sigma_{k}, \sigma_{k-1}, x_{k}, \alpha$}
    \State $d_{k} \leftarrow
           \dfrac{x_{k} - \delta_{\theta}(x_{k},\sigma_{k})}
           {\sigma_{k}}$
    \State $x_{k-1} \leftarrow x_{k} +
           (\sigma_{k-1} - \hat{\sigma}_{k})\,d_{k}$
    \If{$\sigma_{k} \neq 0$}
        \State $d_{k-1} \leftarrow
        \dfrac{x_{k-1} - \delta_{\theta}(x_{k-1},\sigma_{k-1})}{\sigma_{k-1}}$
        \State $x_{k-1} \leftarrow x_{k} +
        (\sigma_{k-1} - \sigma_{k})\left(\tfrac{1}{2}d_{k} + \tfrac{1}{2}d_{k-1}\right)$
    \EndIf
    \State $x_{k-1} \leftarrow x_{k-1} - \alpha \nabla_{\hat{x}_{k}} \log \Tilde{p}_\theta (y \mid \hat{x}_{k})$
    \State \textbf{return} $x_{k-1}$
\end{algorithmic}
\end{algorithm}

\section{Chemistry dataset generation and model training} \label{app:dataset_generation} 

Table~\ref{tab:parameter_range} gives the summary of the ranges parameters were sampled from when generating the dataset. Table~\ref{tab:reaction_networks} gives the Arrhenius parameters used when computing the reaction networks and Table~\ref{tab:scaling factors} gives the diffusion scaling factors for each experiment. We would like to note that the generated dataset therefore is the ground truth from which we compare our solutions against. 

\begin{table*}[t]
\centering
\footnotesize

% ------------
% Subtable 1
% ------------
\begin{subtable}
\centering
\caption{Summary of reduced chemical mechanisms and sampled parameter ranges. $[\cdot, \cdot]$ denote the ranges the parameters were uniformly sampled from.}
\label{tab:parameter_range}
\begin{tabular}{lllccc}
\toprule
Mechanism & Species & Feed Species & $T$ (K) & $U_{\mathrm{avg}}$ (m/s) & $\log_{10} D_{\mathrm{base}}$ \\
\midrule
NO-O$_3$ to NO$_2$ & NO, O$_3$, NO$_2$, O$_2$ & NO, O$_3$ & $[270,330]$ & $[0.01, 0.5]$ & $[-6.5, -5.2]$ \\
Water-gas shift & CO, H$_2$O, CO$_2$, H$_2$ & CO, H$_2$O & $[500,1000]$ & $[0.01, 0.5]$  & $[-6.5, -5.2]$ \\
Hydrogen peroxide decomposition & H$_2$O$_2$, H$_2$O, O$_2$ & H$_2$O$_2$ & $[290, 600]$ & $[0.01, 0.5]$  & $[-6.5, -5.2]$ \\
Two-step methane oxidation & CH$_4$, O$_2$, CO, CO$_2$, H$_2$O & CH$_4$, O$_2$ & $[700, 1500]$ & $[0.01, 0.5]$ & $[-6.5, -5.2]$ \\
Ammonia oxidation & NH$_3$, O$_2$, NO, NO$_2$, H$_2$O & NH$_3$, O$_2$ & $[600, 1200]$ & $[0.01, 0.5]$  & $[-6.5, -5.2]$ \\
Hydrogen oxidation subset & H$_2$, O$_2$, OH, H$_2$O, H, HO$_2$ & H$_2$, O$_2$ & $[800, 1800]$ & $[0.01, 0.5]$  & $[-6.5, -5.2]$ \\
\bottomrule
\end{tabular}
\end{subtable}

\vspace{1em}

% ------------
% Subtable 2
% ------------
\begin{subtable}
\centering
\caption{Reaction networks and Arrhenius parameters used in the dataset generation process.}
\label{tab:reaction_networks}
\begin{tabular}{p{5.2cm} p{5.2cm} cc}
\toprule
Mechanism & Reaction $l$ & $A_l$ & $E_{a,l}$ (J/mol) \\
\midrule
NO-O$_3$ to NO$_2$ & $\mathrm{NO + O_3 \rightarrow NO_2 + O_2}$ & $2.5 \times 10^{6}$ & $1.2 \times 10^{4}$ \\
\midrule
Water-gas shift & $\mathrm{CO + H_2O \rightarrow CO_2 + H_2}$ & $1.5 \times 10^{3}$ & $6.5 \times 10^{4}$ \\ 
& $\mathrm{CO_2 + H_2 \rightarrow CO + H_2O}$ & $4.0 \times 10^{2}$ & $7.0 \times 10^{4}$ \\
\midrule
Hydrogen peroxide decomposition & $\mathrm{2\,H_2O_2 \rightarrow 2\,H_2O + O_2}$ & $2.0 \times 10^{2}$ & $5.5 \times 10^{4}$ \\
\midrule
Two-step methane oxidation & $\mathrm{2\,CH_4 + 3\,O_2 \rightarrow 2\,CO + 4\,H_2O}$ & $8.0 \times 10^{3}$ & $1.2 \times 10^{5}$ \\
& $\mathrm{2\,CO + O_2 \rightarrow 2\,CO_2}$ & $2.5 \times 10^{4}$ & $8.5 \times 10^{4}$ \\
\midrule
Ammonia oxidation & $\mathrm{4\,NH_3 + 5\,O_2 \rightarrow 4\,NO + 6\,H_2O}$ & $6.0 \times 10^{3}$ & $9.5 \times 10^{4}$ \\
& $\mathrm{2\,NO + O_2 \rightarrow 2\,NO_2}$ & $1.2 \times 10^{4}$ & $6.0 \times 10^{4}$ \\
\midrule
Hydrogen oxidation subset & $\mathrm{H_2 + O_2 \rightarrow 2\,OH}$ & $1.0 \times 10^{5}$ & $1.4 \times 10^{5}$ \\
& $\mathrm{H_2 + OH \rightarrow H_2O + H}$ & $8.0 \times 10^{5}$ & $2.0 \times 10^{4}$ \\
& $\mathrm{H + O_2 \rightarrow HO_2}$ & $2.0 \times 10^{6}$ & $1.5 \times 10^{4}$ \\
& $\mathrm{HO_2 + H \rightarrow 2\,OH}$ & $1.5 \times 10^{6}$ & $1.0 \times 10^{4}$ \\
\bottomrule
\end{tabular}
\end{subtable}

\vspace{1em}

% ------------
% Subtable 3
% ------------
\begin{subtable}
\centering
\caption{Species-dependent diffusion scaling factors used in the dataset generation process.}
\label{tab:scaling factors}
\begin{tabular}{ll}
\toprule
Mechanism & Diffusion factors $f_s$ \\
\midrule
NO-O$_3$ to NO$_2$ & NO: 1.00,\; O$_3$: 0.95,\; NO$_2$: 0.92,\; O$_2$: 1.08 \\
Water-gas shift & CO: 1.03,\; H$_2$O: 0.90,\; CO$_2$: 0.88,\; H$_2$: 1.35 \\
Hydrogen peroxide decomposition & H$_2$O$_2$: 0.85,\; H$_2$O: 1.00,\; O$_2$: 1.20 \\
Two-step methane oxidation & CH$_4$: 1.20,\; O$_2$: 1.05,\; CO: 1.00,\; CO$_2$: 0.88,\; H$_2$O: 0.92 \\
Ammonia oxidation & NH$_3$: 1.08,\; O$_2$: 1.03,\; NO: 1.00,\; NO$_2$: 0.93,\; H$_2$O: 0.90 \\
Hydrogen oxidation subset & H$_2$: 1.60,\; O$_2$: 1.05,\; OH: 1.25,\; H$_2$O: 0.90,\; H: 1.80,\; HO$_2$: 0.95 \\
\bottomrule
\end{tabular}
\end{subtable}

\end{table*}

As mentioned previously, the models are trained on all of the time snapshots within each trajectory and thus has the opportunity to learn the spatiotemporal structure, including both the long and short-term dynamics that is induced by the logarithmic spacing of the chemical reactions. Each model takes approximately 4.5 hours to train (1 million image passes per model). At inference time, snapshots are reconstructed with temporal coherence enforced through a PDE residual-based likelihood, coupling consecutive time steps via the governing equations. The learned spatial prior reconstructs each time step, while the residual coupling ensures consistency with the system's temporal dynamics. This deliberate design keeps the generative model flexible and data-driven while physical constraints in the likelihood enforce temporal consistency. Each time step takes 1.5 minutes to reconstruct meaning each trajectory takes $\approx$45 minutes to reconstruct. This further motivates work for a single model that encompasses multiple chemical reactions and also a method which can reconstruct multiple time steps at the same time to reduce the time taken for the sequential reconstruction. 

\section{Supplementary Results}

\subsection{Absolute Error Tables} \label{app:relative_error_extras}

We report relative percentage errors defined as
\[
\mathrm{RPE}(x_{\text{pred}}, x_{\text{true}})
=
100 \times \frac{\lvert x_{\text{pred}} - x_{\text{true}} \rvert}{\max\big(\lvert x_{\text{true}} \rvert, \phi\big)},
\]
where $\phi$ is a small positive constant introduced to ensure numerical stability in regions where the ground truth approaches zero. 

Table~\ref{tab:realchem_relative_error_combined_v2} reports results using $\phi = 10^{-6}$, while Table~\ref{tab:realchem_relative_error_lower_max} shows results with a smaller threshold $\phi = 10^{-12}$. As expected, reducing $\phi$ leads to substantially larger relative errors. This behaviour arises in regions where chemical species concentrations are very small in physical space, causing the relative error to be dominated by the denominator.

Importantly, these large relative errors do not correspond to significant degradation in reconstruction quality. As shown in Figure~\ref{fig:outlet_comparison_h2o2} and Appendices~\ref{app:field_recon_graphs} and \ref{app:outlet_concentration_graphs}, the methods still accurately recover both the concentration fields and outlet quantities. The elevated errors are therefore primarily an artefact of the metric in low-concentration regimes rather than a failure of the model.

\begin{table*}[t]
\centering
\footnotesize
\caption{Gas-phase reaction kinetics full-field and outlet relative percentage errors across all experiments and methods in transformed (normalised) and physical spaces. $\phi = 10^{-6}$. }
\label{tab:realchem_relative_error_combined_v2}
\begin{tabular}{ccccc}
\toprule
& \multicolumn{2}{c}{\textbf{Transformed Space}} & \multicolumn{2}{c}{\textbf{Physical Space}} \\
\cmidrule(lr){2-3} \cmidrule(lr){4-5}
\textbf{Method} & \textbf{Rel. (\%)} & \textbf{Outlet (\%)} & \textbf{Rel. (\%)} & \textbf{Outlet (\%)} \\
\midrule

\multicolumn{5}{c}{\textbf{H$_2$O$_2$ Decomposition}} \\
\midrule
GEM     & $\bm{3.55 \pm 1.65}$ & $0.743 \pm 0.725$ & $0.543 \pm 0.343$ & $\bm{0.345 \pm 0.217}$ \\
SOSaG   & $3.63 \pm 1.87$ & $\bm{0.619 \pm 0.429}$ & $\bm{0.428 \pm 0.097}$ & $0.362 \pm 0.186$ \\
ODE     & $55.3 \pm 58.3$ & $18.5 \pm 12.9$ & $(2.68 \pm 3.32)\times10^{2}$ & $(0.73 \pm 1.44)\times10^{3}$ \\
DiffPDE & $41.7 \pm 60.7$ & $18.4 \pm 7.2$ & $(2.47 \pm 1.93)\times10^{2}$ & $(4.12 \pm 8.66)\times10^{2}$ \\

\midrule

\multicolumn{5}{c}{\textbf{NO + O$_3$ $\rightarrow$ NO$_2$}} \\
\midrule
GEM     & $\bm{7.3 \pm 10.1}$ & $0.909 \pm 0.668$ & $1.77 \pm 0.32$ & $1.45 \pm 0.69$ \\
SOSaG   & $9.2 \pm 17.6$ & $\bm{0.673 \pm 0.307}$ & $\bm{1.65 \pm 0.36}$ & $\bm{1.32 \pm 0.62}$ \\
ODE     & $66.9 \pm 18.5$ & $45.2 \pm 25.0$ & $(8.67 \pm 9.07)\times10^{2}$ & $(4.97 \pm 8.44)\times10^{3}$ \\
DiffPDE & $58.2 \pm 24.2$ & $33.2 \pm 21.8$ & $(3.51 \pm 2.89)\times10^{2}$ & $(3.48 \pm 8.64)\times10^{3}$ \\

\midrule

\multicolumn{5}{c}{\textbf{Ammonia Oxidation}} \\
\midrule
GEM     & $\bm{6.63 \pm 3.01}$ & $\bm{4.82 \pm 5.49}$ & $1.54 \pm 0.55$ & $2.06 \pm 2.82$ \\
SOSaG   & $7.67 \pm 3.69$ & $5.30 \pm 4.07$ & $\bm{1.36 \pm 0.55}$ & $\bm{1.83 \pm 1.76}$ \\
ODE     & $53.9 \pm 20.7$ & $52.7 \pm 25.8$ & $44.5 \pm 58.4$ & $20.9 \pm 21.1$ \\
DiffPDE & $55.1 \pm 27.0$ & $46.4 \pm 18.3$ & $70.6 \pm 93.0$ & $(1.86 \pm 5.05)\times10^{2}$ \\

\midrule

\multicolumn{5}{c}{\textbf{Hydrogen Oxidation Subset}} \\
\midrule
GEM     & $3.54 \pm 1.07$ & $1.74 \pm 1.40$ & $\bm{4.47 \pm 4.43}$ & $92.0 \pm 284.0$ \\
SOSaG   & $\bm{3.16 \pm 1.32}$ & $\bm{1.65 \pm 2.97}$ & $5.8 \pm 13.0$ & $\bm{1.77 \pm 2.45}$ \\
ODE     & $55.8 \pm 30.2$ & $68.6 \pm 52.8$ & $(3.91 \pm 2.14)\times10^{2}$ & $(1.93 \pm 1.15)\times10^{3}$ \\
DiffPDE & $60.3 \pm 50.4$ & $77.7 \pm 83.2$ & $(4.42 \pm 3.53)\times10^{2}$ & $(2.95 \pm 1.85)\times10^{3}$ \\

\midrule

\multicolumn{5}{c}{\textbf{2 Step Methane Oxidation}} \\
\midrule
GEM     & $\bm{6.17 \pm 3.97}$ & $\bm{1.46 \pm 1.02}$ & $\bm{1.04 \pm 0.93}$ & $\bm{0.617 \pm 0.722}$ \\
SOSaG   & $9.3 \pm 13.8$ & $5.2 \pm 12.0$ & $3.52 \pm 7.10$ & $72.0 \pm 221.0$ \\
ODE     & $(1.28 \pm 1.51)\times10^{2}$ & $71.2 \pm 46.8$ & $(4.24 \pm 8.37)\times10^{2}$ & $(2.64 \pm 3.68)\times10^{3}$ \\
DiffPDE & $(1.11 \pm 1.23)\times10^{2}$ & $52.3 \pm 23.6$ & $(1.62 \pm 1.12)\times10^{2}$ & $(1.94 \pm 1.99)\times10^{3}$ \\

\midrule

\multicolumn{5}{c}{\textbf{Water Gas Shift}} \\
\midrule
GEM     & $3.45 \pm 1.39$ & $\bm{1.54 \pm 1.78}$ & $1.03 \pm 0.44$ & $\bm{0.946 \pm 0.614}$ \\
SOSaG   & $\bm{3.02 \pm 1.11}$ & $1.84 \pm 2.35$ & $\bm{0.947 \pm 0.380}$ & $1.06 \pm 0.72$ \\
ODE     & $27.3 \pm 7.9$ & $20.0 \pm 22.1$ & $(1.12 \pm 0.54)\times10^{3}$ & $(5.07 \pm 4.16)\times10^{2}$ \\
DiffPDE & $27.9 \pm 7.1$ & $19.8 \pm 16.2$ & $(1.06 \pm 0.81)\times10^{3}$ & $(4.57 \pm 5.45)\times10^{2}$ \\

\bottomrule
\end{tabular}
\end{table*}

\begin{table*}[t]
\centering
\footnotesize
\caption{Gas-phase reaction kinetics full-field and outlet relative percentage errors across all experiments and methods in transformed (normalised) and physical spaces. $\phi = 10^{-12}$. }
\label{tab:realchem_relative_error_lower_max}
\begin{tabular}{ccccc}
\toprule
& \multicolumn{2}{c}{\textbf{Transformed Space}} & \multicolumn{2}{c}{\textbf{Physical Space}} \\
\cmidrule(lr){2-3} \cmidrule(lr){4-5}
\textbf{Method} & \textbf{Rel. (\%)} & \textbf{Outlet (\%)} & \textbf{Rel. (\%)} & \textbf{Outlet (\%)} \\
\midrule

\multicolumn{5}{c}{\textbf{H$_2$O$_2$ Decomposition}} \\
\midrule
GEM     & $\bm{3.99 \pm 1.85}$ & $0.743 \pm 0.725$ & $\bm{(5.11 \pm 0.98)\times10^{3}}$ & $\bm{(1.63 \pm 0.28)\times10^{4}}$ \\
SOSaG   & $5.04 \pm 4.85$ & $\bm{0.619 \pm 0.429}$ & $(6.21 \pm 0.59)\times10^{3}$ & $(1.98 \pm 0.31)\times10^{4}$ \\
ODE     & $63.4 \pm 58.8$ & $18.5 \pm 12.9$ & $(1.49 \pm 2.22)\times10^{8}$ & $(0.72 \pm 1.44)\times10^{9}$ \\
DiffPDE & $43.3 \pm 60.0$ & $18.4 \pm 7.2$ & $(1.70 \pm 1.48)\times10^{8}$ & $(1.92 \pm 2.24)\times10^{8}$ \\

\midrule

\multicolumn{5}{c}{\textbf{NO + O$_3$ $\rightarrow$ NO$_2$}} \\
\midrule
GEM     & $\bm{7.3 \pm 10.1}$ & $0.909 \pm 0.668$ & $(2.77 \pm 8.30)\times10^{4}$ & $\bm{(1.97 \pm 0.63)\times10^{2}}$ \\
SOSaG   & $9.2 \pm 17.6$ & $\bm{0.673 \pm 0.307}$ & $\bm{(2.00 \pm 5.35)\times10^{4}}$ & $(2.27 \pm 0.48)\times10^{2}$ \\
ODE     & $66.9 \pm 18.5$ & $45.2 \pm 25.0$ & $(3.08 \pm 2.66)\times10^{8}$ & $(2.52 \pm 2.75)\times10^{9}$ \\
DiffPDE & $58.2 \pm 24.2$ & $33.2 \pm 21.8$ & $(2.80 \pm 2.90)\times10^{8}$ & $(3.38 \pm 8.67)\times10^{9}$ \\

\midrule

\multicolumn{5}{c}{\textbf{Ammonia Oxidation}} \\
\midrule
GEM     & $\bm{6.63 \pm 3.01}$ & $\bm{4.82 \pm 5.49}$ & $(1.69 \pm 4.23)\times10^{5}$ & $(2.17 \pm 0.40)\times10^{4}$ \\
SOSaG   & $7.67 \pm 3.69$ & $5.30 \pm 4.07$ & $\bm{(4.27 \pm 8.32)\times10^{4}}$ & $\bm{(1.86 \pm 0.37)\times10^{4}}$ \\
ODE     & $53.9 \pm 20.7$ & $52.7 \pm 25.8$ & $(3.11 \pm 6.14)\times10^{7}$ & $(0.60 \pm 1.40)\times10^{7}$ \\
DiffPDE & $55.1 \pm 27.0$ & $46.4 \pm 18.3$ & $(2.76 \pm 4.19)\times10^{7}$ & $(0.84 \pm 1.79)\times10^{7}$ \\

\midrule

\multicolumn{5}{c}{\textbf{Hydrogen Oxidation Subset}} \\
\midrule
GEM     & $3.54 \pm 1.08$ & $1.74 \pm 1.40$ & $\bm{(1.75 \pm 2.98)\times10^{6}}$ & $(0.90 \pm 2.85)\times10^{8}$ \\
SOSaG   & $\bm{3.17 \pm 1.33}$ & $\bm{1.65 \pm 2.97}$ & $(2.14 \pm 6.59)\times10^{6}$ & $\bm{(3.63 \pm 0.87)\times10^{4}}$ \\
ODE     & $56.2 \pm 31.5$ & $68.6 \pm 52.8$ & $(2.92 \pm 1.93)\times10^{8}$ & $(1.80 \pm 1.08)\times10^{9}$ \\
DiffPDE & $61.3 \pm 53.4$ & $77.7 \pm 83.2$ & $(3.36 \pm 3.05)\times10^{8}$ & $(2.24 \pm 1.89)\times10^{9}$ \\

\midrule

\multicolumn{5}{c}{\textbf{2 Step Methane Oxidation}} \\
\midrule
GEM     & $\bm{6.35 \pm 4.28}$ & $\bm{1.46 \pm 1.02}$ & $\bm{(1.93 \pm 5.03)\times10^{5}}$ & $\bm{(1.67 \pm 0.16)\times10^{4}}$ \\
SOSaG   & $9.6 \pm 14.4$ & $5.2 \pm 12.0$ & $(2.19 \pm 6.83)\times10^{6}$ & $(0.70 \pm 2.22)\times10^{8}$ \\
ODE     & $(1.38 \pm 1.75)\times10^{2}$ & $71.2 \pm 46.8$ & $(1.81 \pm 2.63)\times10^{8}$ & $(2.54 \pm 3.58)\times10^{9}$ \\
DiffPDE & $(1.22 \pm 1.49)\times10^{2}$ & $52.3 \pm 23.6$ & $(1.02 \pm 0.78)\times10^{8}$ & $(1.70 \pm 1.92)\times10^{9}$ \\

\midrule

\multicolumn{5}{c}{\textbf{Water Gas Shift}} \\
\midrule
GEM     & $3.45 \pm 1.40$ & $\bm{1.54 \pm 1.78}$ & $(7.05 \pm 3.31)\times10^{3}$ & $\bm{(1.14 \pm 0.19)\times10^{4}}$ \\
SOSaG   & $\bm{3.02 \pm 1.11}$ & $1.84 \pm 2.35$ & $\bm{(6.07 \pm 1.65)\times10^{3}}$ & $(1.25 \pm 0.18)\times10^{4}$ \\
ODE     & $27.3 \pm 7.9$ & $20.0 \pm 22.1$ & $(8.57 \pm 3.93)\times10^{8}$ & $(5.03 \pm 4.15)\times10^{8}$ \\
DiffPDE & $27.9 \pm 7.1$ & $19.8 \pm 16.2$ & $(8.11 \pm 6.54)\times10^{8}$ & $(4.33 \pm 5.46)\times10^{8}$ \\

\bottomrule
\end{tabular}
\end{table*}

\FloatBarrier
\subsection{Outlet Estimation Figures} \label{app:outlet_concentration_graphs}
\begin{figure*}
    \centering
    \includegraphics[width=\textwidth]{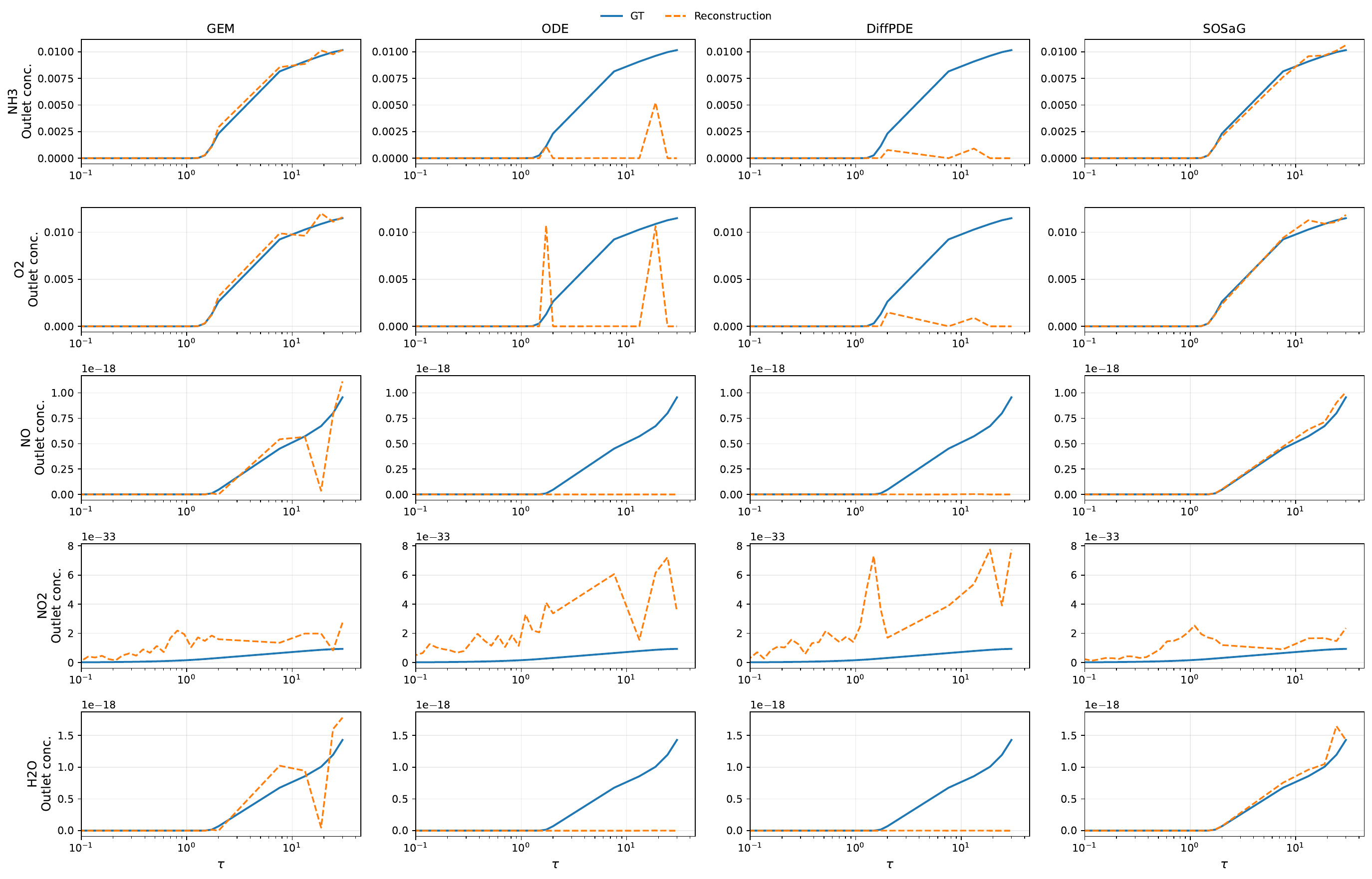}
    \caption{Species outlet comparison for Ammonia Oxidation.}
    \label{fig:outlet_comparison_ammonia}
\end{figure*}

\begin{figure*}
    \centering
    \includegraphics[width=\textwidth]{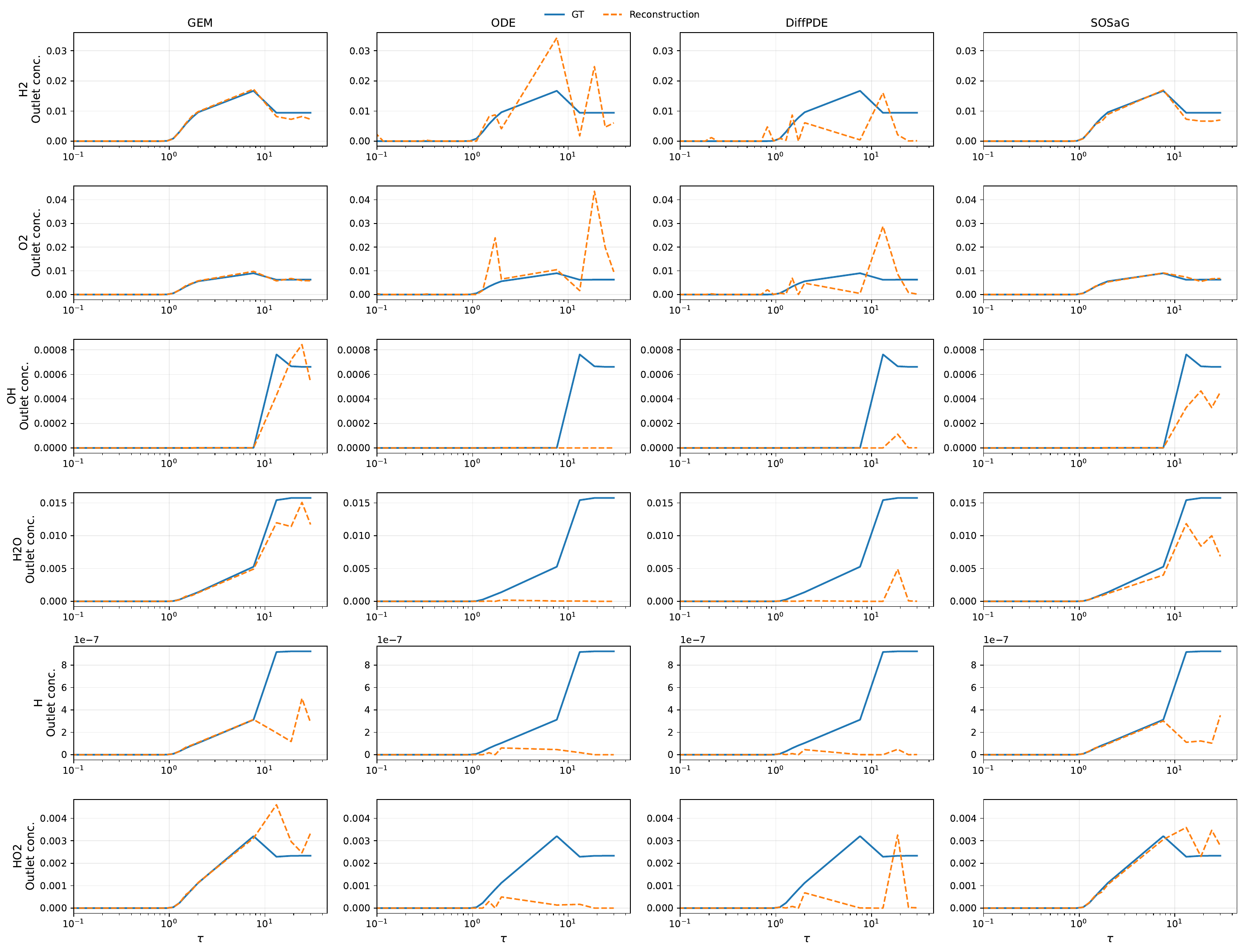}
    \caption{Species outlet comparison for Hydrogen Oxidation Subset.}
    \label{fig:outlet_comparison_hydrogen}
\end{figure*}

\begin{figure*}
    \centering
    \includegraphics[width=\textwidth]{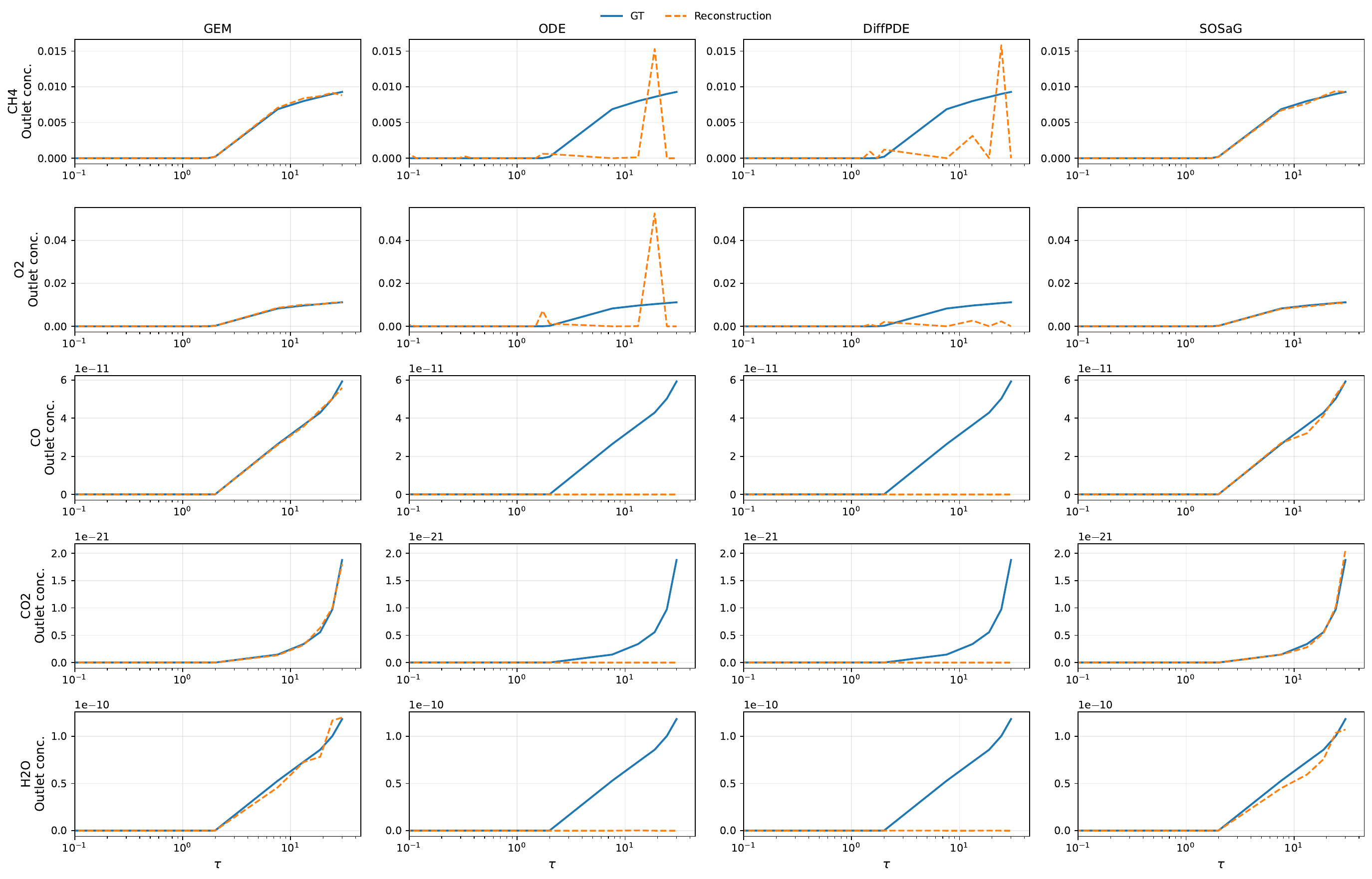}
    \caption{Species outlet comparison for Two-step Methane Oxidation.}
    \label{fig:outlet_comparison_methane}
\end{figure*}

\begin{figure*}
    \centering
    \includegraphics[width=\textwidth]{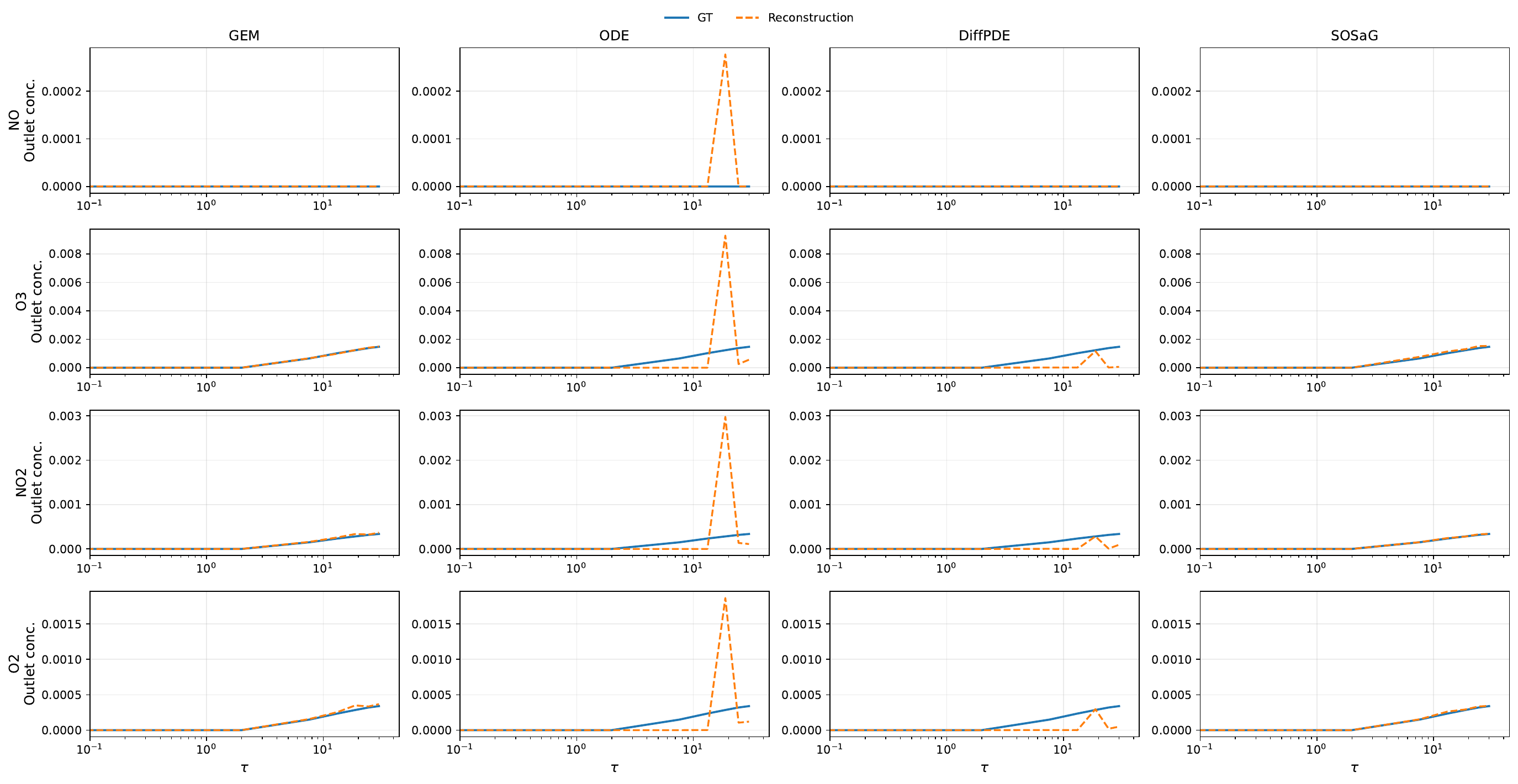}
    \caption{Species outlet comparison for NO + O$_3 \rightarrow$ NO$_2$.}
    \label{fig:outlet_comparison_no_o3}
\end{figure*}

\begin{figure*}
    \centering
    \includegraphics[width=\textwidth]{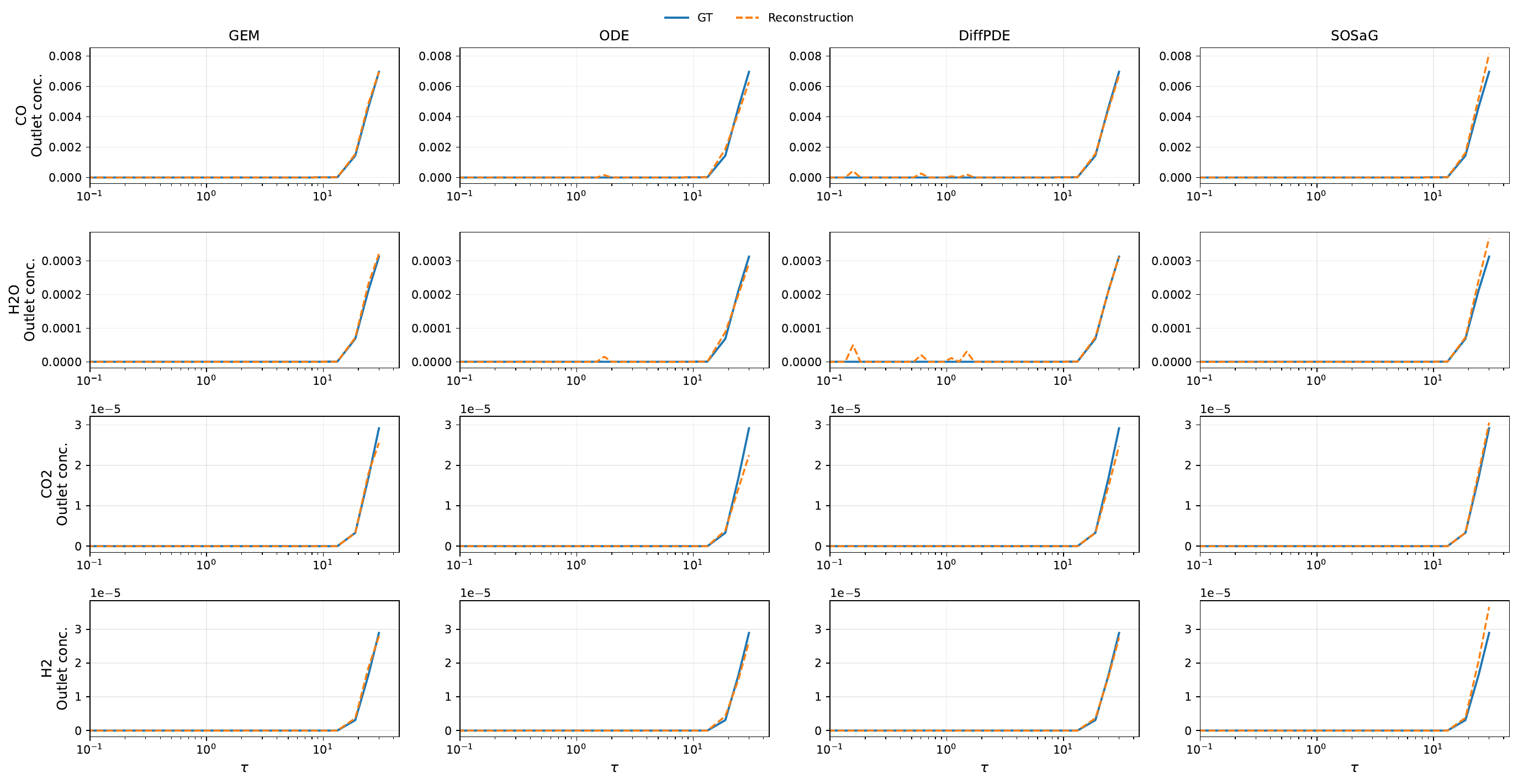}
    \caption{Species outlet comparison for Water Gas Shift.}
    \label{fig:outlet_comparison_water}
\end{figure*}

Figures~\ref{fig:outlet_comparison_ammonia}--\ref{fig:outlet_comparison_water} give the estimated species concentration at the outlet from $\tau = 0.01 \rightarrow 30$ seconds for all methods for the same example run.

\FloatBarrier
\subsection{Field Reconstruction Figures} \label{app:field_recon_graphs}
\begin{landscape}

\begin{figure}[p]
\centering

\includegraphics[width=1.4\textwidth]{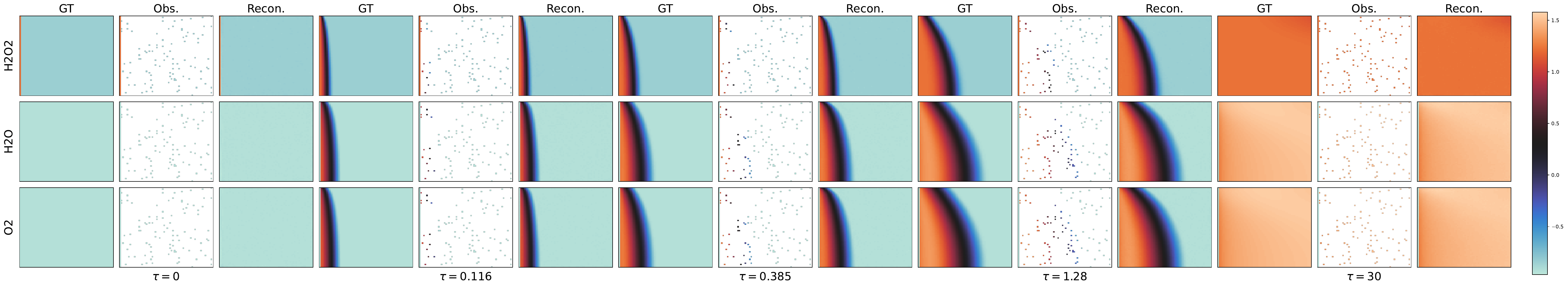}

\vspace{1cm}

\includegraphics[width=1.4\textwidth]{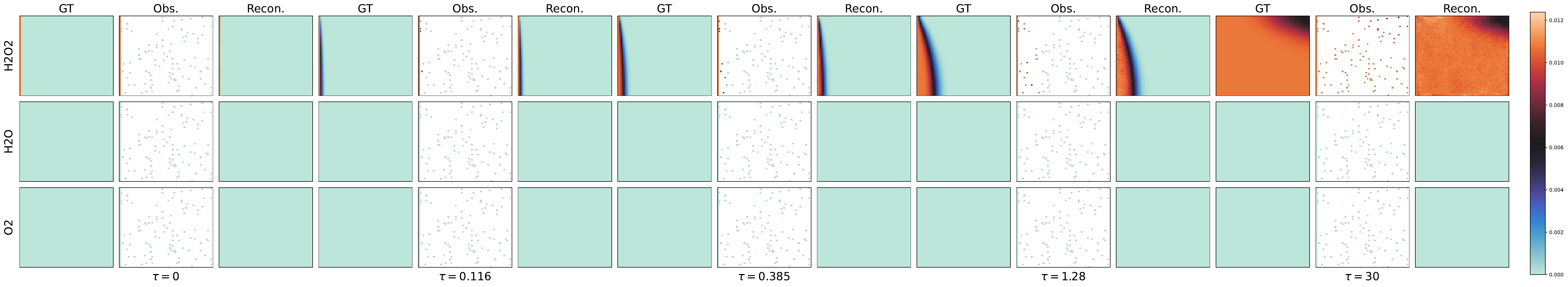}

\caption{H$_2$O$_2$ Decomposition reconstruction panel for the GEM method for select PDE times. Top: Normalised. Bottom: Physical.}
\label{fig:h2o2_recon_gem}
\end{figure}

\end{landscape}

\begin{landscape}

\begin{figure}[p]
\centering

\includegraphics[width=1.4\textwidth]{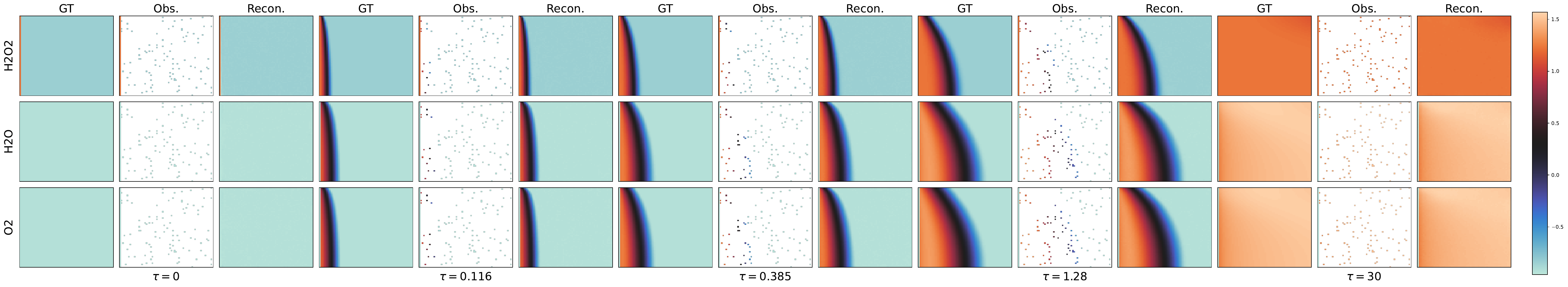}

\vspace{1cm}

\includegraphics[width=1.4\textwidth]{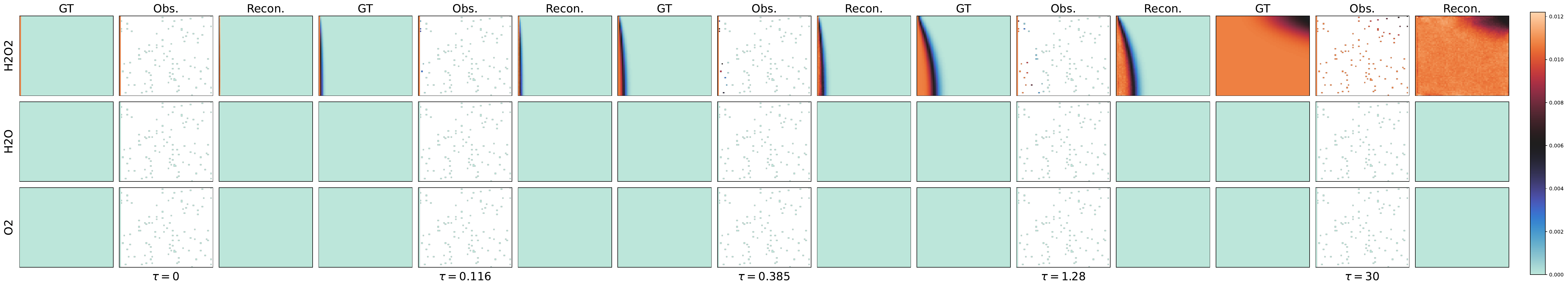}

\caption{H$_2$O$_2$ Decomposition reconstruction panel for the SOSaG method for select PDE times. Top: Normalised. Bottom: Physical.}
\label{fig:h2o2_recon_sosag}
\end{figure}

\end{landscape}

\begin{landscape}

\begin{figure}[p]
\centering

\includegraphics[width=1.4\textwidth]{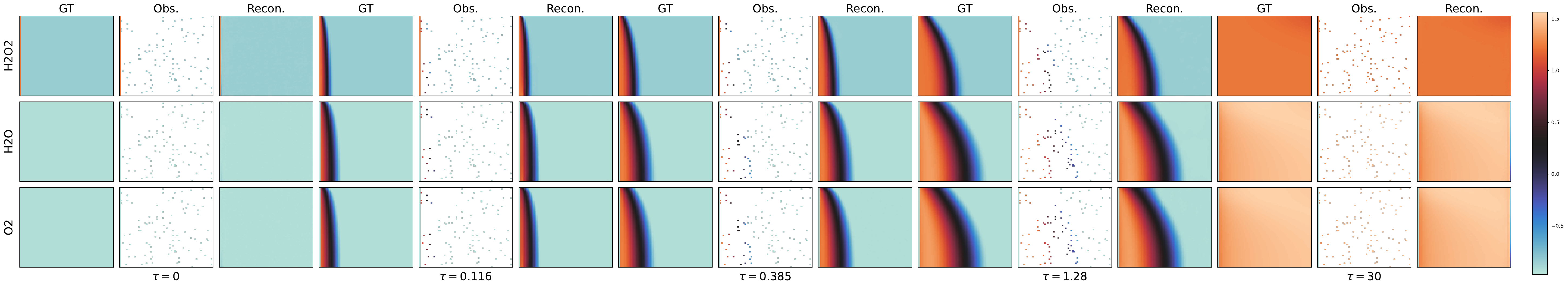}

\vspace{1cm}

\includegraphics[width=1.4\textwidth]{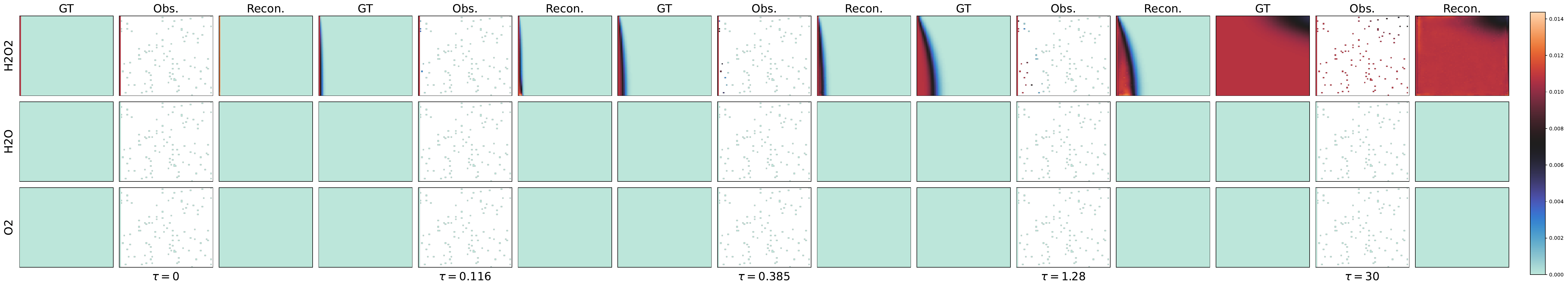}

\caption{H$_2$O$_2$ Decomposition reconstruction panel for the ODE method for select PDE times. Top: Normalised. Bottom: Physical.}
\label{fig:h2o2_recon_ode}
\end{figure}

\end{landscape}

\begin{landscape}

\begin{figure}[p]
\centering

\includegraphics[width=1.4\textwidth]{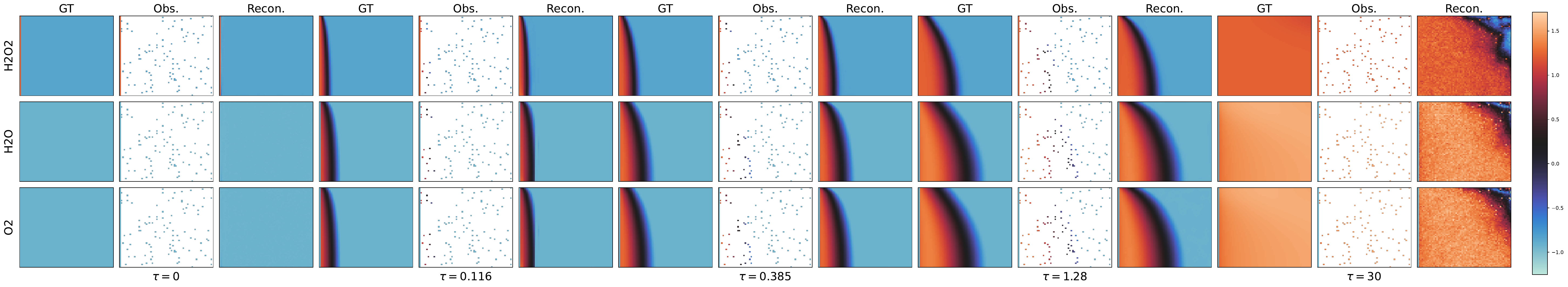}

\vspace{1cm}

\includegraphics[width=1.4\textwidth]{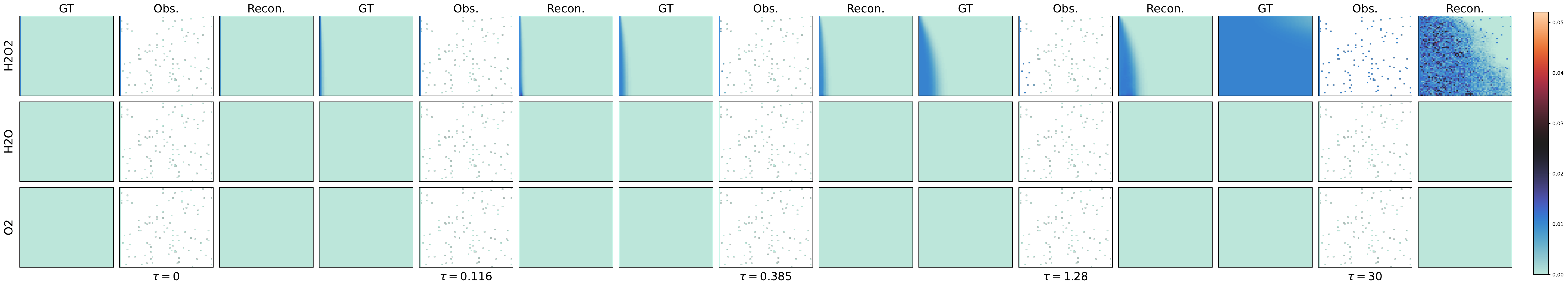}

\caption{H$_2$O$_2$ Decomposition reconstruction panel for the DiffPDE method for select PDE times. Top: Normalised. Bottom: Physical.}
\label{fig:h2o2_recon_diffpde}
\end{figure}

\end{landscape}

\begin{landscape}

\begin{figure}[p]
\centering

\includegraphics[width=1.4\textwidth]{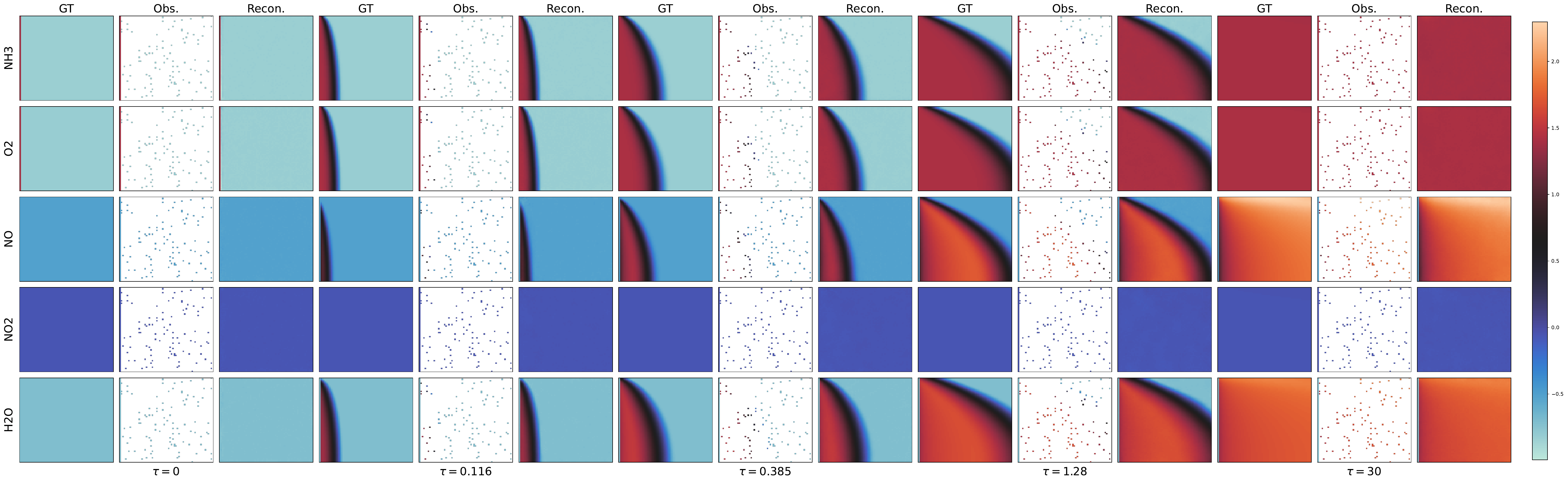}

\vspace{1cm}

\includegraphics[width=1.4\textwidth]{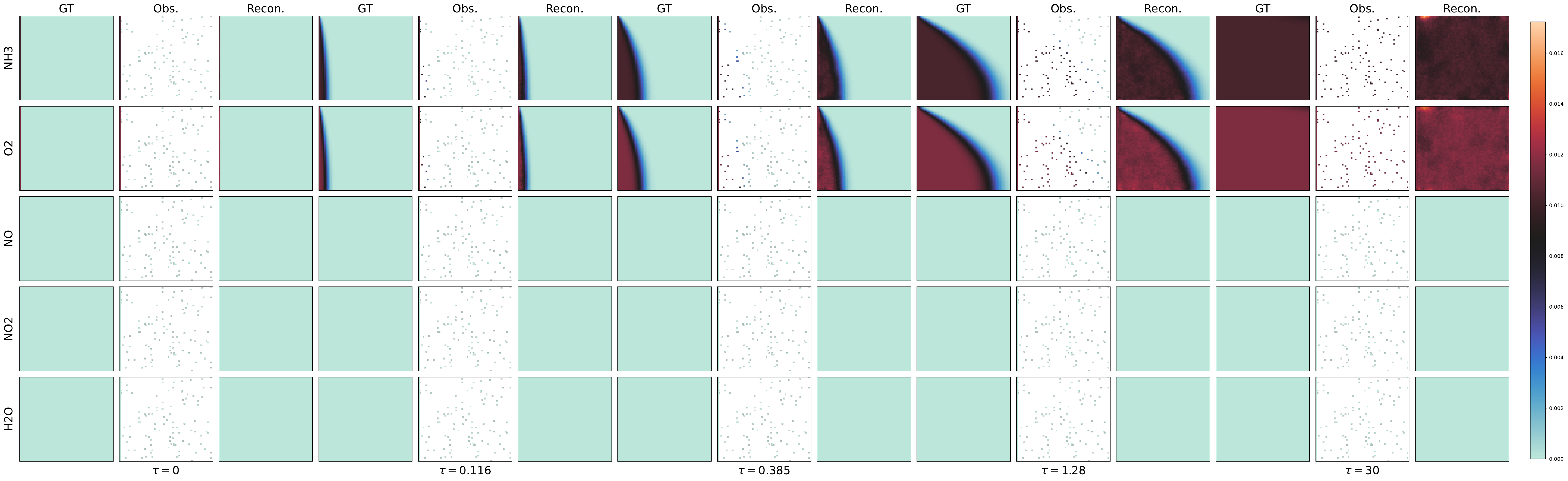}

\caption{Ammonia Oxidation reconstruction panel for the GEM method for select PDE times. Top: Normalised. Bottom: Physical.}
\label{fig:ammonia_oxidation_recon_gem}
\end{figure}

\end{landscape}

\begin{landscape}

\begin{figure}[p]
\centering

\includegraphics[width=1.4\textwidth]{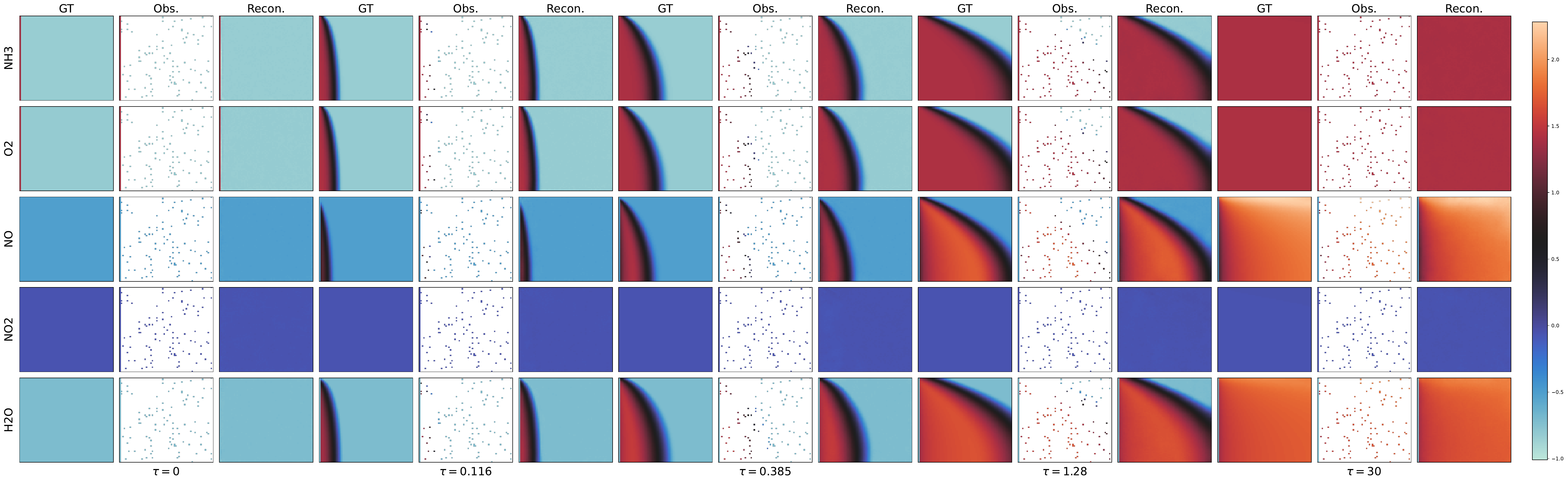}

\vspace{1cm}

\includegraphics[width=1.4\textwidth]{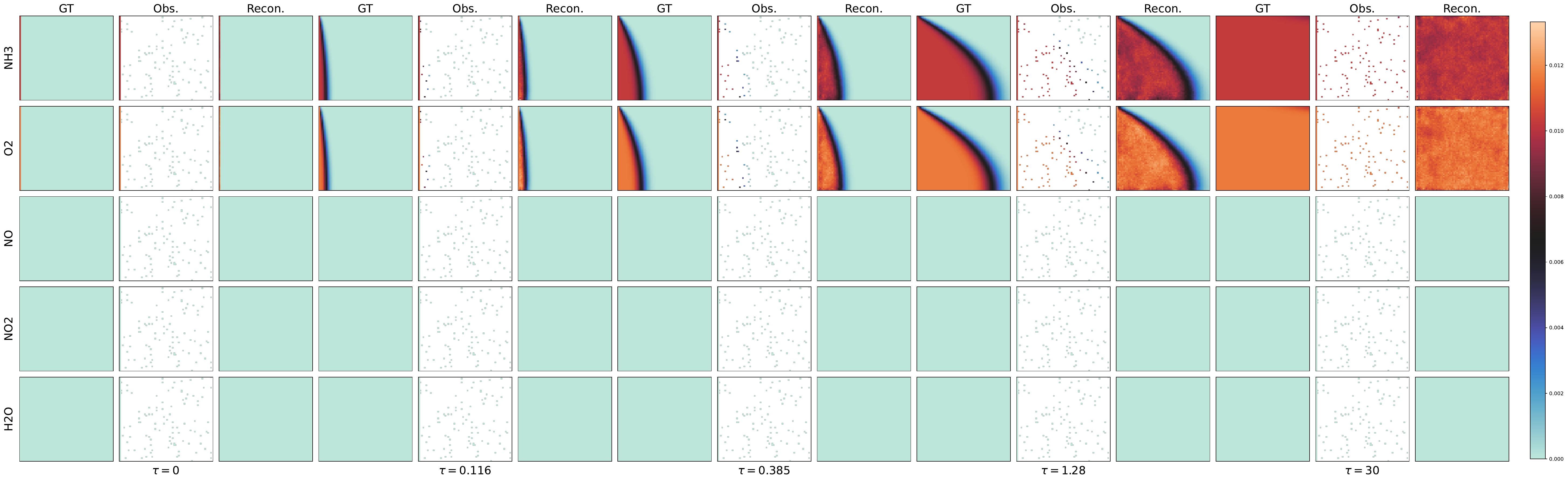}

\caption{Ammonia Oxidation reconstruction panel for the SOSaG method for select PDE times. Top: Normalised. Bottom: Physical.}
\label{fig:ammonia_oxidation_recon_sosag}
\end{figure}

\end{landscape}

\begin{landscape}

\begin{figure}[p]
\centering

\includegraphics[width=1.4\textwidth]{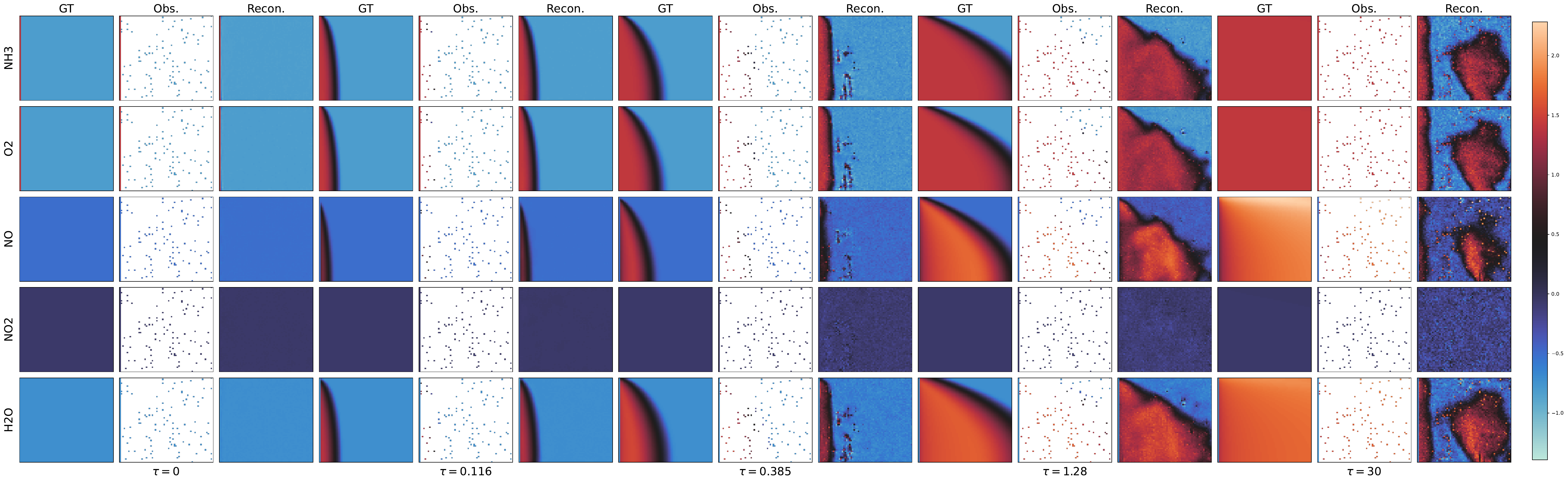}

\vspace{1cm}

\includegraphics[width=1.4\textwidth]{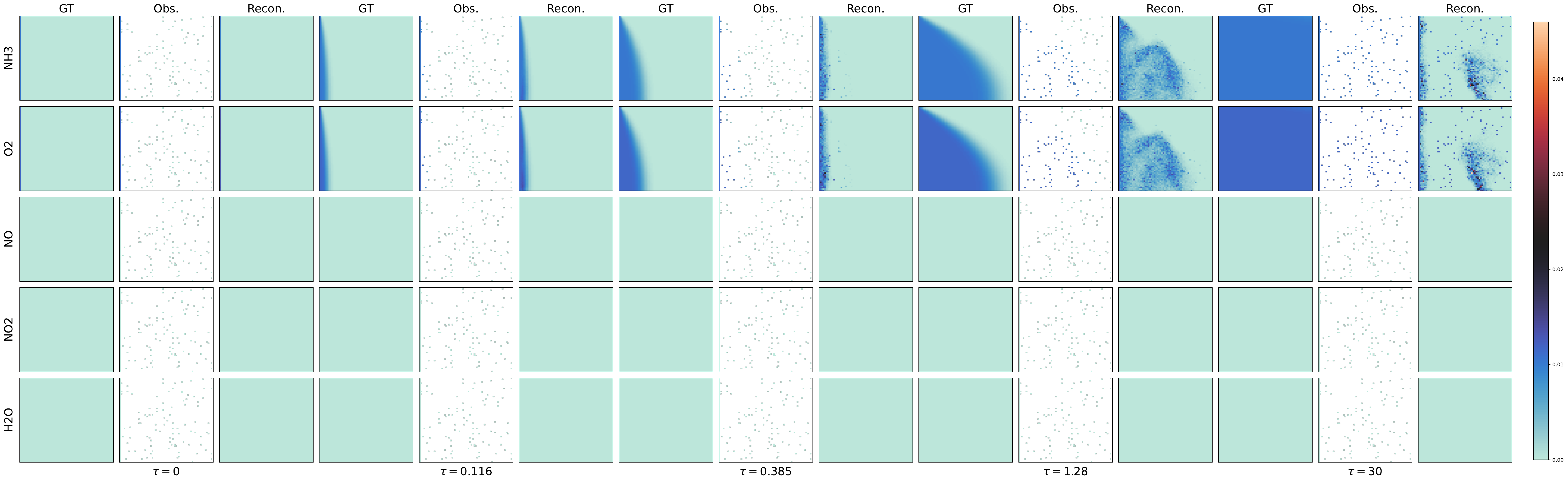}

\caption{Ammonia Oxidation reconstruction panel for the ODE method for select PDE times. Top: Normalised. Bottom: Physical.}
\label{fig:ammonia_oxidation_recon_ode}
\end{figure}

\end{landscape}

\begin{landscape}

\begin{figure}[p]
\centering

\includegraphics[width=1.4\textwidth]{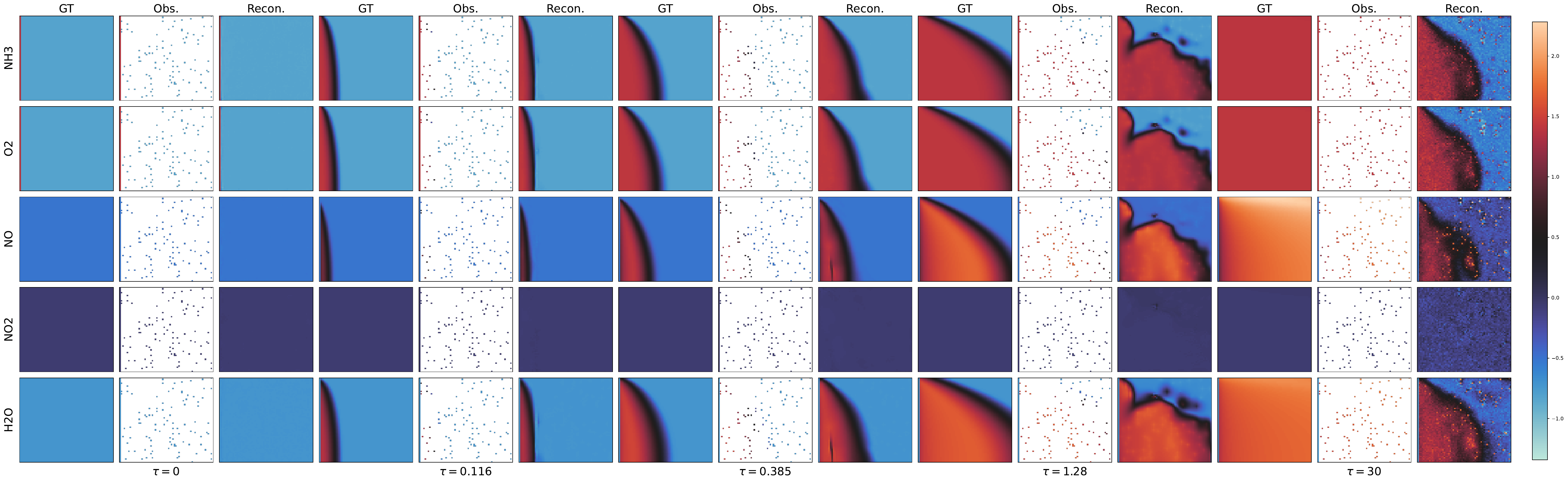}

\vspace{1cm}

\includegraphics[width=1.4\textwidth]{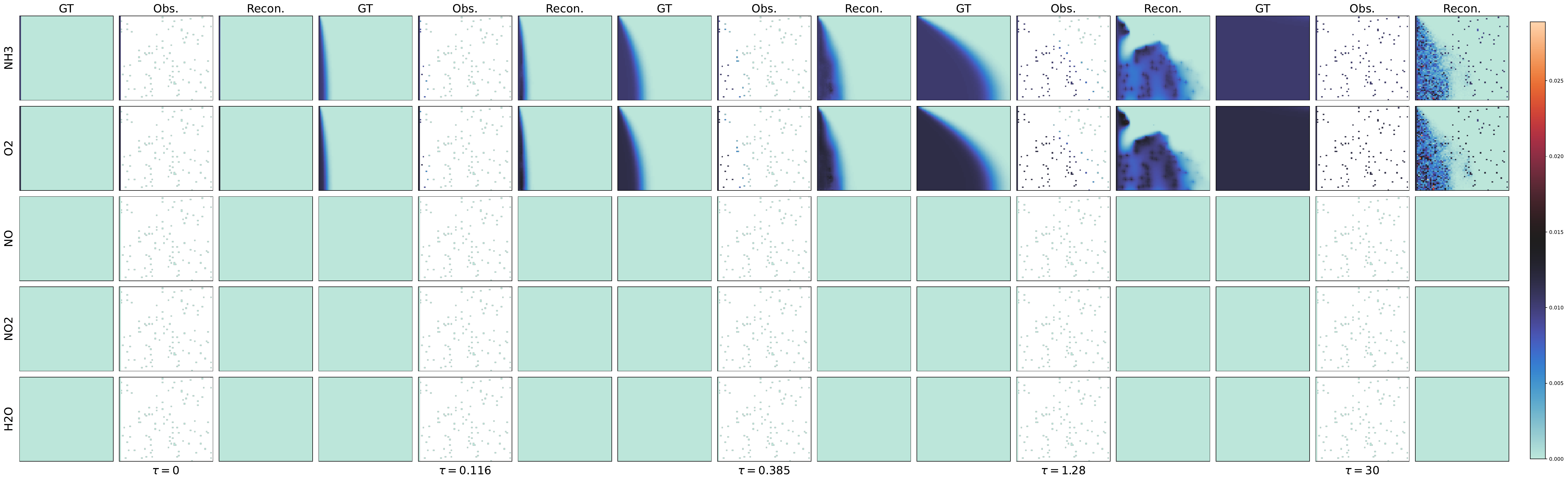}

\caption{Ammonia Oxidation reconstruction panel for the DiffPDE method for select PDE times. Top: Normalised. Bottom: Physical.}
\label{fig:ammonia_oxidation_recon_diffpde}
\end{figure}

\end{landscape}

\begin{landscape}

\begin{figure}[p]
\centering

\includegraphics[width=1.2\textwidth]{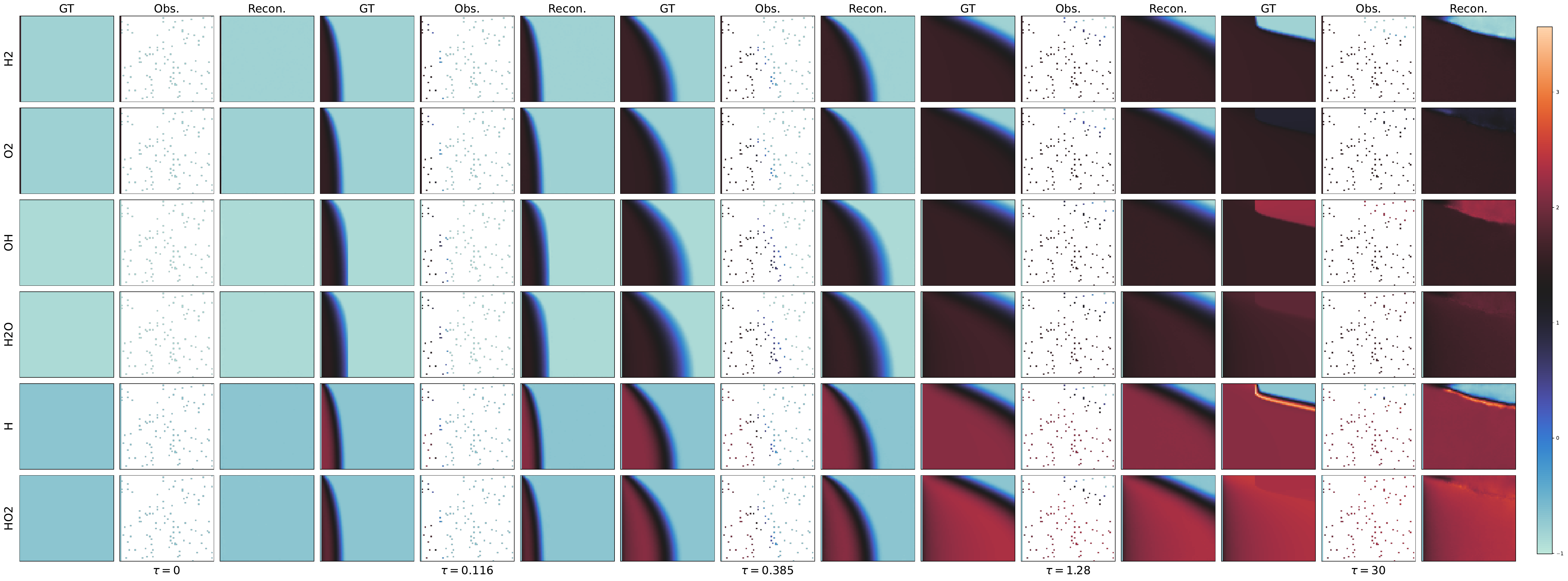}

\vspace{1cm}

\includegraphics[width=1.2\textwidth]{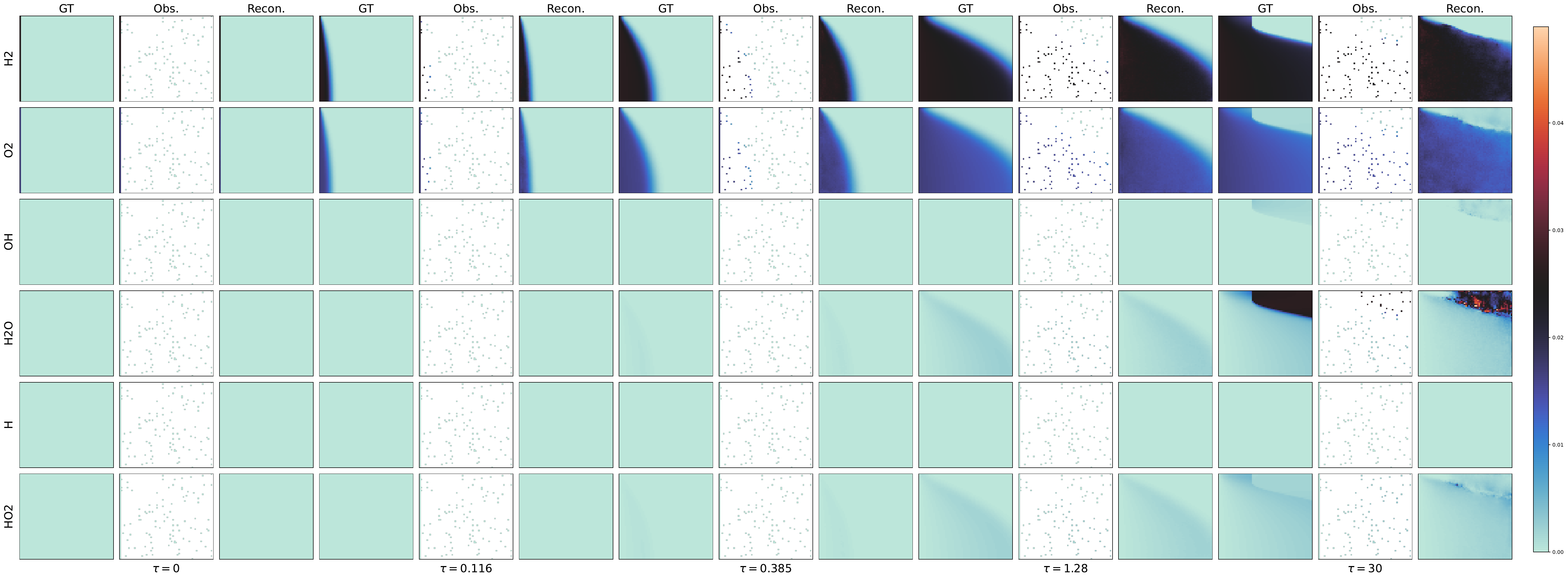}

\caption{Hydrogen Oxidation Subset reconstruction panel for the GEM method for select PDE times. Top: Normalised. Bottom: Physical.}
\label{fig:hydrogen_oxidation_recon_gem}
\end{figure}

\end{landscape}

\begin{landscape}

\begin{figure}[p]
\centering

\includegraphics[width=1.2\textwidth]{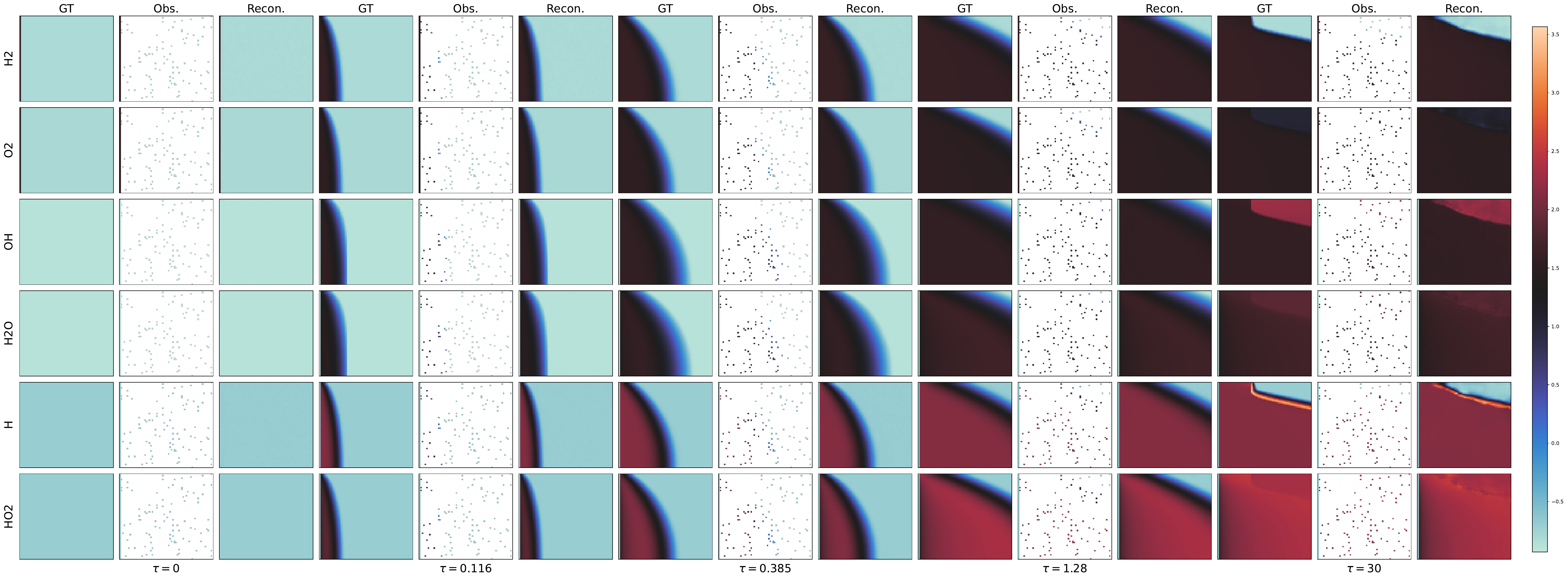}

\vspace{1cm}

\includegraphics[width=1.2\textwidth]{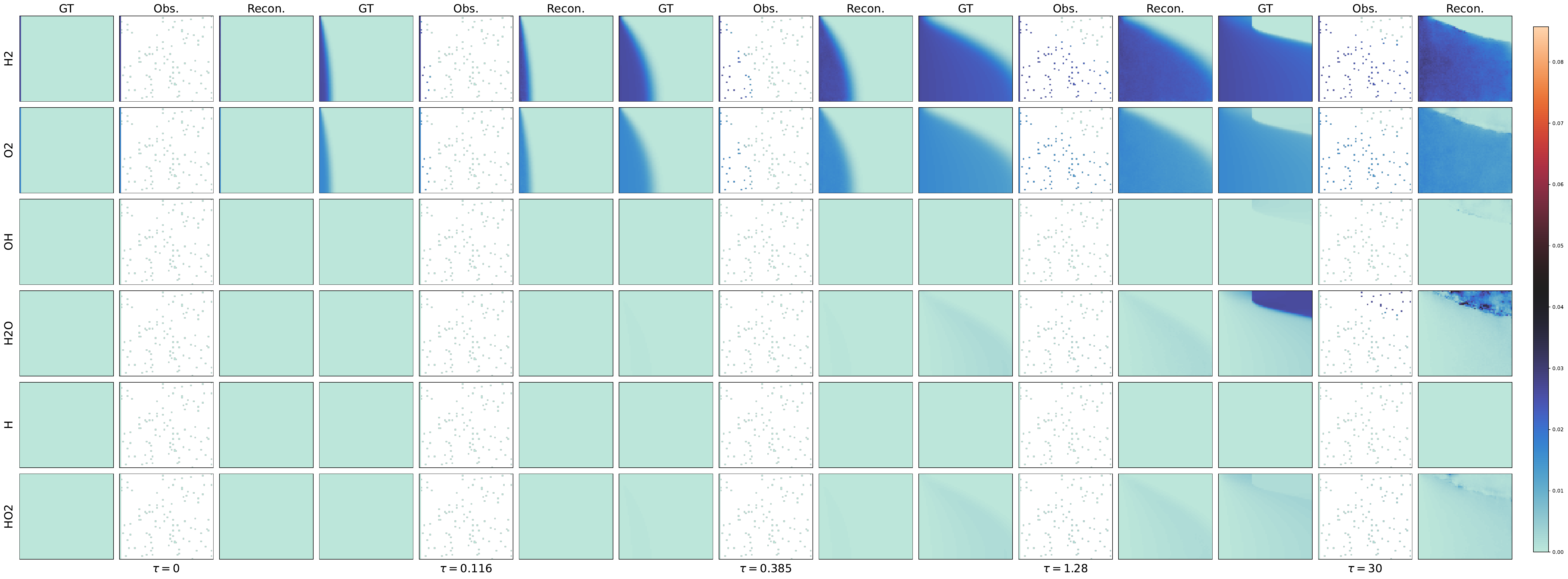}

\caption{Hydrogen Oxidation Subset reconstruction panel for the SOSaG method for select PDE times. Top: Normalised. Bottom: Physical.}
\label{fig:hydrogen_oxidation_recon_sosag}
\end{figure}

\end{landscape}

\begin{landscape}

\begin{figure}[p]
\centering

\includegraphics[width=1.2\textwidth]{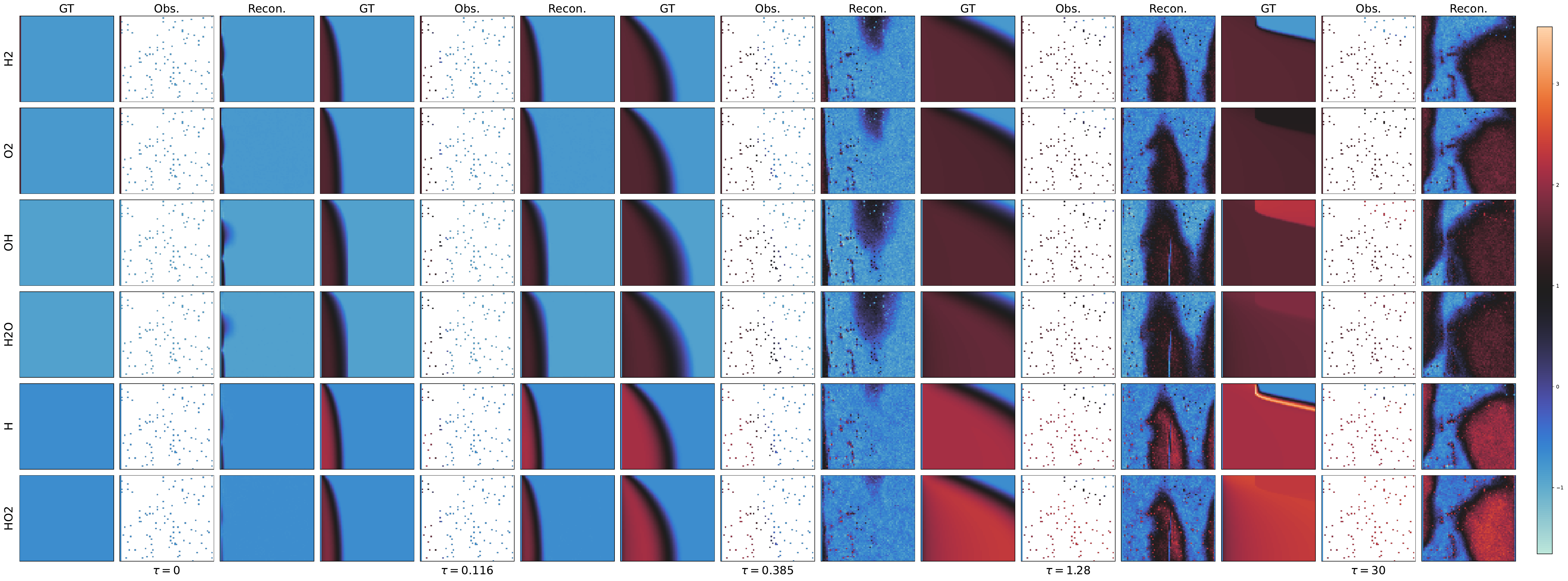}

\vspace{1cm}

\includegraphics[width=1.2\textwidth]{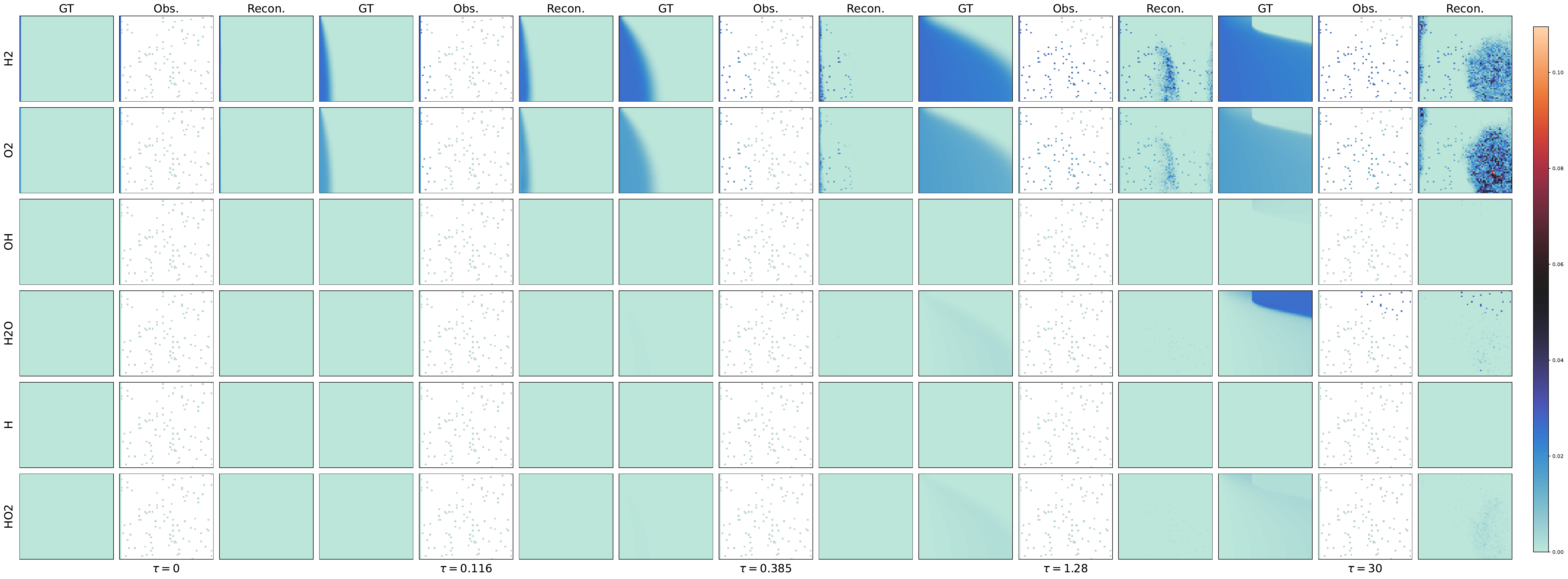}

\caption{Hydrogen Oxidation Subset reconstruction panel for the ODE method for select PDE times. Top: Normalised. Bottom: Physical.}
\label{fig:hydrogen_oxidation_recon_ode}
\end{figure}

\end{landscape}

\begin{landscape}

\begin{figure}[p]
\centering

\includegraphics[width=1.2\textwidth]{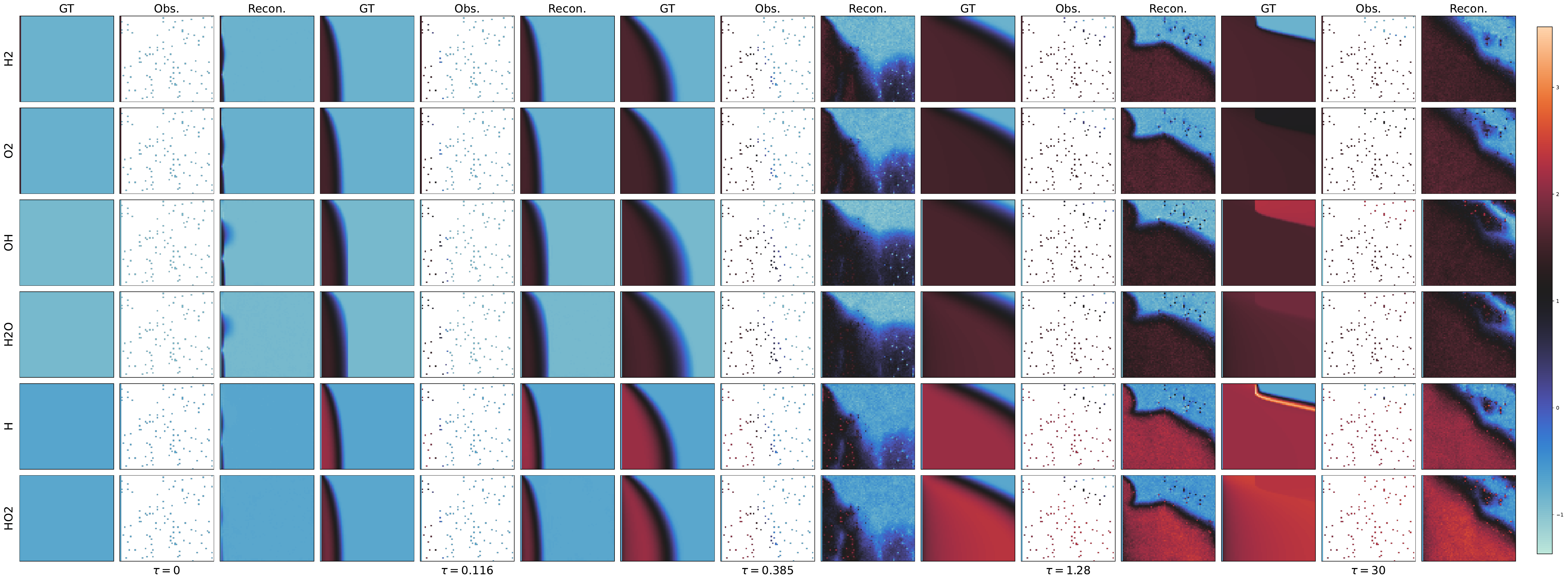}

\vspace{1cm}

\includegraphics[width=1.2\textwidth]{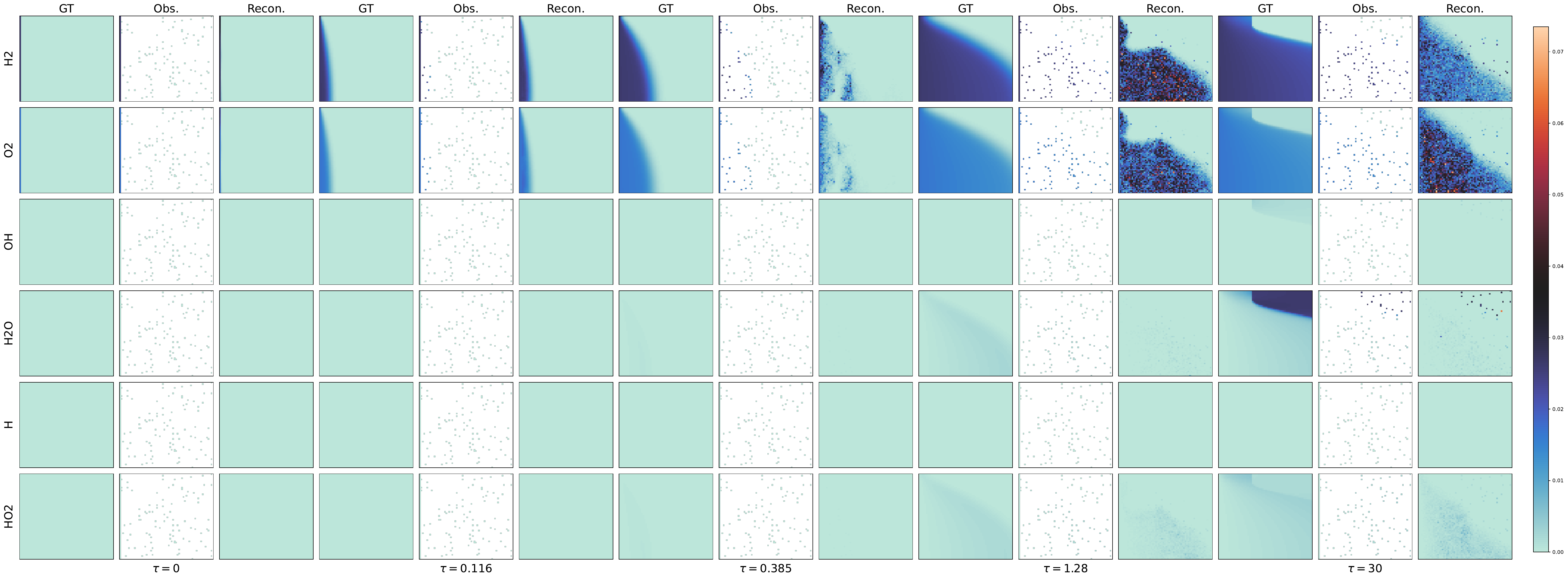}

\caption{Hydrogen Oxidation Subset reconstruction panel for the DiffPDE method for select PDE times. Top: Normalised. Bottom: Physical.}
\label{fig:hydrogen_oxidation_recon_diffpde}
\end{figure}

\end{landscape}

\begin{landscape}

\begin{figure}[p]
\centering

\includegraphics[width=1.4\textwidth]{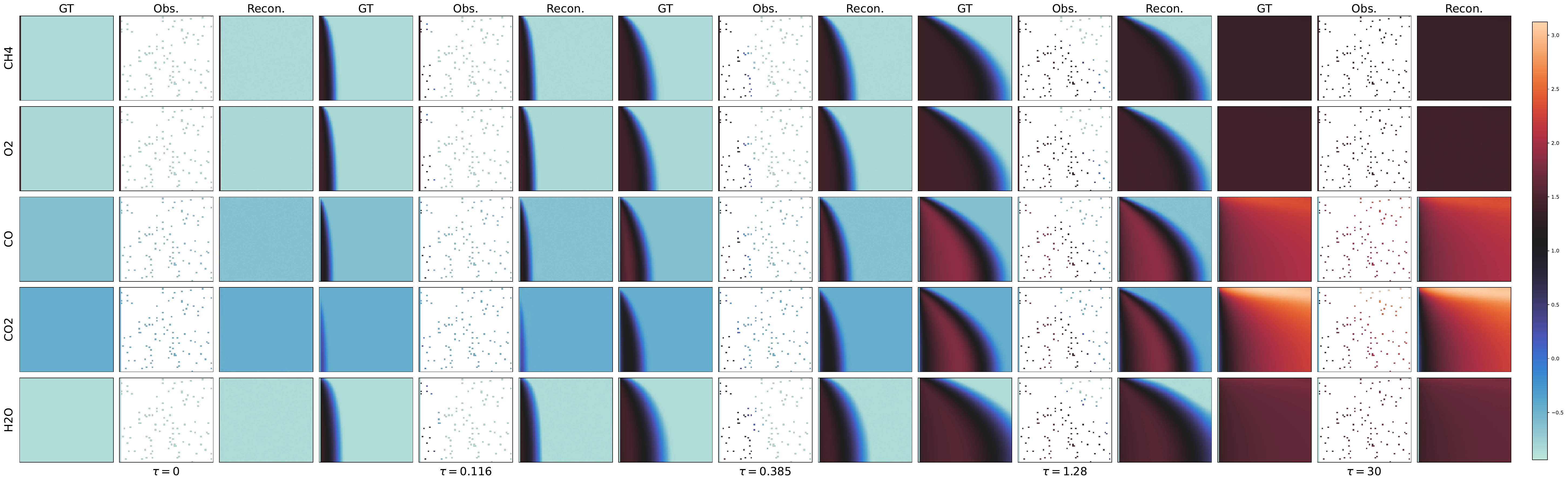}

\vspace{1cm}

\includegraphics[width=1.4\textwidth]{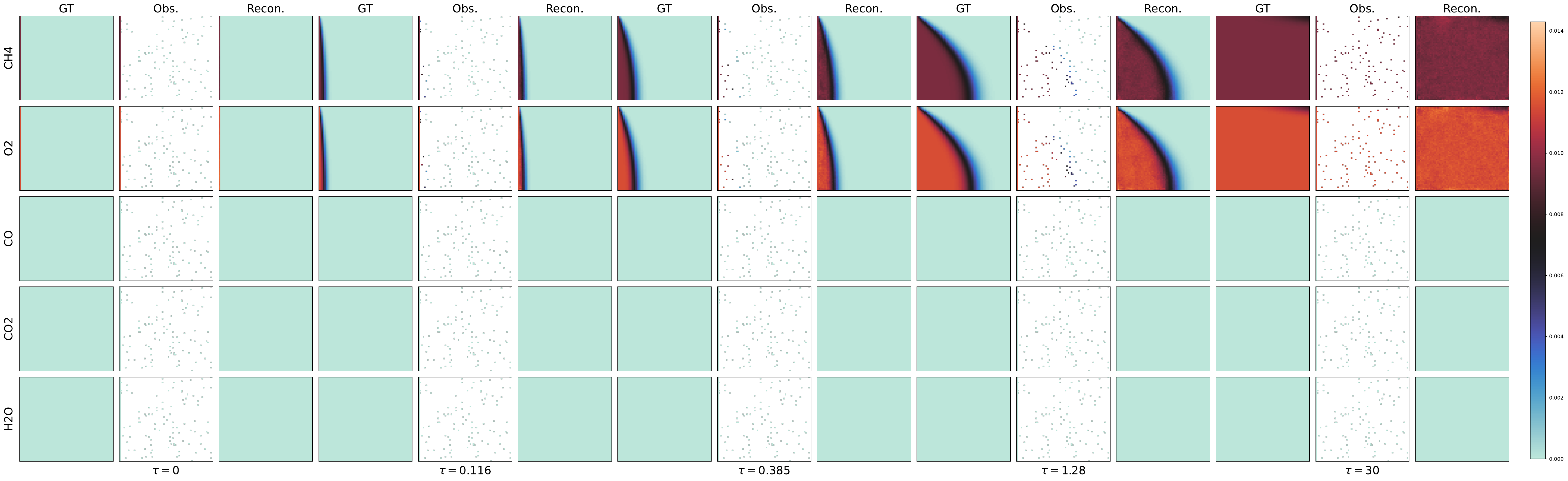}

\caption{Two-step Methane Oxidation Subset reconstruction panel for the GEM method for select PDE times. Top: Normalised. Bottom: Physical.}
\label{fig:methane_oxidation_recon_gem}
\end{figure}

\end{landscape}

\begin{landscape}

\begin{figure}[p]
\centering

\includegraphics[width=1.4\textwidth]{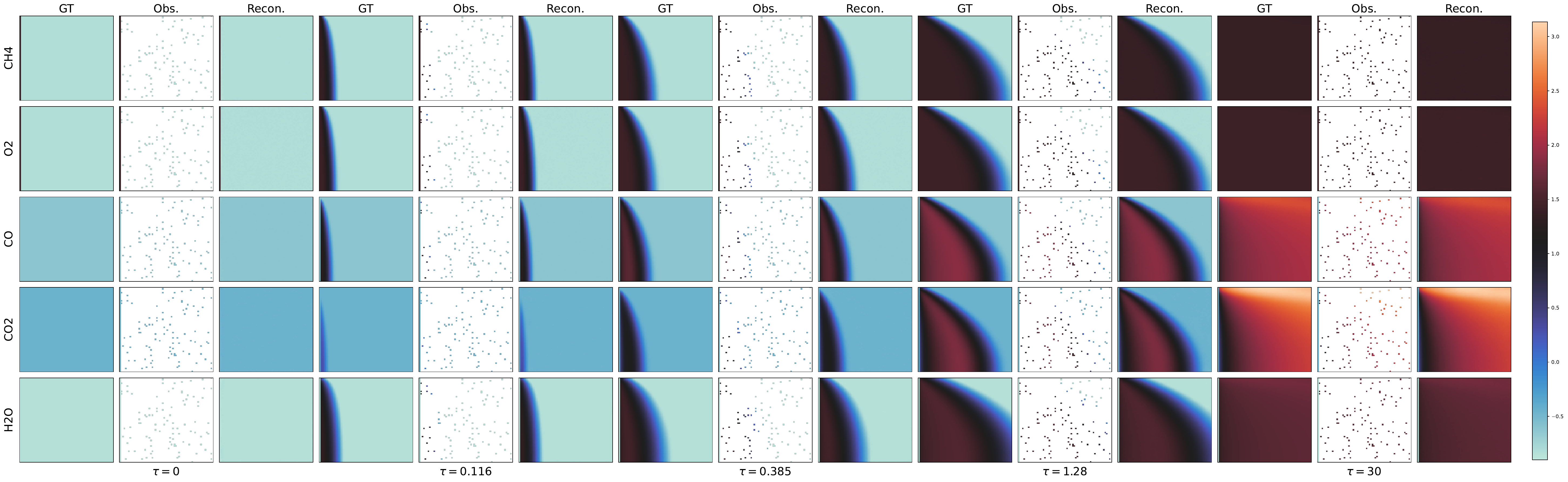}

\vspace{1cm}

\includegraphics[width=1.4\textwidth]{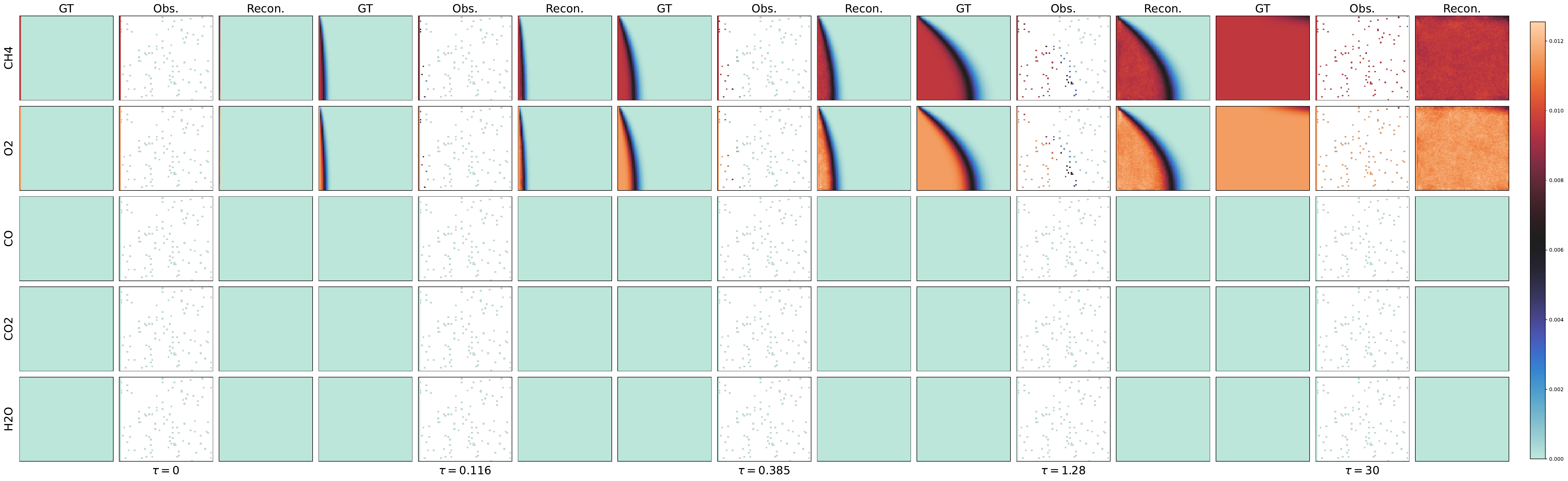}

\caption{Two-step Methane Oxidation Subset reconstruction panel for the SOSaG method for select PDE times. Top: Normalised. Bottom: Physical.}
\label{fig:methane_oxidation_recon_sosag}
\end{figure}

\end{landscape}

\begin{landscape}

\begin{figure}[p]
\centering

\includegraphics[width=1.4\textwidth]{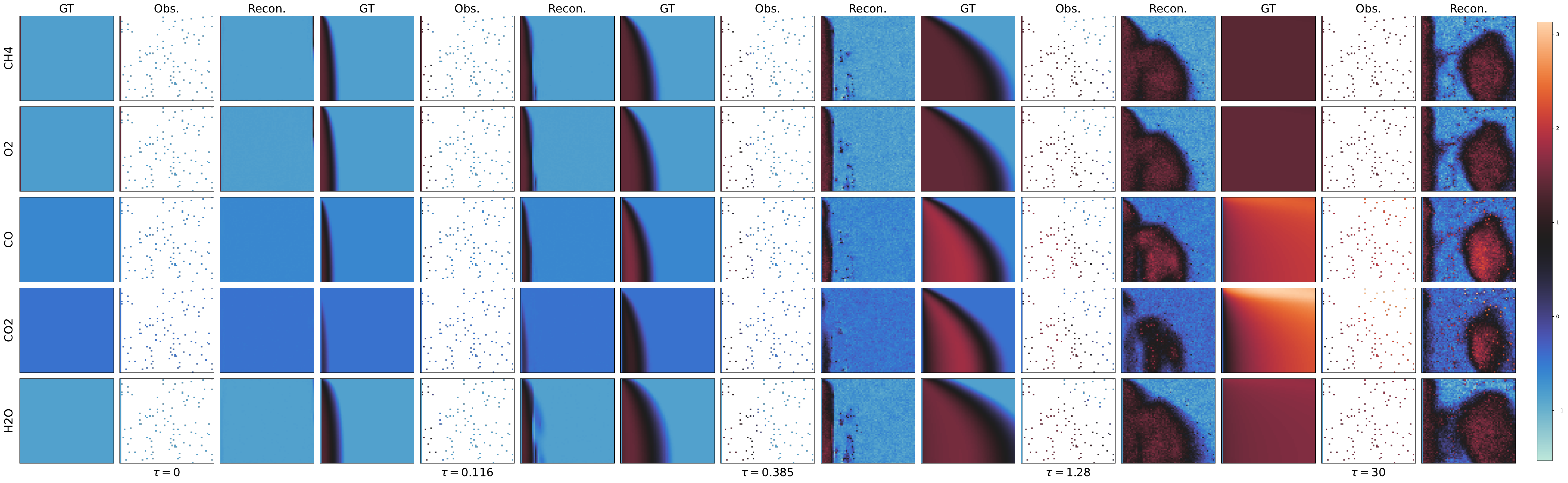}

\vspace{1cm}

\includegraphics[width=1.4\textwidth]{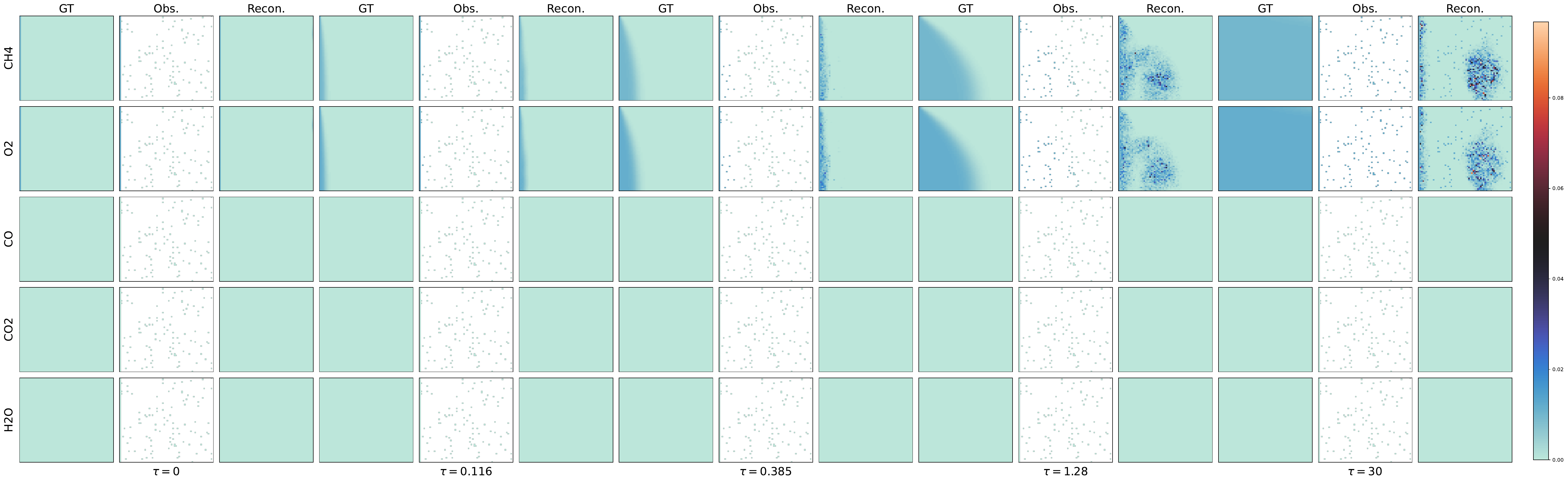}

\caption{Two-step Methane Oxidation Subset reconstruction panel for the ODE method for select PDE times. Top: Normalised. Bottom: Physical.}
\label{fig:methane_oxidation_recon_ode}
\end{figure}

\end{landscape}

\begin{landscape}

\begin{figure}[p]
\centering

\includegraphics[width=1.4\textwidth]{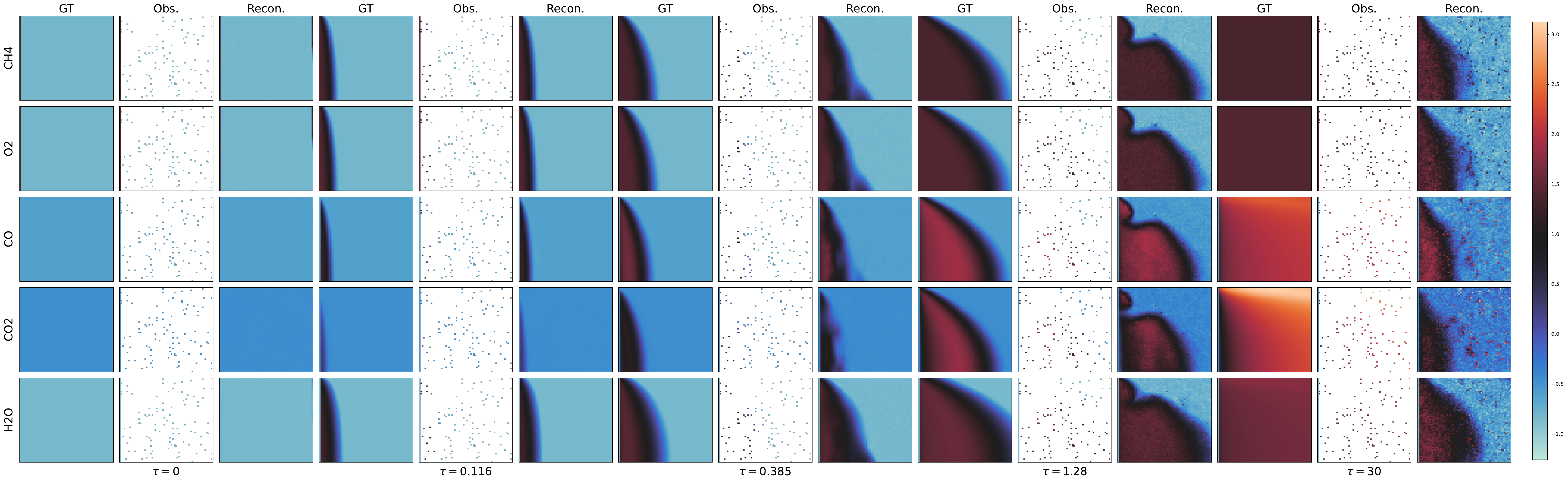}

\vspace{1cm}

\includegraphics[width=1.4\textwidth]{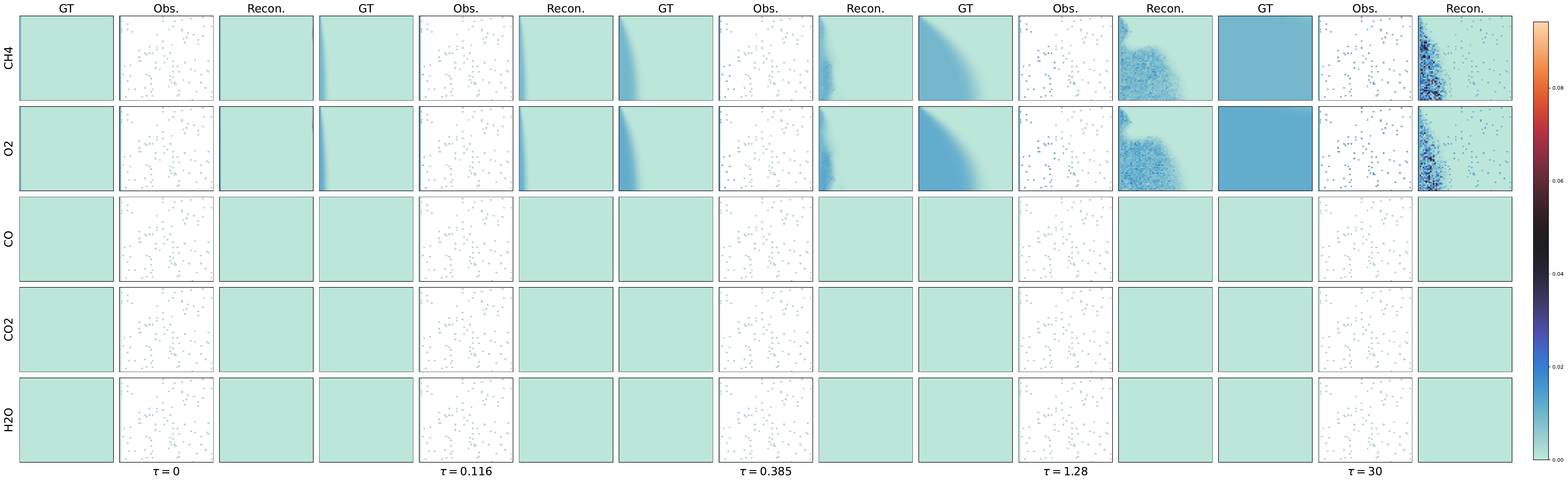}

\caption{Two-step Methane Oxidation Subset reconstruction panel for the DiffPDE method for select PDE times. Top: Normalised. Bottom: Physical.}
\label{fig:methane_oxidation_recon_diffpde}
\end{figure}

\end{landscape}

\begin{landscape}

\begin{figure}[p]
\centering

\includegraphics[width=1.4\textwidth]{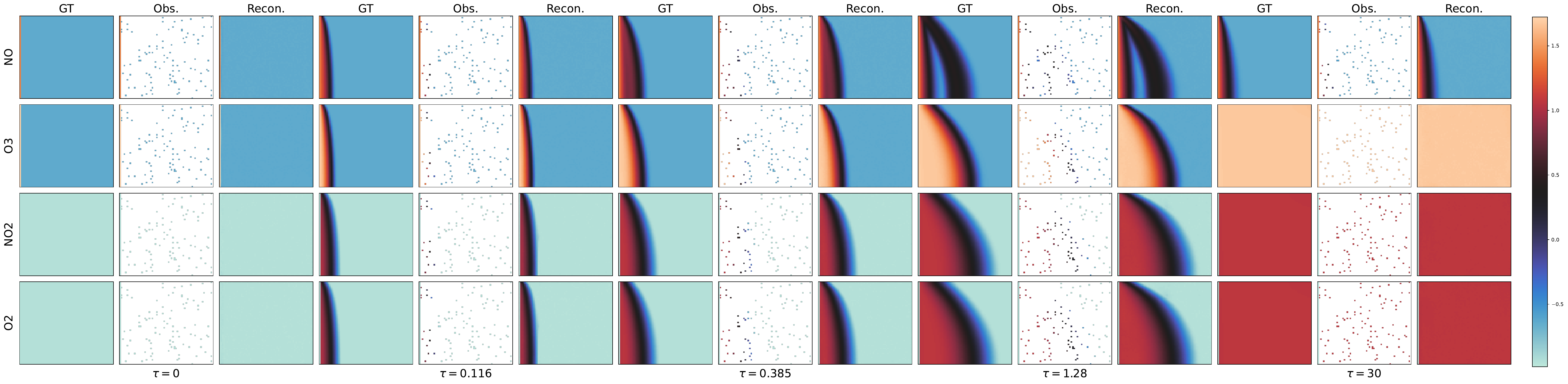}

\vspace{1cm}

\includegraphics[width=1.4\textwidth]{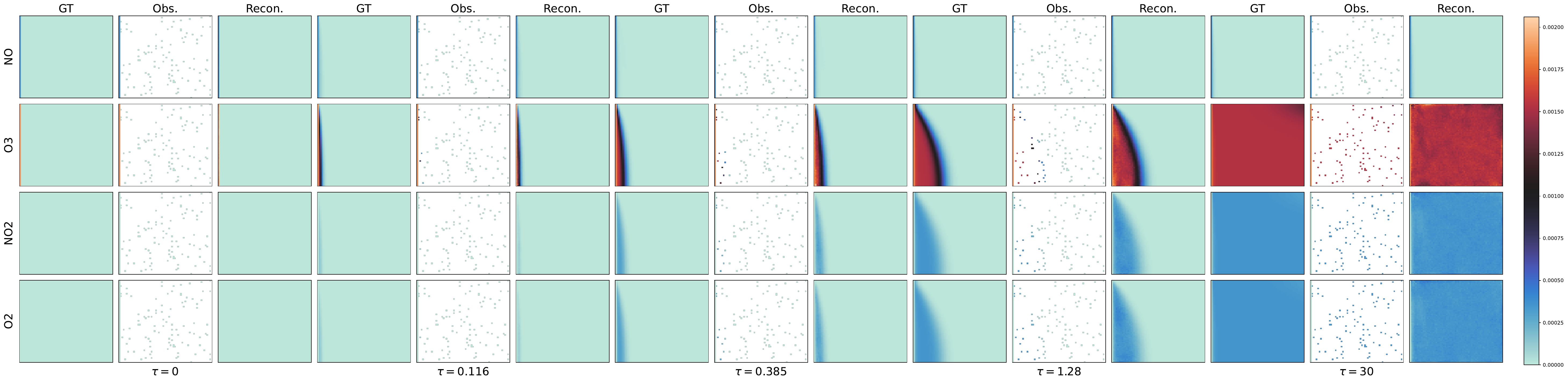}

\caption{NO + O$_3 \rightarrow$ NO$_2$ reconstruction panel for the GEM method for select PDE times. Top: Normalised. Bottom: Physical.}
\label{fig:no_o3_recon_gem}
\end{figure}

\end{landscape}

\begin{landscape}

\begin{figure}[p]
\centering

\includegraphics[width=1.4\textwidth]{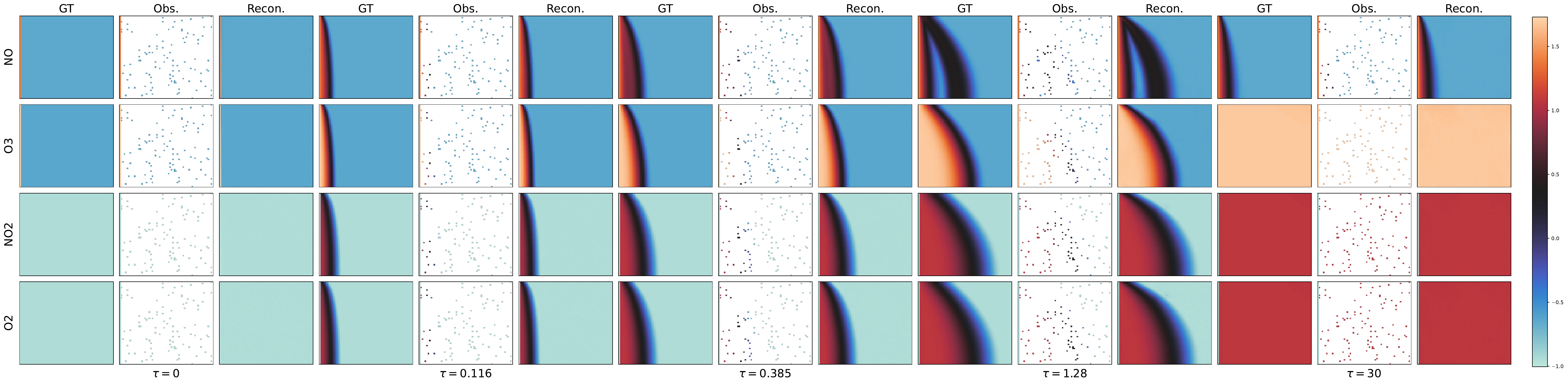}

\vspace{1cm}

\includegraphics[width=1.4\textwidth]{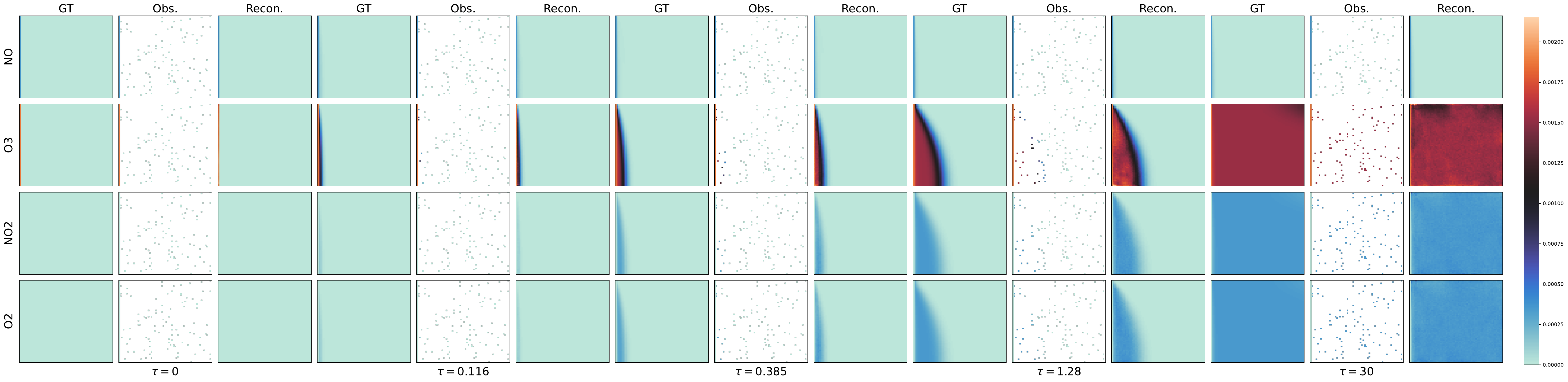}

\caption{NO + O$_3 \rightarrow$ NO$_2$ reconstruction panel for the SOSaG method for select PDE times. Top: Normalised. Bottom: Physical.}
\label{fig:no_o3_recon_sosag}
\end{figure}

\end{landscape}

\begin{landscape}

\begin{figure}[p]
\centering

\includegraphics[width=1.4\textwidth]{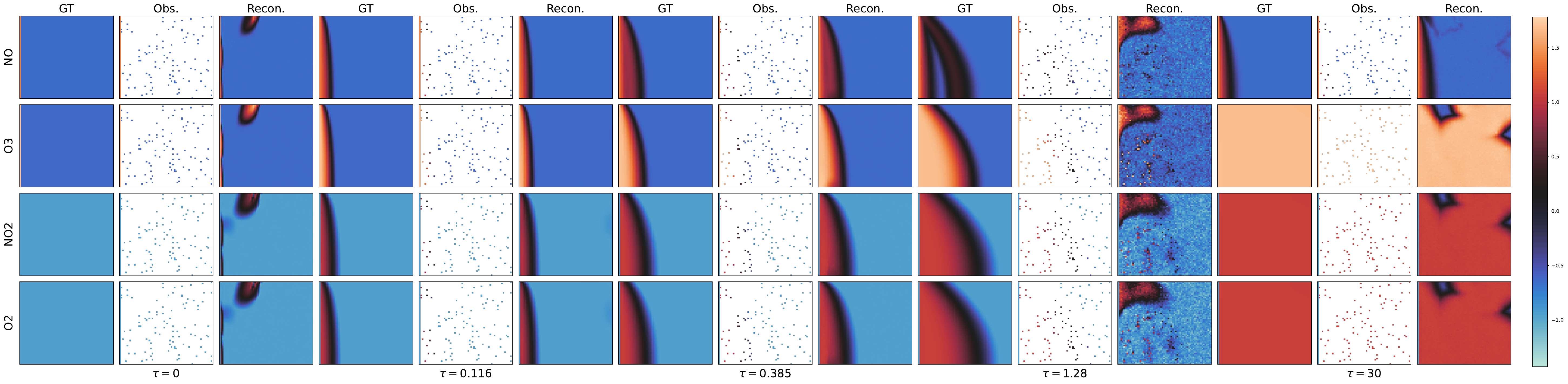}

\vspace{1cm}

\includegraphics[width=1.4\textwidth]{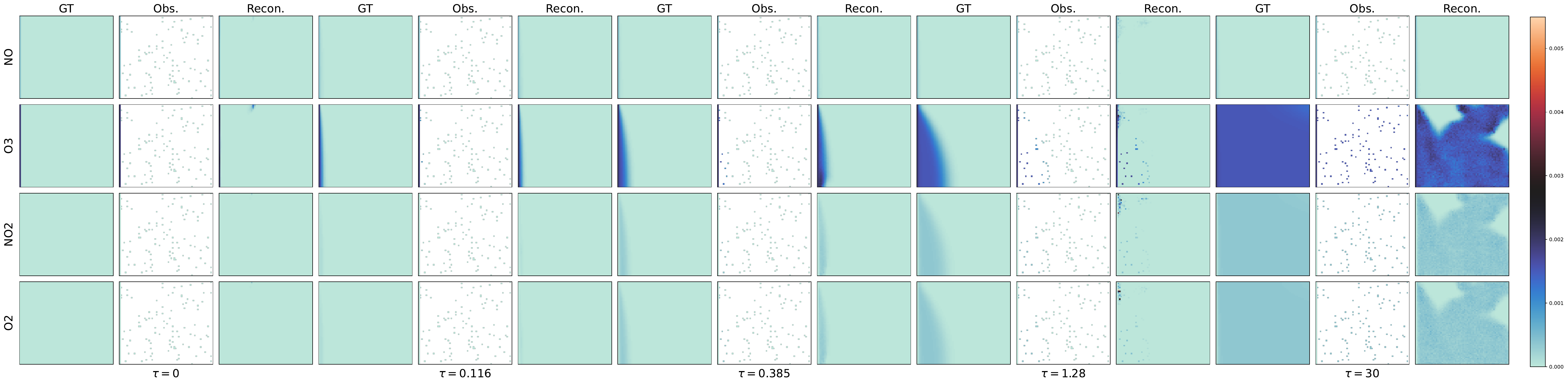}

\caption{NO + O$_3 \rightarrow$ NO$_2$ reconstruction panel for the ODE method for select PDE times. Top: Normalised. Bottom: Physical.}
\label{fig:no_o3_recon_ode}
\end{figure}

\end{landscape}

\begin{landscape}

\begin{figure}[p]
\centering

\includegraphics[width=1.4\textwidth]{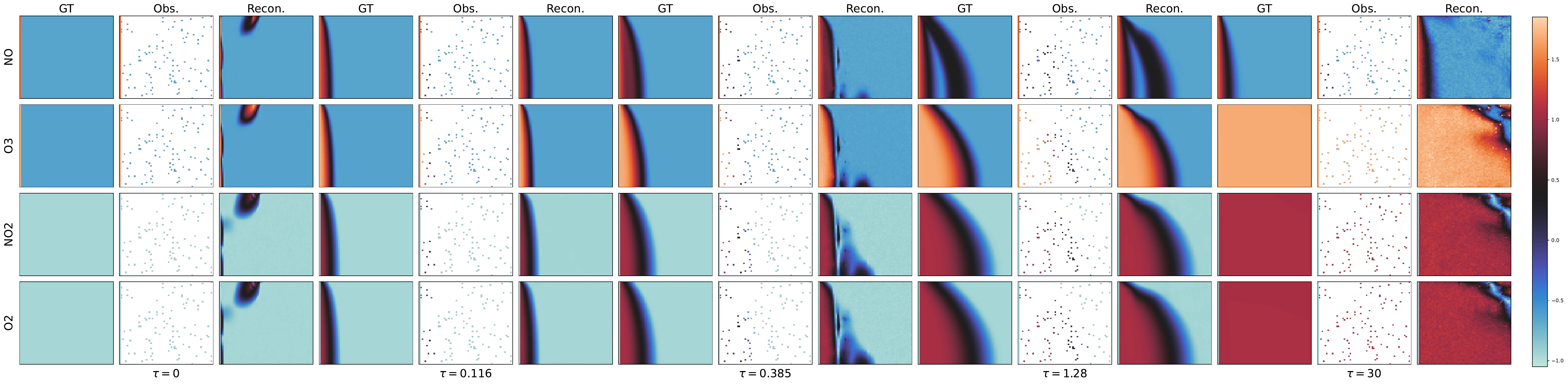}

\vspace{1cm}

\includegraphics[width=1.4\textwidth]{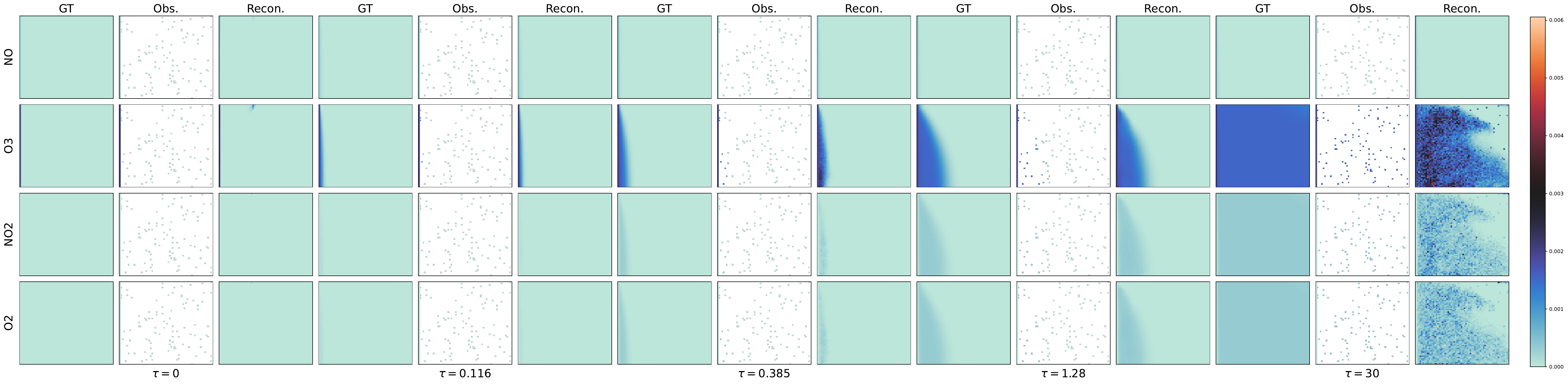}

\caption{NO + O$_3 \rightarrow$ NO$_2$ reconstruction panel for the DiffPDE method for select PDE times. Top: Normalised. Bottom: Physical.}
\label{fig:no_o3_recon_diffpde}
\end{figure}

\end{landscape}

\begin{landscape}

\begin{figure}[p]
\centering

\includegraphics[width=1.4\textwidth]{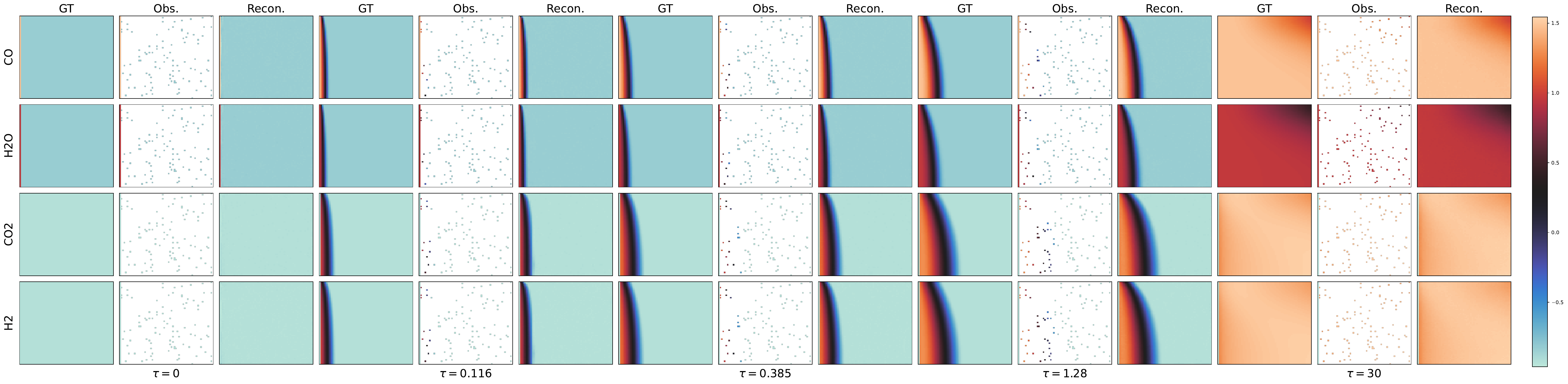}

\vspace{1cm}

\includegraphics[width=1.4\textwidth]{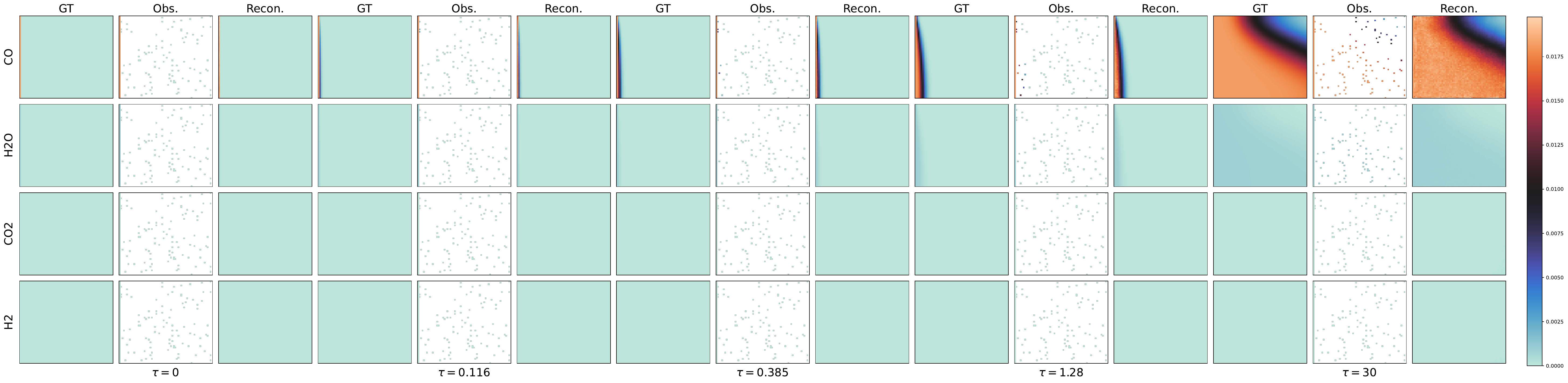}

\caption{Water Gas Shift reconstruction panel for the GEM method for select PDE times. Top: Normalised. Bottom: Physical.}
\label{fig:water_gas_shift_recon_gem}
\end{figure}

\end{landscape}

\begin{landscape}

\begin{figure}[p]
\centering

\includegraphics[width=1.4\textwidth]{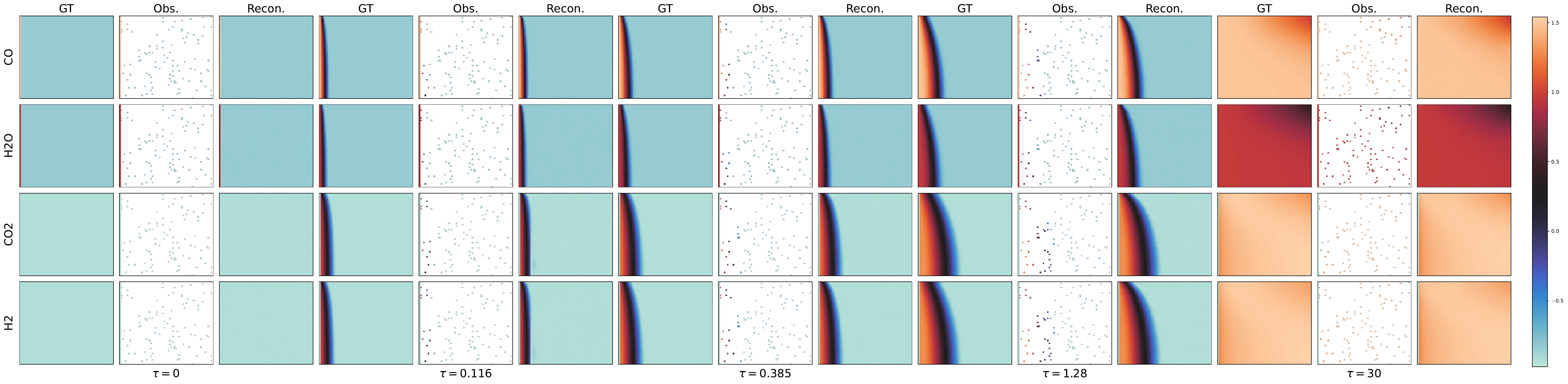}

\vspace{1cm}

\includegraphics[width=1.4\textwidth]
{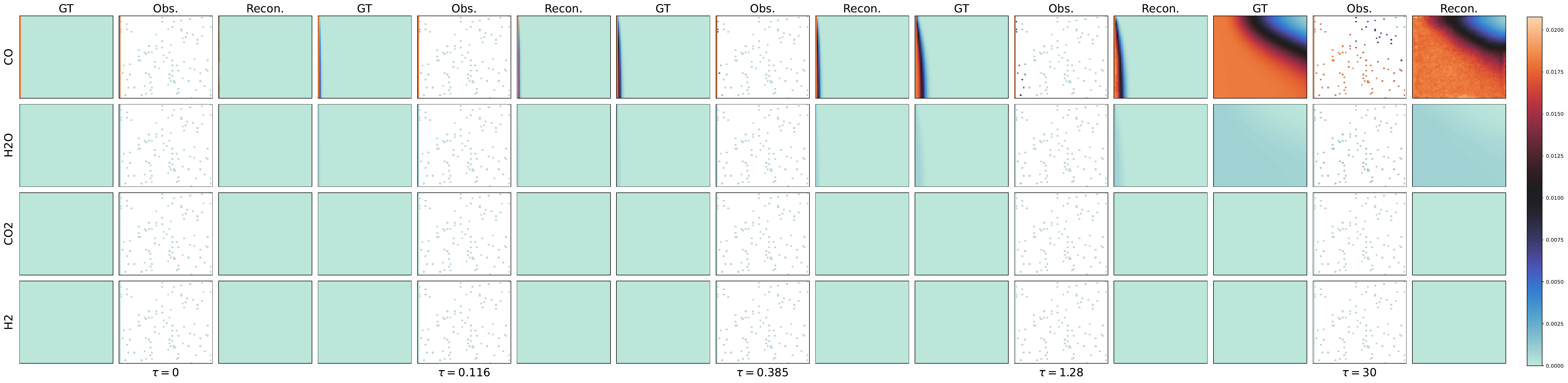}

\caption{Water Gas Shift reconstruction panel for the SOSaG method for select PDE times. Top: Normalised. Bottom: Physical.}
\label{fig:water_gas_shift_recon_sosag}
\end{figure}

\end{landscape}

\begin{landscape}

\begin{figure}[p]
\centering

\includegraphics[width=1.4\textwidth]{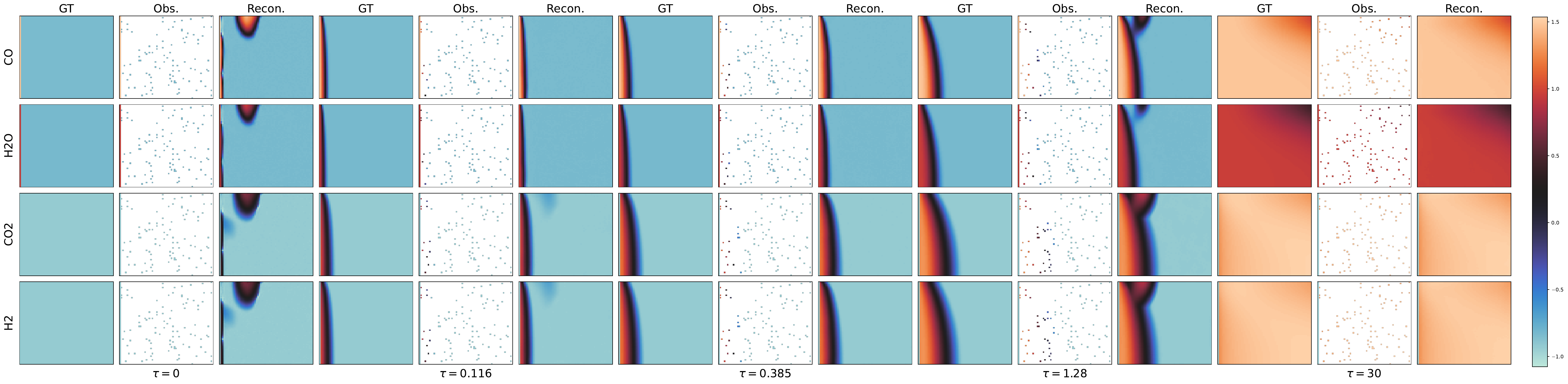}

\vspace{1cm}

\includegraphics[width=1.4\textwidth]{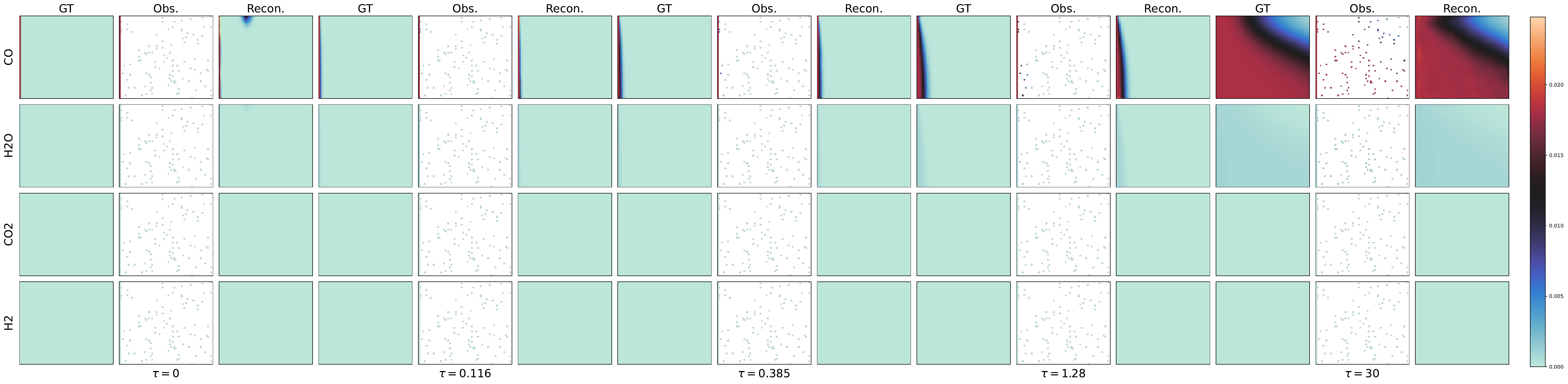}

\caption{Water Gas Shift reconstruction panel for the ODE method for select PDE times. Top: Normalised. Bottom: Physical.}
\label{fig:water_gas_shift_recon_ode}
\end{figure}

\end{landscape}

\begin{landscape}

\begin{figure}[p]
\centering

\includegraphics[width=1.4\textwidth]{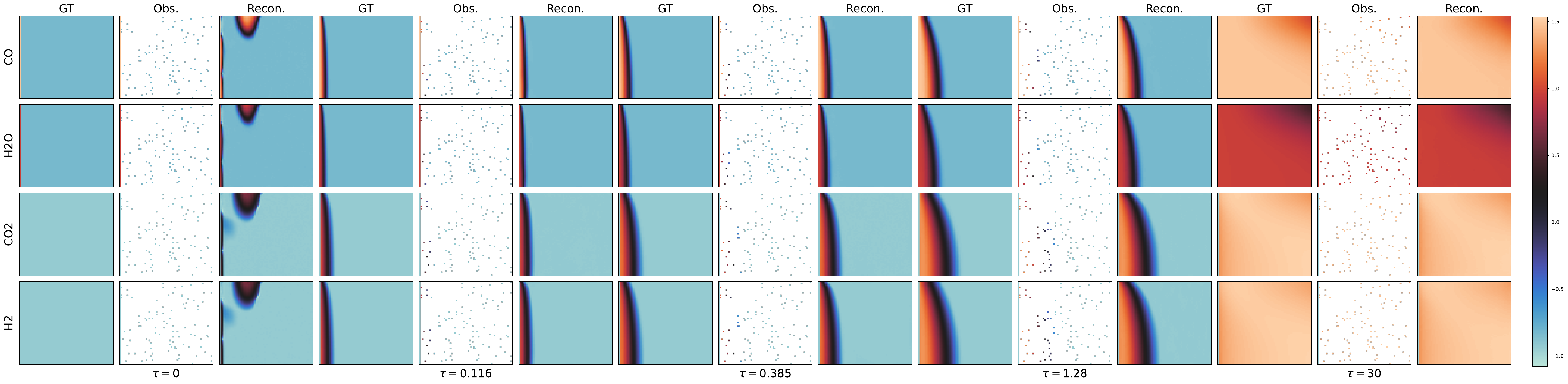}

\vspace{1cm}

\includegraphics[width=1.4\textwidth]{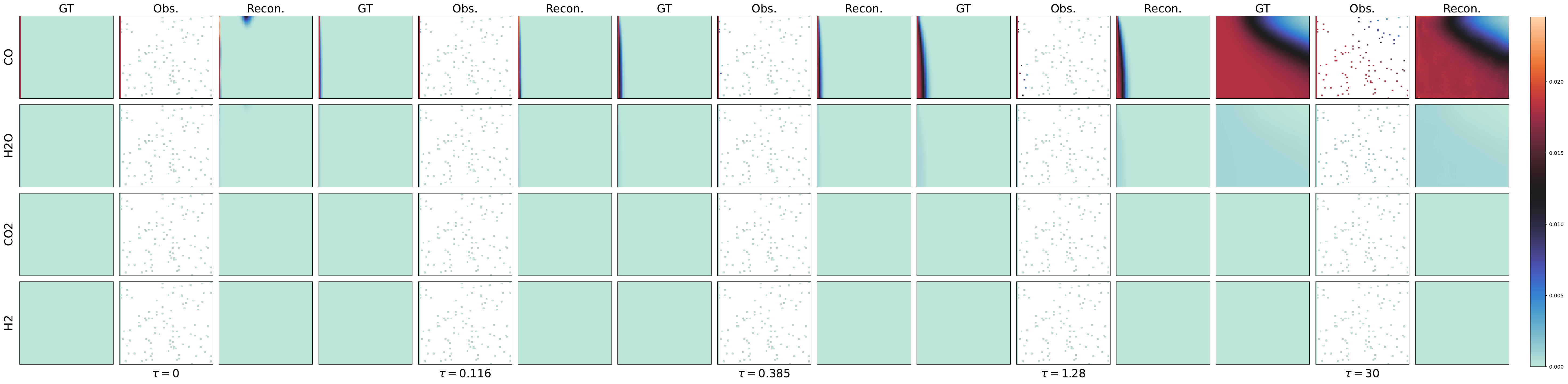}

\caption{Water Gas Shift reconstruction panel for the DiffPDE method for select PDE times. Top: Normalised. Bottom: Physical.}
\label{fig:water_gas_shift_recon_diffpde}
\end{figure}

\end{landscape}

Figures~\ref{fig:h2o2_recon_gem}--\ref{fig:water_gas_shift_recon_diffpde} show an example field reconstructions at selected time snapshots for each method. All methods use the same random seed for the example run.

\section{Error Maps} \label{app:error_map}

\begin{figure*}[!t]
    \centering
    \includegraphics[width=1.0\linewidth]{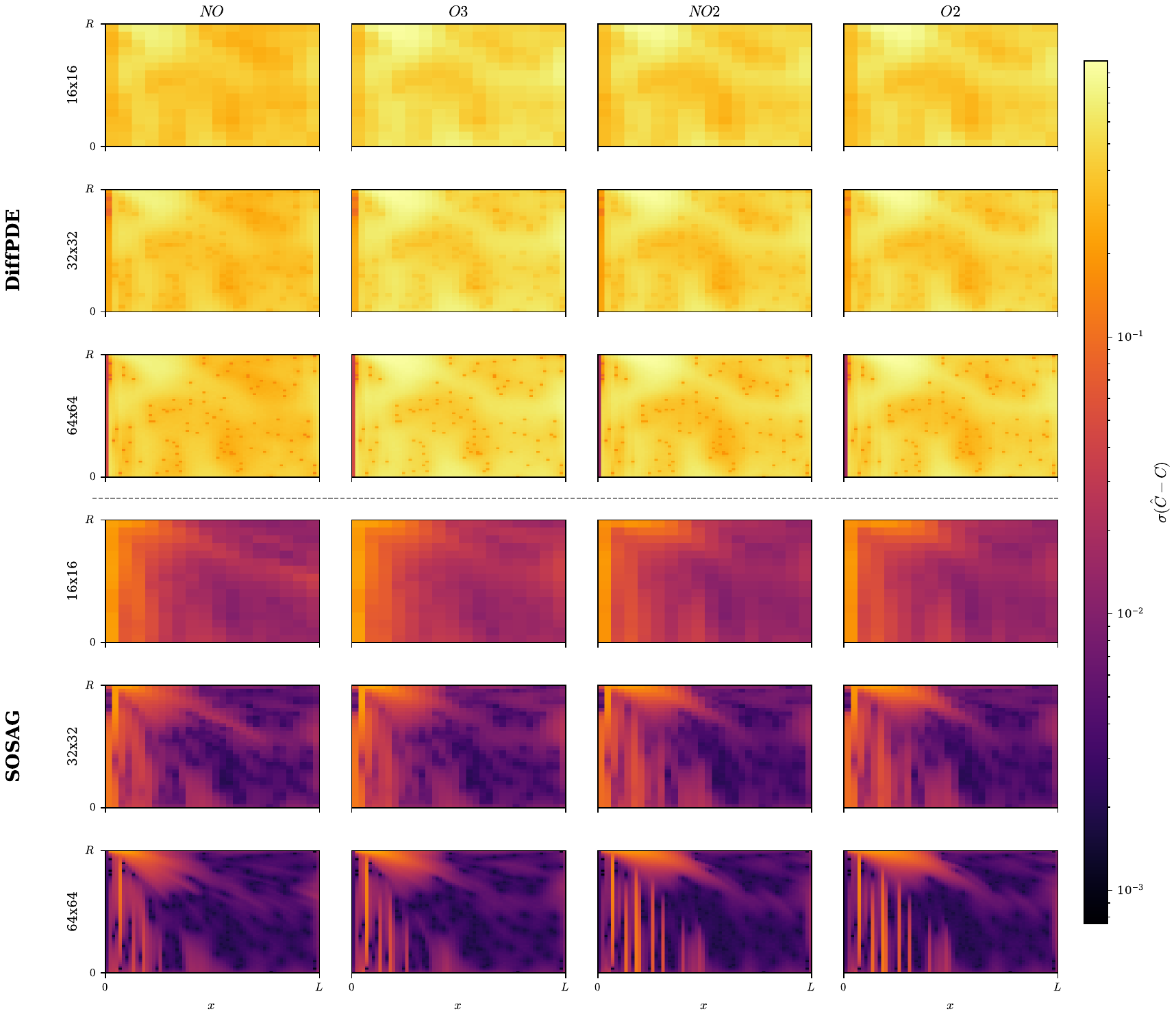}
    \caption{Map of the standard deviation of reconstruction errors for the DiffPDE and SOSaG methods on the NO-O$_3$ to NO$_2$ dataset for differing resolutions. The standard deviation is calculated per pixel over all timesteps across the 10 test set trajectories. SOSaG produces a lower error variability compared to DiffPDE.}
    \label{fig:NO_error_map}
\end{figure*}

\begin{figure*}[!t]
    \centering
    \includegraphics[width=1.0\linewidth]{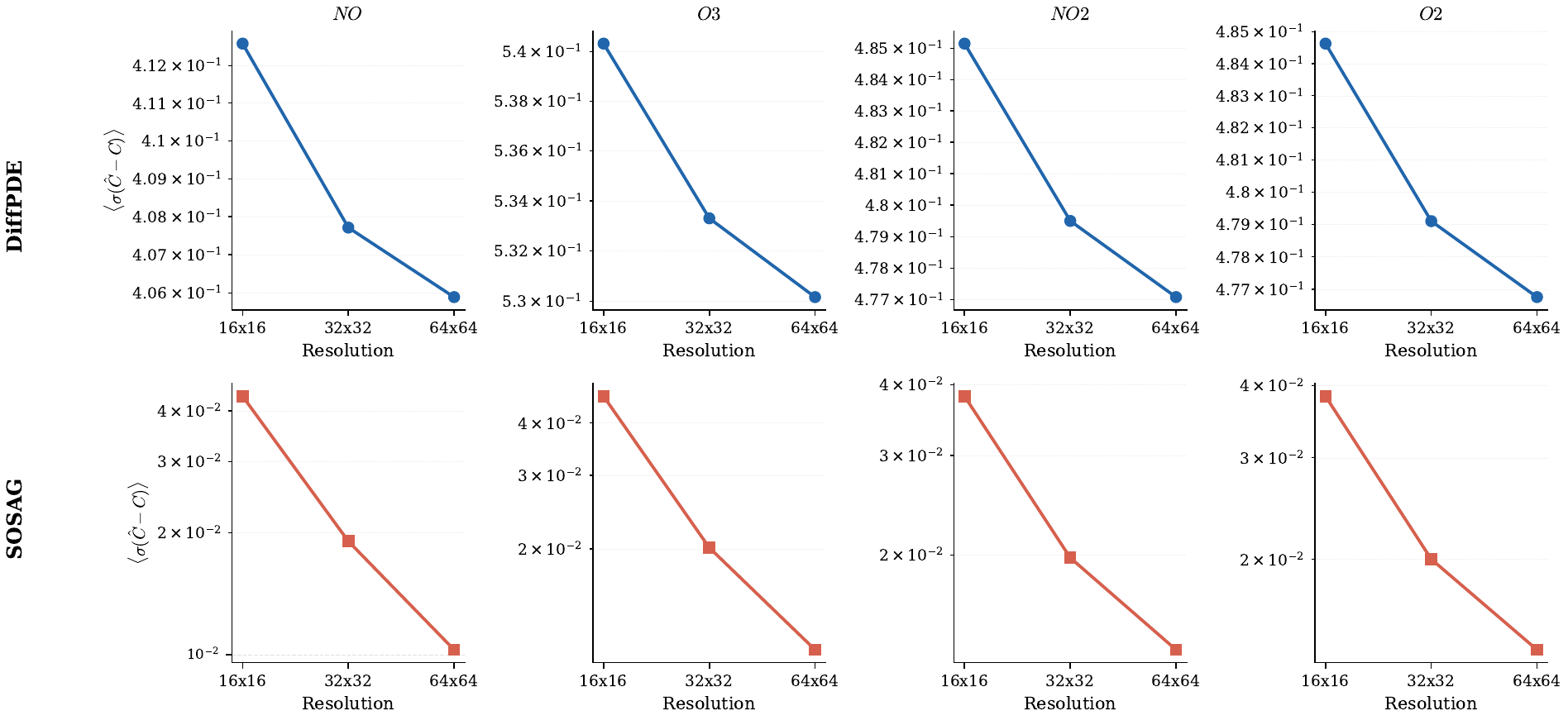}
    \caption{Line plots of the standard deviation of reconstruction errors for the DiffPDE and SOSaG methods on the NO-O$_3$ to NO$_2$ dataset for different resolutions. We notice that the error decreases as the resolution increases, an expected result. The stand deviation in the error drops at a faster rate when using SOSaG compared to DiffPDE.}
    \label{fig:NO_line_map}
\end{figure*}

\end{document}